\documentclass{aa}

\usepackage{graphicx}
\usepackage{lscape}
\usepackage{subfigure}
\usepackage{natbib}
\usepackage[varg]{txfonts}
\usepackage{longtable}
\usepackage{booktabs}
\usepackage[normalem]{ulem} 
\usepackage{epstopdf}
\usepackage{xcolor}
\usepackage{soul}
\usepackage{makecell}
\usepackage{multicol,tabularx,capt-of}

\usepackage{scalerel}
\usepackage{tikz}
\usetikzlibrary{svg.path}

\definecolor{orcidlogocol}{HTML}{A6CE39}
\tikzset{
  orcidlogo/.pic={
    \fill[orcidlogocol] svg{M256,128c0,70.7-57.3,128-128,128C57.3,256,0,198.7,0,128C0,57.3,57.3,0,128,0C198.7,0,256,57.3,256,128z};
    \fill[white] svg{M86.3,186.2H70.9V79.1h15.4v48.4V186.2z}
                 svg{M108.9,79.1h41.6c39.6,0,57,28.3,57,53.6c0,27.5-21.5,53.6-56.8,53.6h-41.8V79.1z M124.3,172.4h24.5c34.9,0,42.9-26.5,42.9-39.7c0-21.5-13.7-39.7-43.7-39.7h-23.7V172.4z}
                 svg{M88.7,56.8c0,5.5-4.5,10.1-10.1,10.1c-5.6,0-10.1-4.6-10.1-10.1c0-5.6,4.5-10.1,10.1-10.1C84.2,46.7,88.7,51.3,88.7,56.8z};
  }
}

\newcommand\orcidicon[1]{\href{https://orcid.org/#1}{\mbox{\scalerel*{
\begin{tikzpicture}[yscale=-1,transform shape]
\pic{orcidlogo};
\end{tikzpicture}
}{|}}}}

\usepackage{hyperref} 
\hypersetup{
    colorlinks=true,
    citecolor=blue,
    linkcolor=blue,
    urlcolor=blue,
    }
\makeatletter
\renewcommand*\aa@pageof{, page \thepage{} of \pageref*{LastPage}}
\makeatother

\defcitealias{decicco19}{D19}
\defcitealias{decicco15}{D15}

\bibpunct{(}{)}{;}{a}{}{,} 
\usepackage{amstext} 
\definecolor{cadmiumred}{rgb}{0.89, 0.0, 0.13}

\definecolor{ste}{rgb}{0., 0.26, 0.15}

\newcommand{\orcid}[1]{\href{https://orcid.org/#1}{\includegraphics[width=8pt]{figures/orcid.png}}}

\begin{document} 
   \title{A structure function analysis of VST-COSMOS AGN\thanks{Observations were provided by the ESO programs 088.D-4013, 092.D-0370, and 094.D-0417 (PI G. Pignata).}\mbox{$^,$}\thanks{Table 6 is only available in electronic format the CDS via anonymous ftp to cdsarc.u-strasbg.fr (130.79.128.5) or via http://cdsweb.u-strasbg.fr/cgi-bin/qcat?J/A+A/}}

   \author{D. De Cicco\inst{1,2,3\orcidicon{0000-0001-7208-5101}}, 
   F. E. Bauer\inst{1,2,4\orcidicon{0000-0002-8686-8737}}, 
   M. Paolillo\inst{3,5,6\orcidicon{0000-0003-4210-7693}}, 
   P. S\'{a}nchez-S\'{a}ez\inst{7,2,1\orcidicon{0000-0003-0820-4692}}, 
   W. N. Brandt\inst{8,9,10\orcidicon{0000-0002-0167-2453}},
   F. Vagnetti\inst{11,12 \orcidicon{0000-0002-6689-9317}},
   G. Pignata\inst{13,2\orcidicon{0000-0001-6003-8877}},
   M. Radovich\inst{14\orcidicon{0000-0002-3585-866X}},
   M. Vaccari\inst{15,16,17\orcidicon{0000-0002-6748-0577}}
   }

   \titlerunning{Analysis of the structure function of VST-COSMOS AGN}
   \authorrunning{D. De Cicco et al.}

\institute{Instituto de Astrof\'{i}sica, Pontificia Universidad Cat\'{o}lica de Chile, Av. Vicu\~{n}a    Mackenna 4860, 7820436 Macul, Santiago, Chile    
\\e-mail: demetradecicco@gmail.com, demetra.decicco@unina.it
\and
Millennium Institute of Astrophysics (MAS), Nuncio Monse\~nor Sotero Sanz 100, Of. 104, Providencia, Santiago, Chile     
\and
Dipartimento di Fisica, Università degli Studi di Napoli ``Federico II'', via Cinthia 9, 80126 Napoli, Italy 
\and
Space Science Institute, 4750 Walnut Street, Suite 2015, Boulder, CO 80301, USA 
\and
INFN - Sezione di Napoli, via Cinthia 9, 80126 Napoli, Italy 
\and
INAF - Osservatorio Astronomico di Capodimonte, via Moiariello 16, 80131 Napoli, Italy 
\and
European Southern Observatory, Karl-Schwarzschild-Strasse 2, 85748 Garching bei München, Germany 
\and
Department of Astronomy and Astrophysics, 525 Davey Laboratory, The Pennsylvania State University, University Park, PA 16802, USA 
\and
Institute for Gravitation and the Cosmos, The Pennsylvania State University, University Park, PA 16802, USA	
\and
Department of Physics, 104 Davey Laboratory, The Pennsylvania State University, University Park, PA 16802, USA 
\and
Dipartimento di Fisica, Università di Roma ``Tor Vergata'', via della Ricerca Scientifica 1, 00133 Roma, Italy 
\and
INAF - Istituto di Astrofisica e Planetologia Spaziali, Via del Fosso del Caveliere 100, I-00133 Roma, Italy 
\and
Departamento de Ciencias Fisicas, Universidad Andres Bello, Avda. Republica 252, Santiago, Chile 
\and
INAF - Osservatorio Astronomico di Padova, vicolo dell'Osservatorio 5, I-35122 Padova, Italy 
\and
Inter-university Institute for Data Intensive Astronomy, Department of Physics and Astronomy, University of the Western Cape, 7535 Bellville, Cape Town, South Africa 
\and
Inter-university Institute for Data Intensive Astronomy, Department of Astronomy, University of Cape Town, 7701 Rondebosch, Cape Town, South Africa 
\and
INAF - Istituto di Radioastronomia, via Gobetti 101, 40129 Bologna, Italy 
}

   \date{}
  \abstract
   {We present our sixth work in a series dedicated to variability studies of active galactic nuclei (AGN) based on the survey of the COSMOS field by the VLT Survey Telescope (VST). Its 54 $r$-band visits over 3.3 yr and single-visit depth of 24.6 \emph{r}-band mag make this dataset a valuable scaled-down version that can help forecast the performance of the Rubin Observatory Legacy Survey of Space and Time (LSST).}
   {This work is centered on the analysis of the structure function (SF) of VST-COSMOS AGN, investigating possible differences in its shape and slope related to how the AGN were selected, and explores possible connections between the ensemble variability of AGN and black-hole mass, accretion rate, bolometric luminosity, redshift, and obscuration of the source. Given its features, our dataset opens up the exploration of samples $\sim2$ mag fainter than most of the literature to date.}
   {We identify several samples of AGN - 677 in total - obtained by a variety of selection techniques which partly overlap. Our analysis compares results for the various samples. We split each sample in two based on the median of the physical property of interest, and analyze differences in the shape and slope of the SF, and possible causes.}
   {While the shape of the SF does not change with depth, it is highly affected by the type of AGN (unobscured/obscured) included in the sample. Where a linear region can be identified, we find that the variability amplitude anticorrelates with accretion rate and bolometric luminosity, consistent with previous literature on the topic, while no dependence on black-hole mass emerges from this study. With its longer baseline and denser and more regular sampling, the LSST will allow an improved characterization of the SF and its dependencies on the mentioned physical properties over much larger AGN samples.}
   {}

   
   \maketitle

\section{Introduction}
\label{section:intro}
Variability is a characteristic of active galactic nuclei (AGN) at all wavelengths, observed in the continuum emission as well as in the emission lines. Because of its universal character, it can be used as a selection criterion for AGN; in particular, the search for AGN in multivisit surveys via optical variability has been the subject of a large number of studies over the past decades \citep[e.g.,][and references therein]{macleod12,decicco15,falocco,simm,paula18,sartori,decicco19,poulain,decicco21}. This variability is stochastic and aperiodic, as suggested by their featureless power spectra \citep[e.g.,][]{kelly09}. Optical emission from AGN is generally thought to originate from the accretion disk surrounding the central supermassive black hole (SMBH), and the corresponding variations are most commonly thought to originate from instabilities in the accretion disk and changes in the accretion rate \citep[e.g.,][]{shakura_sunyaev}, though this has been somewhat controversial over the last decades \citep[e.g.,][]{a&t,kawaguchi,hawkins,kimura}. As a consequence, AGN monitoring campaigns are a powerful tool to investigate deeper into the topic. 

One way to quantitatively describe the time dependence of the variability of a sample of AGN is via the structure function (SF; e.g., \citealt{Simonetti, diClemente, kawaguchi, hawkins, deVries05, Bauer, Graham, kozlowski}, and references therein). This is built from the light curves of the sample of sources and is basically a measure of the ensemble rms magnitude difference as a function of the time lag between different visits, where measurements are typically grouped together into bins. Such a statistical approach allows handling large samples of objects and characterization of their overall variability and average properties, and its strength lies in the mutual independence of the different bins over which we measure variability, which does not hold for individual sources \citep[e.g.,][]{deVries05}. 

While the physical process driving the observed red-noise variability is still unknown, the damped random walk (DRW; e.g., \citealt{kelly09}) model has proven to be an effective characterization of AGN light curves. The DRW describes AGN light curves as a stochastic process via an exponential covariance matrix defined by two parameters: a variability amplitude and a characteristic damping timescale. For short timescales the random walk provides a good description of the process, while for longer timescales the variations are damped and tend to an asymptotic amplitude \citep[e.g.,][]{ivezic13,zu}. Some works from the literature \citep[e.g.,][]{macleod10} found the DRW parameters to correlate with properties such as wavelength, AGN luminosity, and black-hole (BH) mass, and speculate about the damping timescale being related to the thermal timescale of the AGN accretion disk \citep[e.g.,][]{kozlowski17,burke}. Nevertheless, as a model DRW is limited by its inability to constrain the underlying physical process, and the fact that the same variability can be described effectively also by other models \citep[e.g.,][]{kozlowski}. In addition, the use of DRW to model AGN light curves is affected by some limitations due to the data, as detailed in \citet{kozlowski17} where, basically, the author proves that the sampled timescales should be at least ten times the damping timescale (1 yr, leading to $> 10$ yr rest-frame length for the light curves) in order to allow a proper modeling and recover accurate DRW parameters. These results are also confirmed by \citet{suberlak}, who investigate the impact of the ratio of the timescale and the light curve baseline by means of simulated light curves, and find that, as this ratio decreases, it allows recovery of unbiased DRW parameters, thus not affecting the analysis of the relations with physical properties of the AGN.

Here we present an analysis of the SF of AGN centered on the 1 sq. deg. area of the Cosmic Evolution Survey (COSMOS; \citealt{scoville07b}) imaged by the VLT Survey Telescope (VST; \citealt{VST}); the dataset was introduced and extensively described in \citet{decicco15} and \citet{decicco19}. Together with \citet{decicco21}, these works are part of a series of five dedicated to variability studies which are relevant in the context of performance forecasting for the Legacy Survey of Space and Time (LSST; see, e.g., \citealt{lsst,ivezich19}), which will be conducted with the Simonyi Survey Telescope at the Vera C. Rubin Observatory. The series illustrates the development of an efficient and automated methodology for the identification of optically variable AGN; in particular, \citet{decicco21} selected a sample of AGN candidates via a random forest (RF; \citealt{Breiman2001}) classifier built using optical variability, optical and near-infrared (NIR) colors, and a morphology indicator as features. A set of multiwavelength diagnostics was used for confirmation, examining the properties imparted by each selection technique. 

In this work we aim to characterize the sample of known AGN in the VST-COSMOS area, which consists of the confirmed sources that we selected via optical variability plus the samples of multiwavelength-selected AGN that we used for the validation of our selection. We therefore want to assess whether the shape of the SF changes when using different AGN samples based on observed and intrinsic properties such as flux, redshift, obscuration, with the aim of understanding their variability properties. 

Via the SF we also aim to investigate possible connections between AGN variability -- in terms of SF amplitude and slope -- and bolometric luminosity $L_{bol}$, BH mass $M_{BH}$, and accretion rate -- as quantified by the Eddington ratio $\lambda_E$ -- onto the SMBH. The existence of these connections has been discussed by several studies in the last decades, especially in the X-rays, which map the AGN inner regions \citep[e.g.,][and references therein]{mcHardy,gonzalez-martin, vagnetti16,paolillo17}. If we manage to determine the connection between AGN variability and $L_{bol}$, $M_{BH}$, and $\lambda_E$ of the central SMBH, we can use the first to characterize the latter, especially in the context of the wide-field surveys to come, such as the LSST. This approach is model independent and helpful when the sampled timescales are too short to adopt, e.g., a DRW-based approach.

A large number of relatively recent studies have been dedicated to this inquiry, but the investigation of the potential relations between variability and $M_{BH}$ remains controversial, as different studies arrive at conflicting conclusions. For example, \citet{macleod10} modeled the optical/UV variability of a sample of $\approx9,000$ quasars in the Stripe 82, adopting a damped random walk model and making use of light curves in the $ugriz$ bands from the Sloan Digital Sky Survey (SDSS; \citealt{sdss}) spanning 10 yr. They investigated possible correlations between two temporal or variability-related parameters -- a characteristic damping timescale and the asymptotic rms variability -- and some physical parameters, and found that both correlate with $M_{BH}$. \citet{wilhite} obtained similar results from their analysis of the ensemble variability of $\approx8,000$ quasars in the SDSS Equatorial Stripe. In this case they focused on the variability amplitude via the SF of their sources. Conversely, \citet{kelly13} analyzed correlations of optical and X-ray variability (via power spectral density, PSD) with luminosity, $M_{BH}$, and $\lambda_E$ for a sample of 39 AGN, and found an anticorrelation with $M_{BH}$ for both wavebands, with a larger scatter characterizing the relation with optical variability amplitude. Meanwhile, \citet{simm} searched for correlations for AGN in the \emph{XMM}-COSMOS catalog from \citet{Brusa}, with AGN optical variability quantified by the excess variance, finding no dependence on $M_{BH}$. \citet{caplar} resorted to characterization using the SF and PSD to analyze optical variability in quasars from the (intermediate) Palomar Transient Factory ([i]PTF) and its dependence on the physical parameters of the central BH, and found a weak correlation of the variability amplitude with $M_{BH}$.

Many of these same works point to $\lambda_E$ as the main driver for AGN variability: \citet{simm} indeed found an anticorrelation with this quantity; \citet{paula18} used data from the QUEST-La Silla AGN variability survey to show that the variability amplitude anticorrelates with $\lambda_E$; and \citet{wilhite}, \citet{macleod10}, and \citet{kelly13} additionally support the existence of this anticorrelation. \citet{wilhite}, \citet{macleod10}, \citet{kelly13} also find an anticorrelation between AGN variability and luminosity, as well as \citet{simm}, who resort to excess variance (EV) and PSD. \citet{laurenti} investigate the optical and X-ray variability properties in a sample of 795 AGN from the Multi-Epoch \emph{XMM} Serendipitous AGN Sample 2 (MEXAS2; \citealt{serafinelli17}, making use of data from the Catalina Surveys Data Release 2 (CSDR2\footnote{\url{http://nesssi.cacr.caltech.edu/DataRelease/}}), and find a strong anticorrelation with luminosity, a weaker anticorrelation with $\lambda_E$, and a modest correlation with $M_{BH}$.

A key novelty of this work is the depth of the analyzed sample, which probes to $r \simeq 23.5$ mag, thus opening exploration of more frequently observed AGN, compared to typically $1.5-2$ mag brighter samples in previous studies.  A relevant feature is that the single-visit depth of VST-COSMOS $r$-band images is $\approx 24.6$ mag for point sources, at a 5$\sigma$ confidence level, hence roughly the same as the LSST typical single-visit depth of 24.7 mag for the same band.

This paper is organized as follows: Section \ref{section:samples} introduces the samples of AGN selected for our analysis on the basis of different properties, while Section \ref{section:sf_analysis}  describes the definition of the SF adopted in this work and presents the SF obtained for the different samples of AGN; Section \ref{section:phys} investigates the connection between optical variability and $M_{BH}$, $\lambda_E$, $L_{bol}$, redshift, and absorption; we summarize our main findings in Section \ref{section:conclusions}. Throughout this paper we will refer to optical variability even when omitting the word ``optical'', unless otherwise stated.  

\section{The sample of VST-COSMOS AGN}
\label{section:samples}
The VST-COSMOS dataset used in this work consists of 54 visits surveying a $\approx1$ sq. deg. area in the $r$ band, spanning a 3.3 yr baseline and spread over three observing seasons, with two gaps. Table \ref{tab:seasons} reports information about the observed (i.e., not redshift-corrected) baseline and the number of visits for the three observing seasons, with basic statistics for the sample of sources analyzed. We refer the reader to \citet{decicco15,decicco19,decicco21} for further details. Magnitudes are in the AB system. 

\begin{table*}[htb]
\caption{Basic statistical information about the three observing seasons of data used in this work.
}\label{tab:seasons}
\begin{center}
\footnotesize
\resizebox{\textwidth}{!}{\begin{tabular}{l c c c c c c c c c}
\toprule
\ & \thead{duration} & \thead{obs. baseline (d)} & \thead{mean obs. \\ baseline (d)} & \thead{median obs. \\ baseline (d)} & \thead{tot. $\#$ \\ of visits} & \thead{mean $\#$ \\ of visits} & \thead{median $\#$\ \\ of visits} & \thead{mean \\ sampling \\ rate (d)} & \thead{median \\ sampling \\ rate (d)}\\
\midrule
\ season 1 & \makecell{Dec. 2011--\\ May 2012} & 151 & 150 & 151 & 26 & 24 & 25 & 6.04 & 3 \\
\ season 2 & \makecell{Dec. 2013--\\ Apr 2014} & 101 & 101 & 101 & 20 & 19 & 20 & 5.32 & 3 \\
\ season 3 & \makecell{Dec. 2014--\\ Mar 2015} & 106 & 103 & 106 &  8 &  7 &  8 & 15.14 & 15 \\
\ seasons 1+2+3 & \makecell{Dec. 2011--\\Mar 2015} & 1,187 & 1,185 & 1,187 & 54 & 51 & 52 & 22.40 & 4 \\
\bottomrule
\end{tabular}}
\end{center}
\footnotesize{\textbf{Notes.} The table reports (\emph{left to right}): maximum observed baseline; mean and median value of the observed baseline for the 677 sources in the \emph{main} sample; total number of visits; mean and median number of visits for the 677 sources in the \emph{main} sample; mean and median sampling rate.}
\end{table*}

\begin{figure*}[t!]
 \centering
\subfigure
            {\includegraphics[width=5cm]{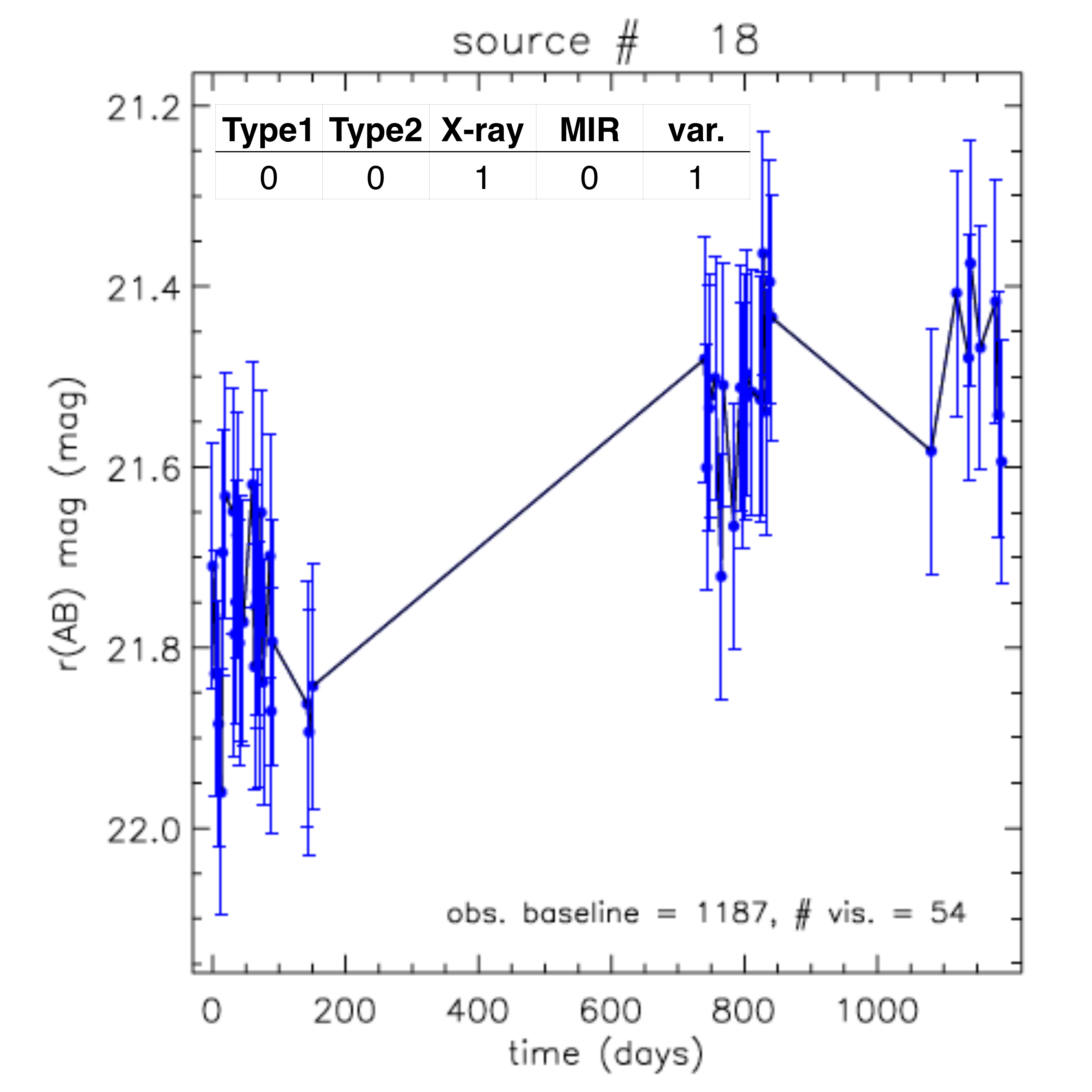}}
\subfigure
            {\includegraphics[width=5cm]{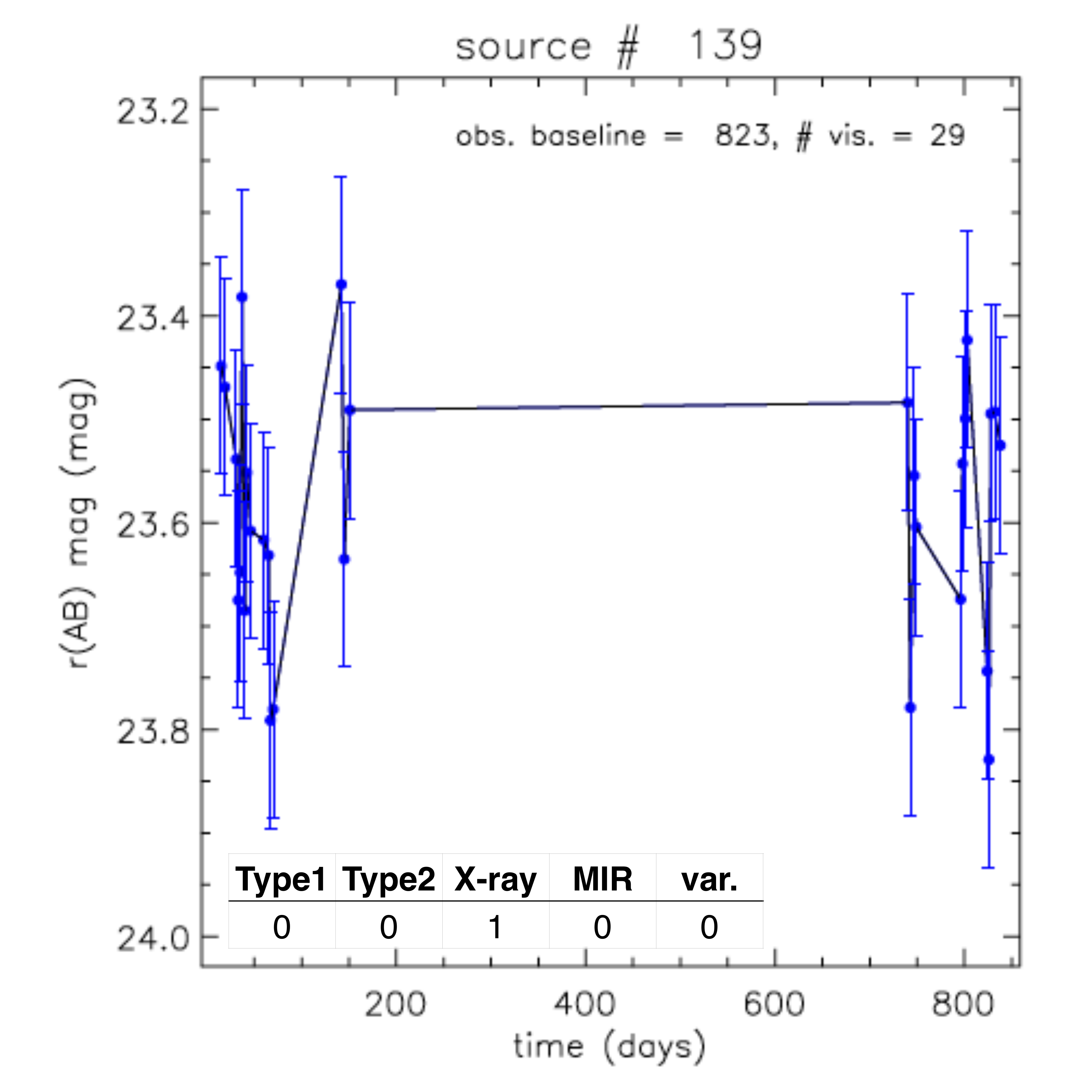}} 
\subfigure
            {\includegraphics[width=5cm]{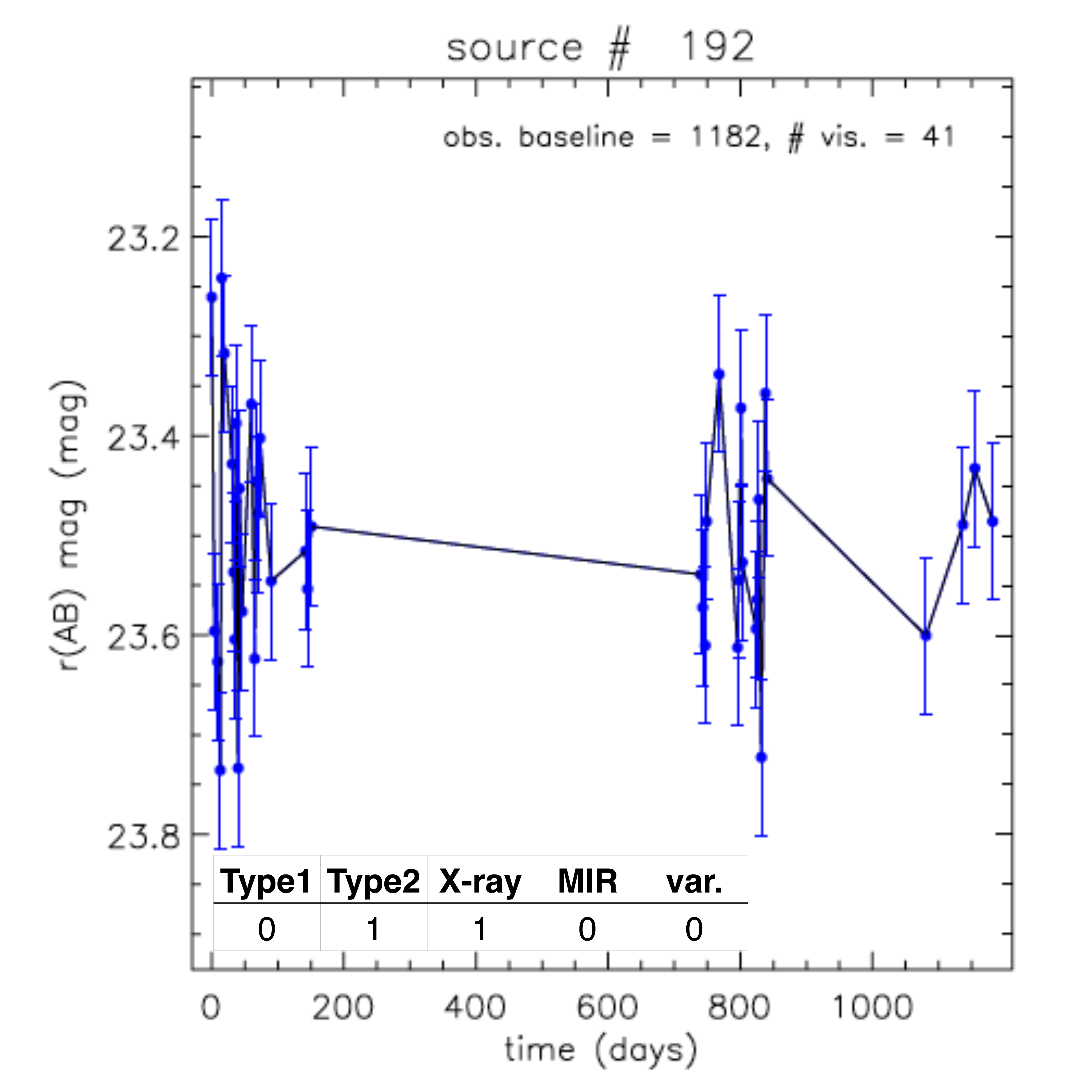}} \\
\subfigure
            {\includegraphics[width=5cm]{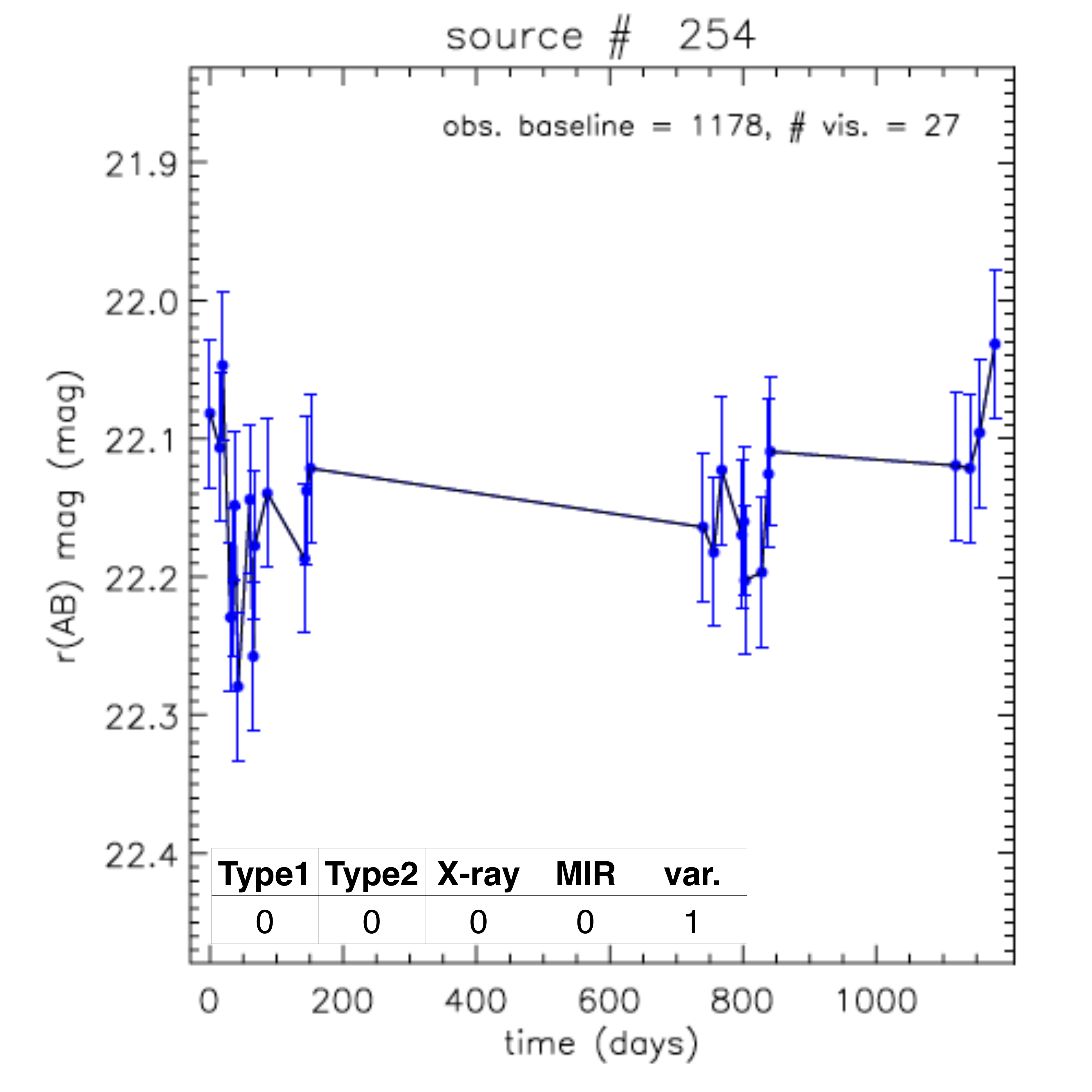}} 
\subfigure
            {\includegraphics[width=5cm]{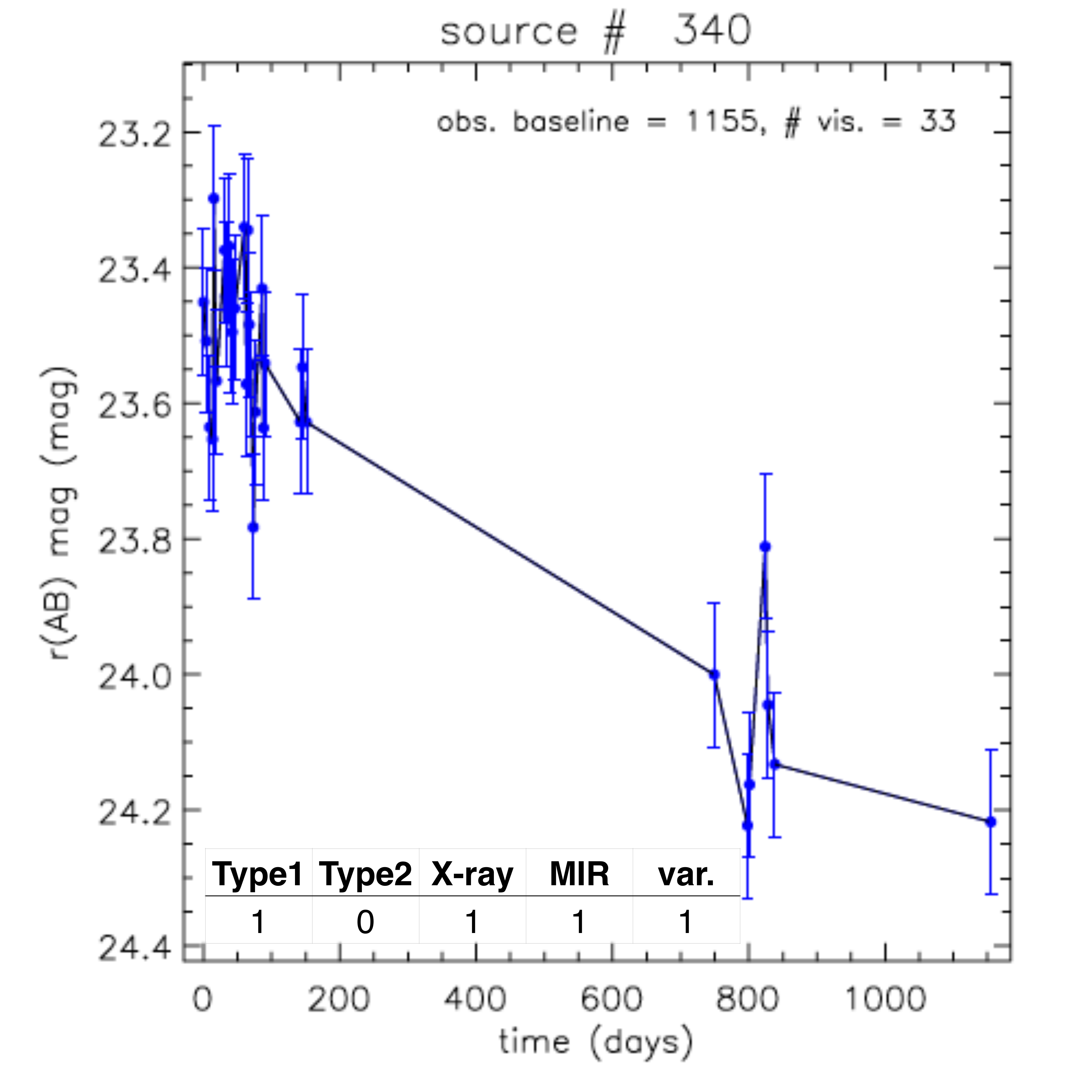}} 
\subfigure
            {\includegraphics[width=5cm]{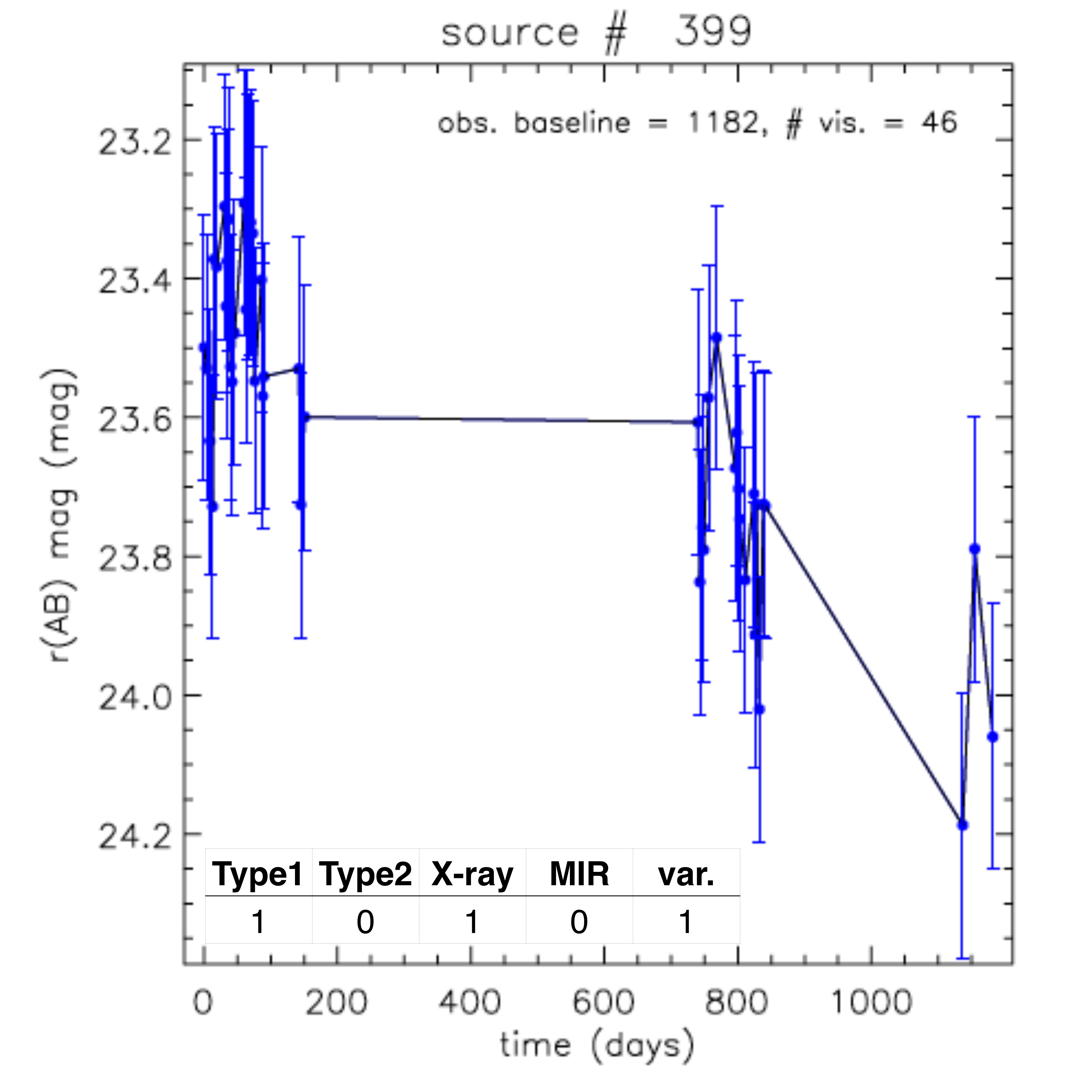}} \\
\subfigure
            {\includegraphics[width=5cm]{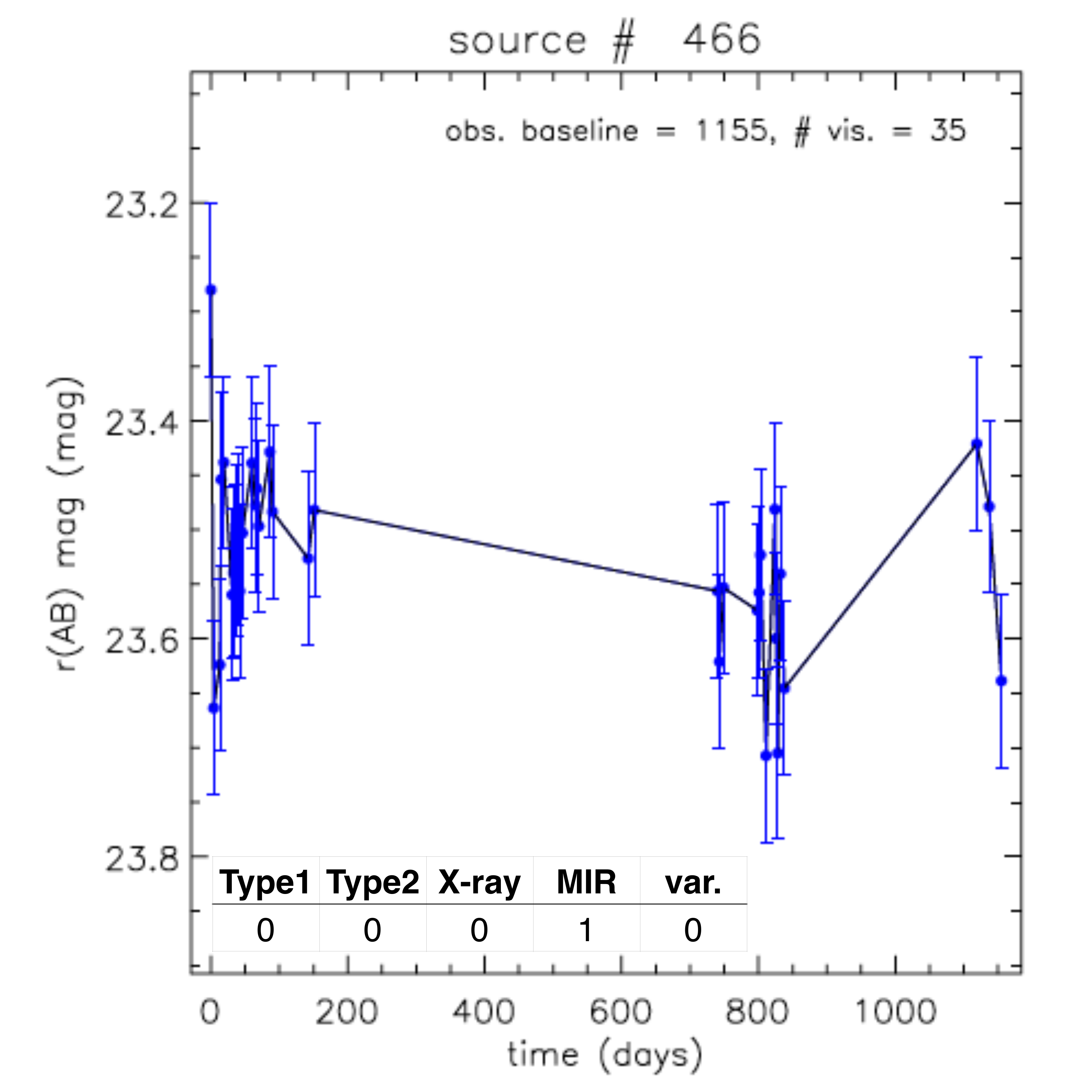}} 
\subfigure
            {\includegraphics[width=5cm]{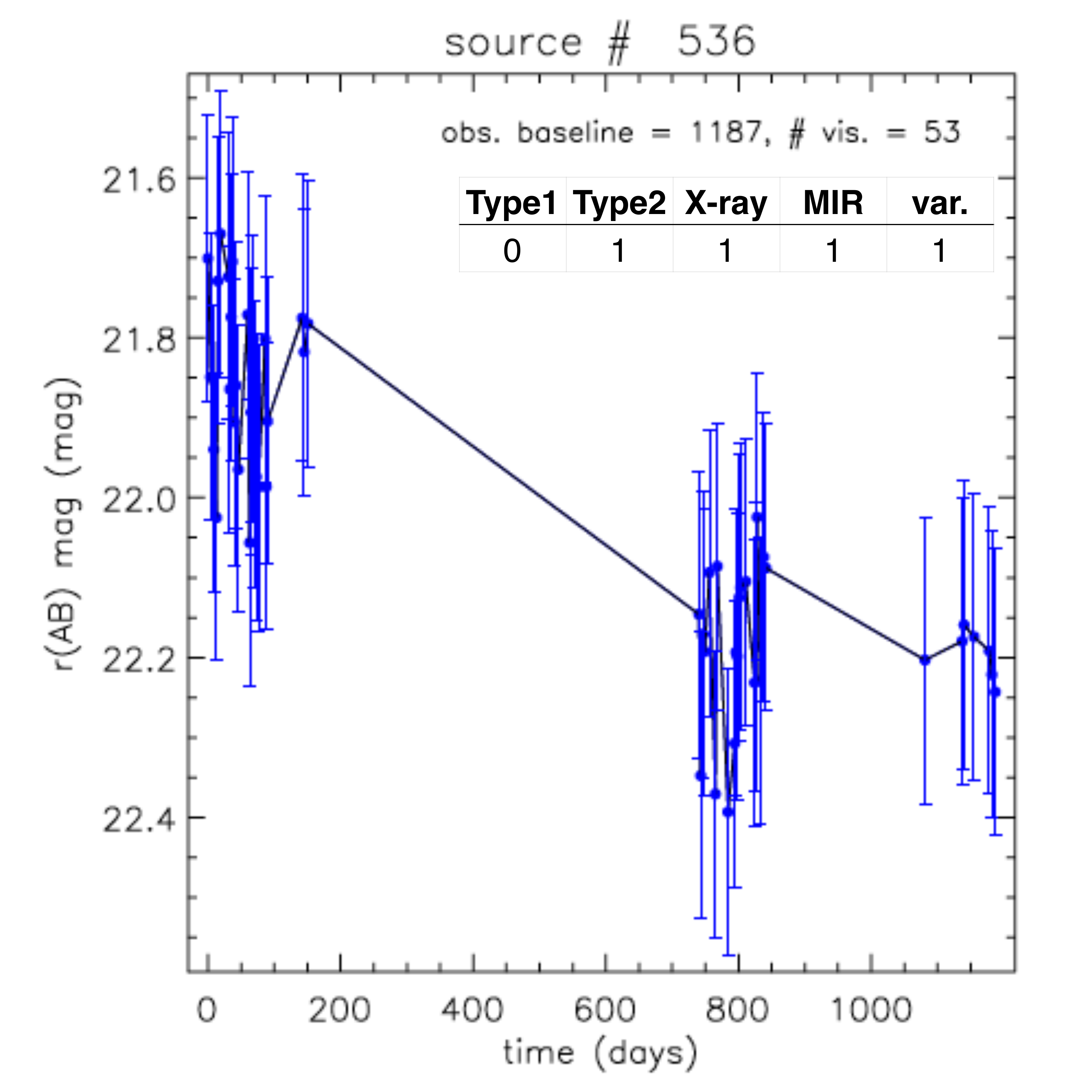}}
\subfigure
            {\includegraphics[width=5cm]{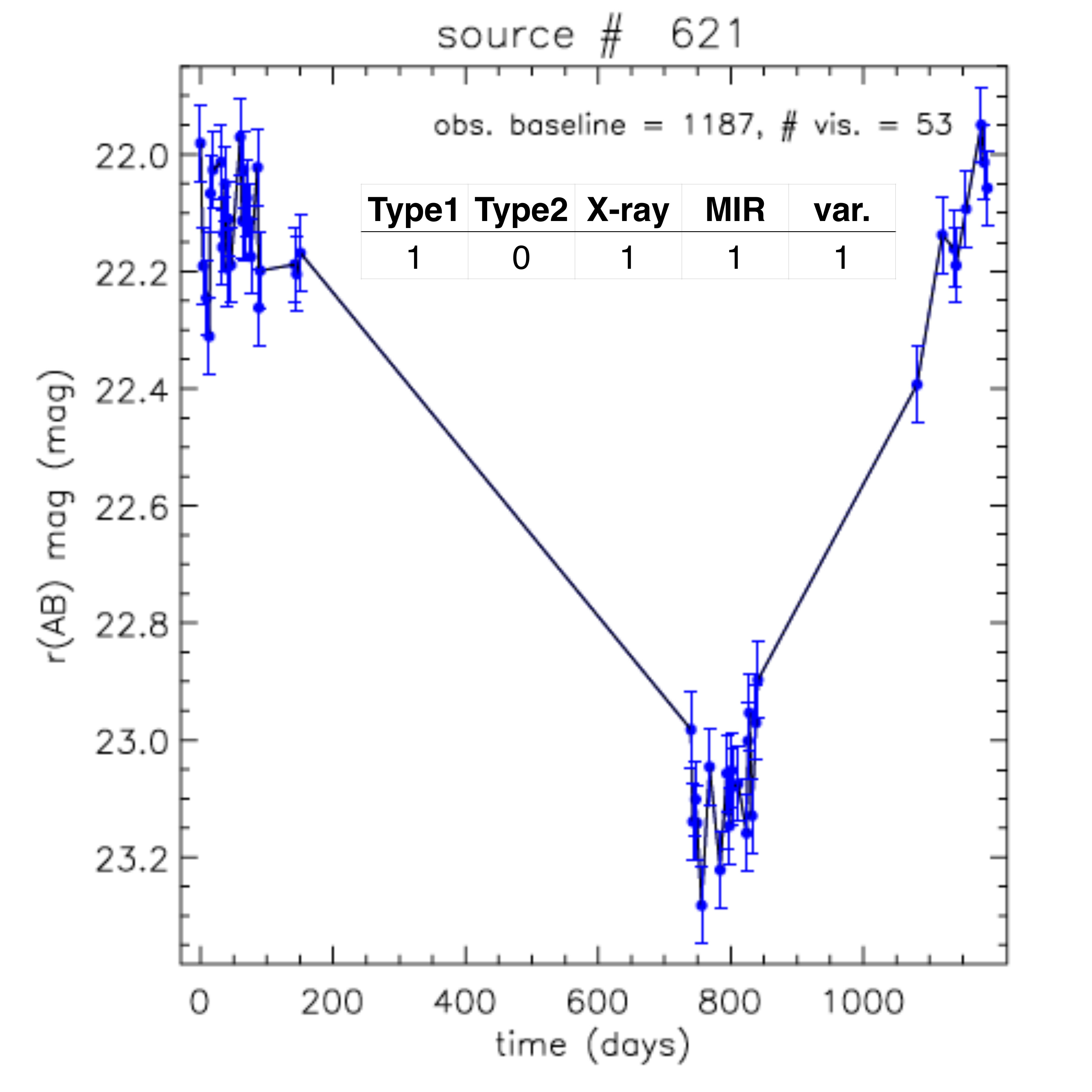}}
   \caption{A selection of light curves from the \emph{main} sample. Source n. 18 is an example of the most commonly observed ones in this sample, where both the length of the observed baseline and the number of visits coincide with the corresponding maxima (see Table \ref{tab:seasons}); source n. 139 is the one with the shortest baseline, while source 254 is one of the two detected in 27 visits (i.e., the minimum number of visits based on our selection threshold; see main text); the other instances correspond to other combinations of observed baseline and number of visits. We note that the light curve of source n. 621 is characterized by the largest magnitude variation over the analyzed baseline. The table in each panel indicates which subsample(s) the corresponding source belongs/does not belong to (1/0).}\label{fig:lc}
   \end{figure*}

Since COSMOS is a widely studied area, a large number of sources are already known to be AGN based on a number of properties and diagnostics. Specifically, limited to the COSMOS area surveyed by the VST, we identified 677 sources confirmed as AGN (hereafter, \emph{main} sample) on the basis of at least one of the following: optical spectroscopy \citep{marchesi}; X-ray to optical flux ratio \citep{Maccacaro}; mid-infrared (MIR) properties \citep{donley}. These diagnostics have been widely used in \citet{decicco15,decicco19,decicco21} to validate the samples of AGN candidates selected via optical variability, which is the remaining selection criterion here adopted. These 677 AGN have counterparts in the sample of VST-COSMOS sources used in \citet{decicco21}, which includes sources with a magnitude $r \leq 23.5$ mag, and a redshift estimate, either spectroscopic (91\% of sources) or photometric (9\% of sources), available from different COSMOS catalogs. Details about the catalogs and on how we assigned a redshift to each source can be found in Sect. 2.1 of \citet{decicco21}. By construction, the catalog contains sources with a minimum of 27 points in their light curves, the maximum being defined by the total number of visits (54); specifically 97\% of the sources in our \emph{main} sample contain at least 40 points in their light curves. Figure \ref{fig:lc} shows a selection of our light curves, reporting for each one the length of the baseline, the number of visits, and the subsample(s) the corresponding source belongs to.

\begin{figure}[tbh]
 \centering
        {\includegraphics[width=\columnwidth]{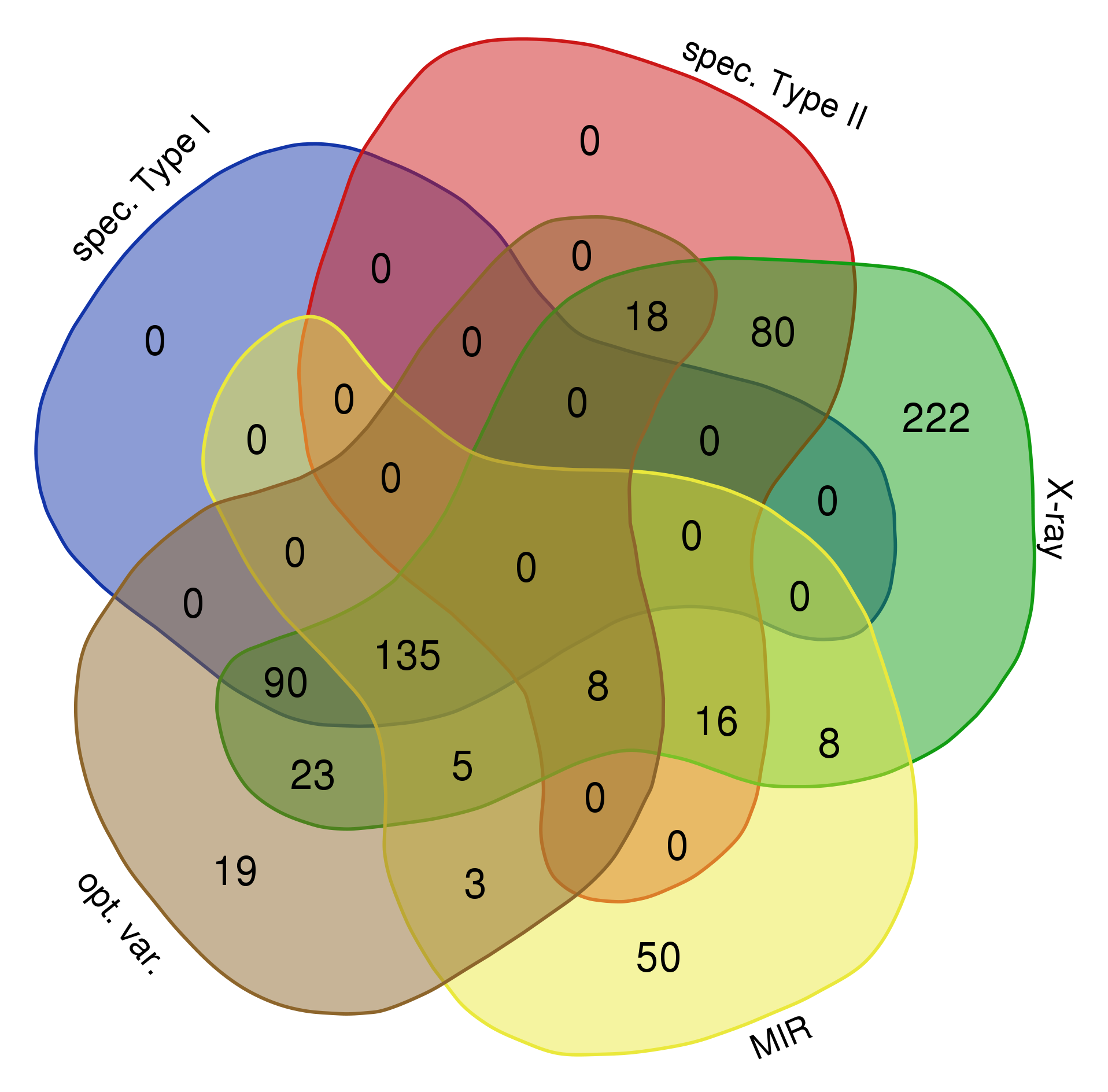}}
   \caption{Venn diagram showing: the subsamples of spectroscopically confirmed Type I and Type II AGN (spec. Type I, 225 sources, and spec. Type II, 122 sources, respectively; these two subsamples, by definition, do not overlap); the sample of sources classified as AGN on the basis of their $X/O$ diagram (X-ray, 605 sources); the subsample of AGN selected after their MIR properties (\citealt{donley}; MIR, 225 sources); and the subsample of objects classified as AGN based on their optical variability (\citealt{decicco21}; opt. var., 301 sources). We caution that the area covered by each region and the corresponding number of sources are not related. These five subsamples of sources combined together constitute the \emph{main} sample of AGN used in this study. This diagram was rendered via \url{http://bioinformatics.psb.ugent.be/webtools/Venn/}.}\label{fig:venn}
   \end{figure}

In what follows we analyze the SF of the main AGN sample and also of five subsamples, which partly overlap each other, with the aim of investigating the dependence on the selection criteria and the resulting variability properties\footnote{Hereafter we will usually refer to these five subsamples including the prefix ``sub'', to stress that they are part of our \emph{main} sample. We will omit the prefix when talking about all six samples together or generically, as in that case we will also be including the \emph{main} sample, which is not a subsample.}. In particular, we will focus on: X-ray selected AGN; Type I and Type II AGN subsamples, defined after the properties of their optical spectra reported in the X-ray catalogs by \citet{marchesi,Brusa}\footnote{In \citet{marchesi} sources are classified as BLAGN if their spectra exhibit at least one broad (FWHM $> 2000$ km s$^{-1}$) emission line, while sources labeled as non-BLAGN could be NLAGN or star-forming galaxies: this can be due to low S/N spectra, or to the lack of disentangling diagnostics based on emission lines in the waveband in which the spectra are obtained. In this case we crossmatch the classification with the one provided in the \emph{XMM}-COSMOS catalog by \citet{Brusa}, where sources are classified as BLAGN, NLAGN, and inactive galaxies. Here the criterion identifying BLAGN is the same as in \citet{marchesi}; the spectra of sources flagged as NLAGN are typically characterized by unresolved high-ionization emission lines with line ratios suggesting AGN activity, while inactive galaxy spectra are generally consistent with those of star-forming or normal galaxies and, when detected in the hard X-rays, they generally have rest-frame luminosity $L_X<2\times10^{42}$ erg\,s$^{-1}$ in that band.}; MIR selected AGN; and AGN identified from their optical variability. Specifically, this last subsample is the one obtained from the analysis presented in \citet{decicco21}. The Venn diagram in Fig. \ref{fig:venn} provides a visual assessment of the overlap among the five subsamples of AGN constituting our \emph{main} sample. We stress that, since the subsamples of spectroscopically confirmed Type I and Type II AGN are drawn from X-ray catalogs (see above), we have X-ray information for each of these sources and hence they are completely included in the X-ray subsample. We also note that the subsample of spectroscopic Type I sources is completely included in the subsample of optically variable AGN as well, since optical variability is highly biased towards this type of AGN \citep[e.g.,][and references therein]{padovani}, and our selection method based on optical variability allowed to retrieve all the sources in the Type I subsample \citep{decicco21}.

In order to unearth possible correlations between variability and some physical SMBH properties in our sample of AGN, we select those sources in the main AGN sample for which an estimate of $M_{BH}$ and $\lambda_E$ are known. We make use of estimates for these quantities from several works, ordered here according to the preference with which they were used:
\begin{itemize}
    \item \citet{lusso12}, where X-ray selected Type I and Type II AGN in COSMOS are investigated; $M_{BH}$ estimates for Type I AGN are virial and were obtained from Mg {\scriptsize{II}} or H{$\beta$} lines (81 matches; $0.25-0.4$ dex uncertainties, depending on the line), while for Type II AGN (106 matches; 0.5 dex average uncertainty) they were obtained via scaling relations, and uncertainties on the estimates of stellar masses, bolometric luminosities, and on the intrinsic scatter in the relation between $M_{BH}$ and stellar mass were taken into account via Monte Carlo simulations; 
    \item \citet{Schulze}, where a sample of X-ray selected Type I AGN is investigated via NIR spectroscopy and virial $M_{BH}$ estimates (80 matches) are obtained from H{$\alpha$}, H{$\beta$}, and Mg {\scriptsize{II}}; 
    \item \citet[][private communication]{rosario}, where the possible dependence of star formation rate on redshift, $M_{BH}$, nuclear $L_{bol}$ is investigated in a sample of spectroscopically observed quasars in COSMOS, out to a redshift $z \sim 2$, and virial $M_{BH}$ estimates (120 matches; 0.3 dex average uncertainty) were obtained from H{$\beta$} and Mg {\scriptsize{II}}; 
    \item \citet{rakshit}, where the spectral properties of a sample of more than 520,000 quasars from the Sloan Digital Sky Survey \citep[SDSS;][]{sdss} Data Release 14 are investigated, and single-visit virial $M_{BH}$ estimates (26 matches; uncertainties $\geq 0.4$ dex) are estimated from H{$\beta$}, Mg {\scriptsize{II}}, and C {\scriptsize{IV}} lines;
    \item \citet{paula18}, where the connection between AGN variability and some physical properties of the central SMBH are investigated, and $M_{BH}$ estimates (11 matches) are obtained via H{$\alpha$}, H{$\beta$}, Mg {\scriptsize{II}}, and C {\scriptsize{IV}} lines.
\end{itemize}
In total we find 264 AGN in the \emph{main} sample with available $M_{BH}$ and $\lambda_E$ estimates. These 264 sources include 80 with multiple estimates. In these cases we compare the various estimates available for each source in pairs, compute the difference and, for each source with more than two estimates, we obtain a mean value for this difference. We then average all the (mean) values, thus obtaining an average difference of 0.28 for $M_{BH}$ (logarithmic values) and of 0.30 for $\lambda_E$.

\begin{figure*}[t]
 \centering
\subfigure
            {\includegraphics[width=6cm]{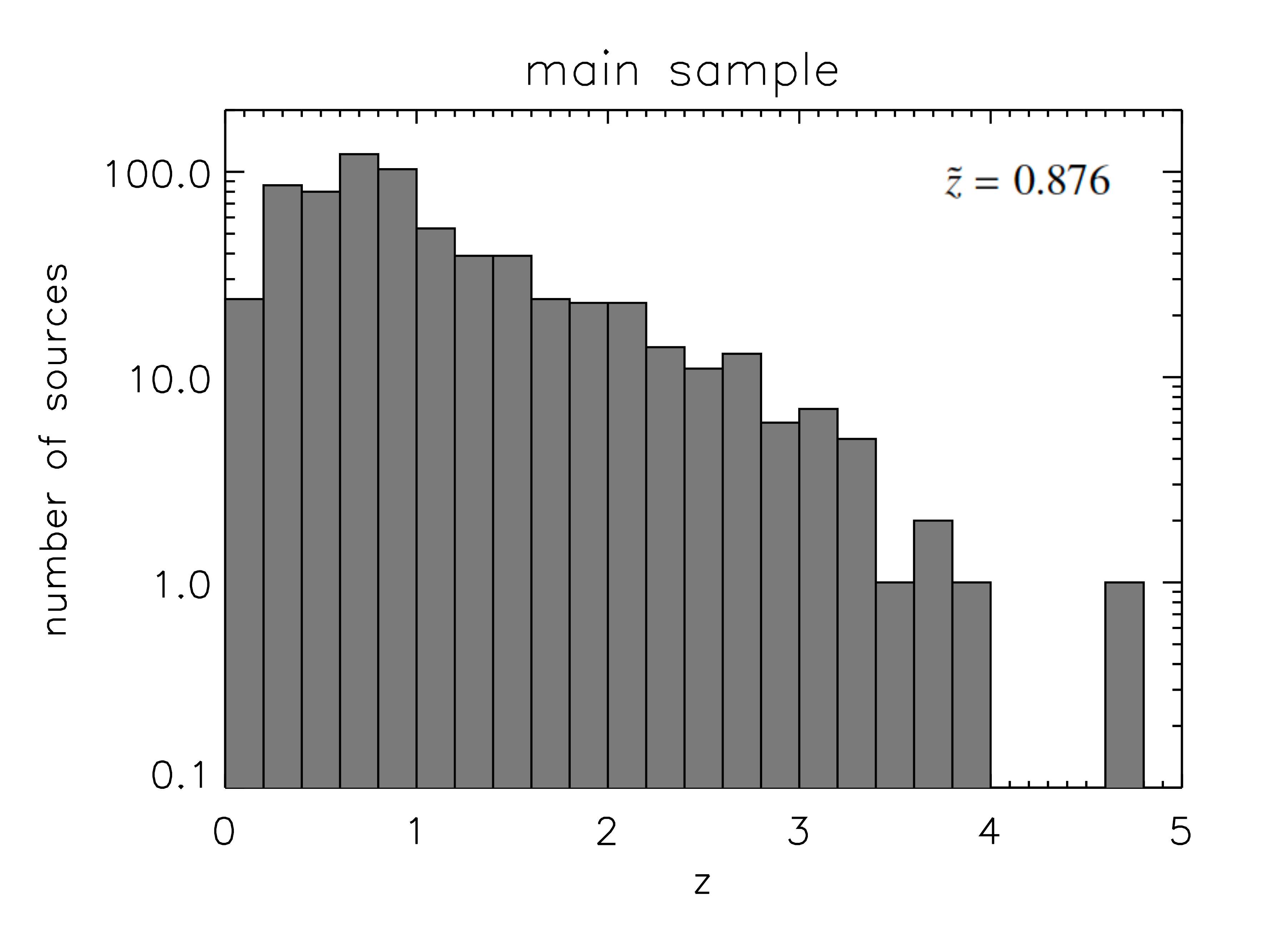}}
\subfigure
            {\includegraphics[width=6cm]{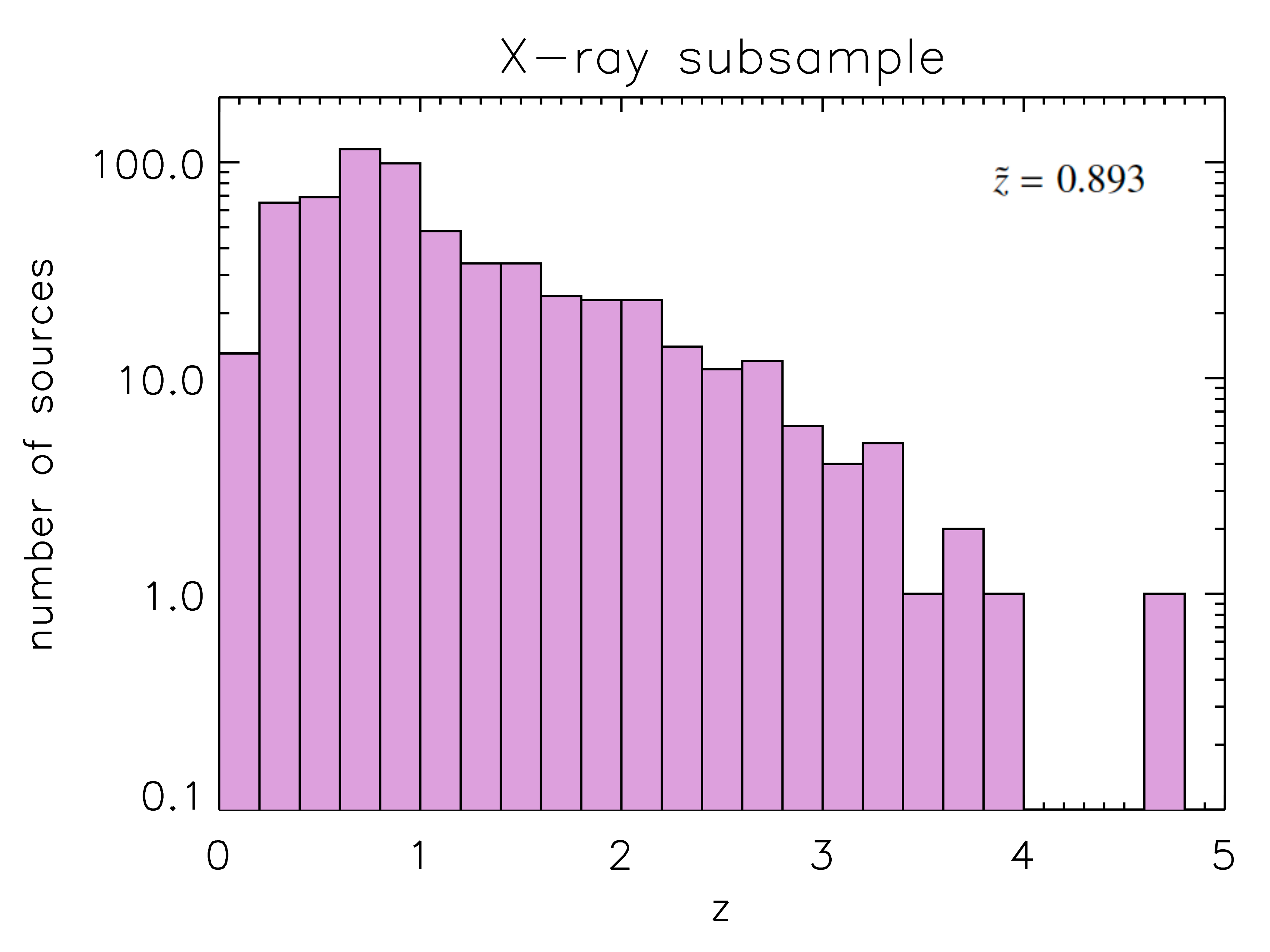}}
\subfigure
            {\includegraphics[width=6cm]{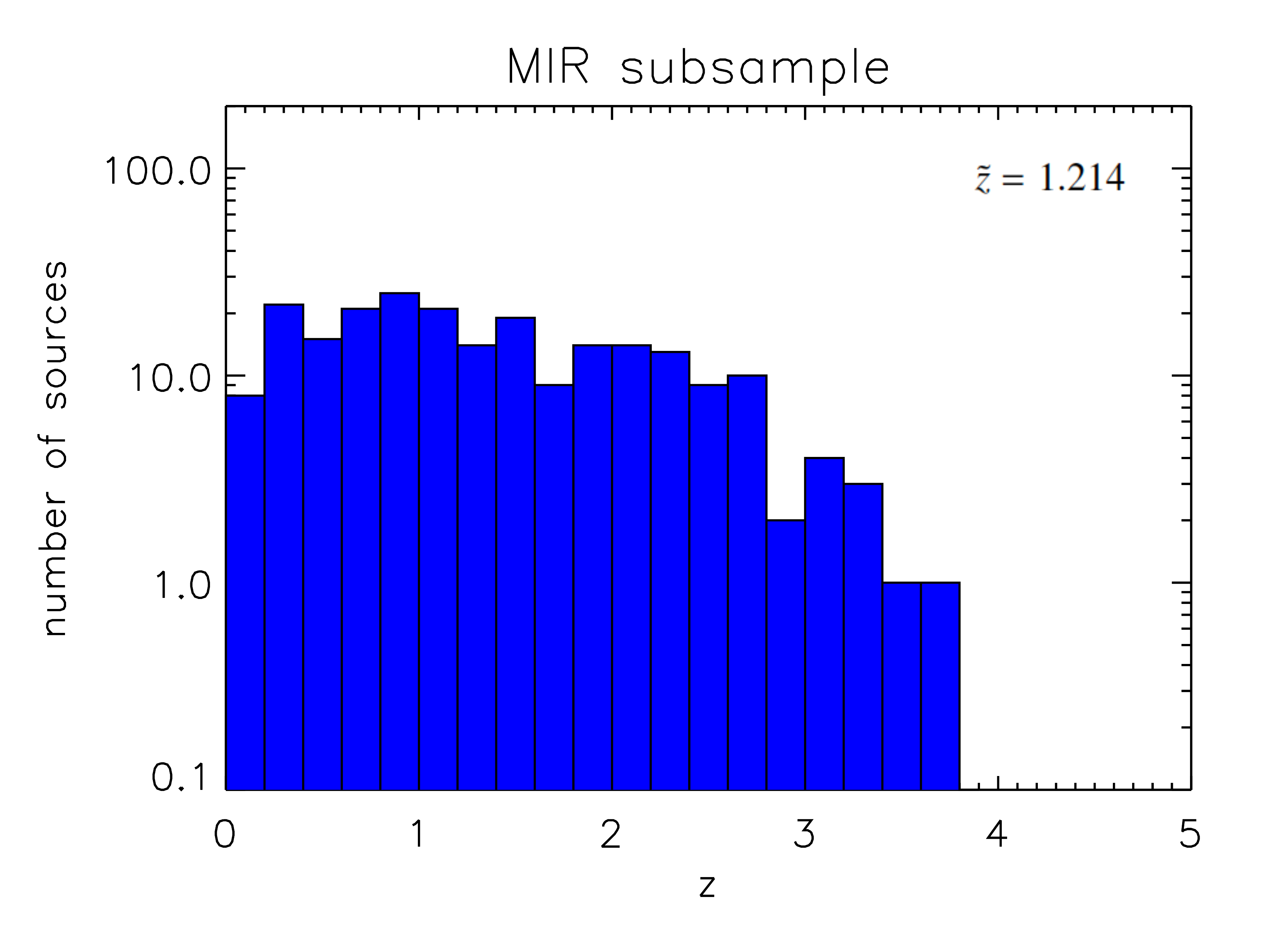}}\\
\subfigure
            {\includegraphics[width=6cm]{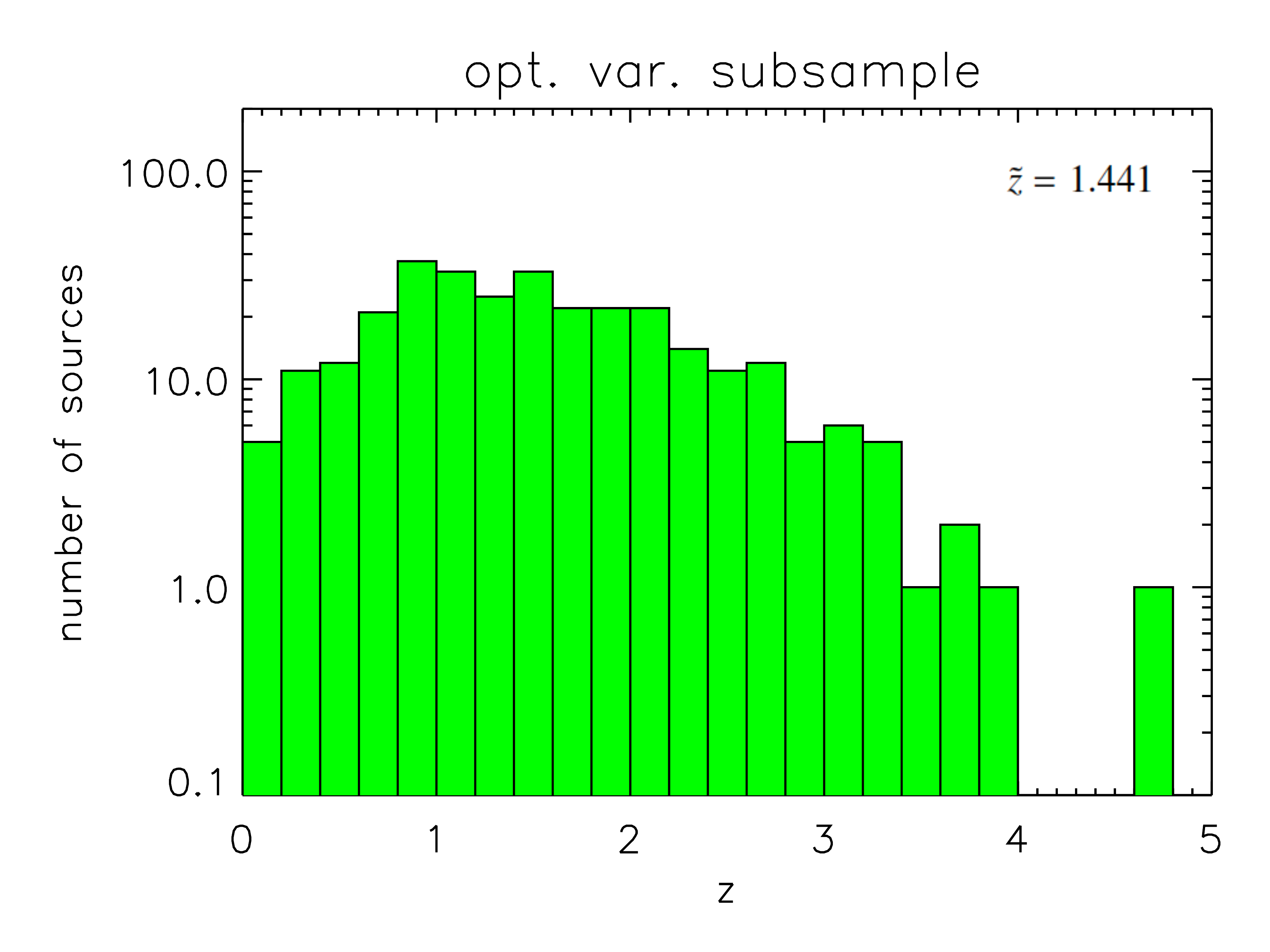}}
\subfigure
            {\includegraphics[width=6cm]{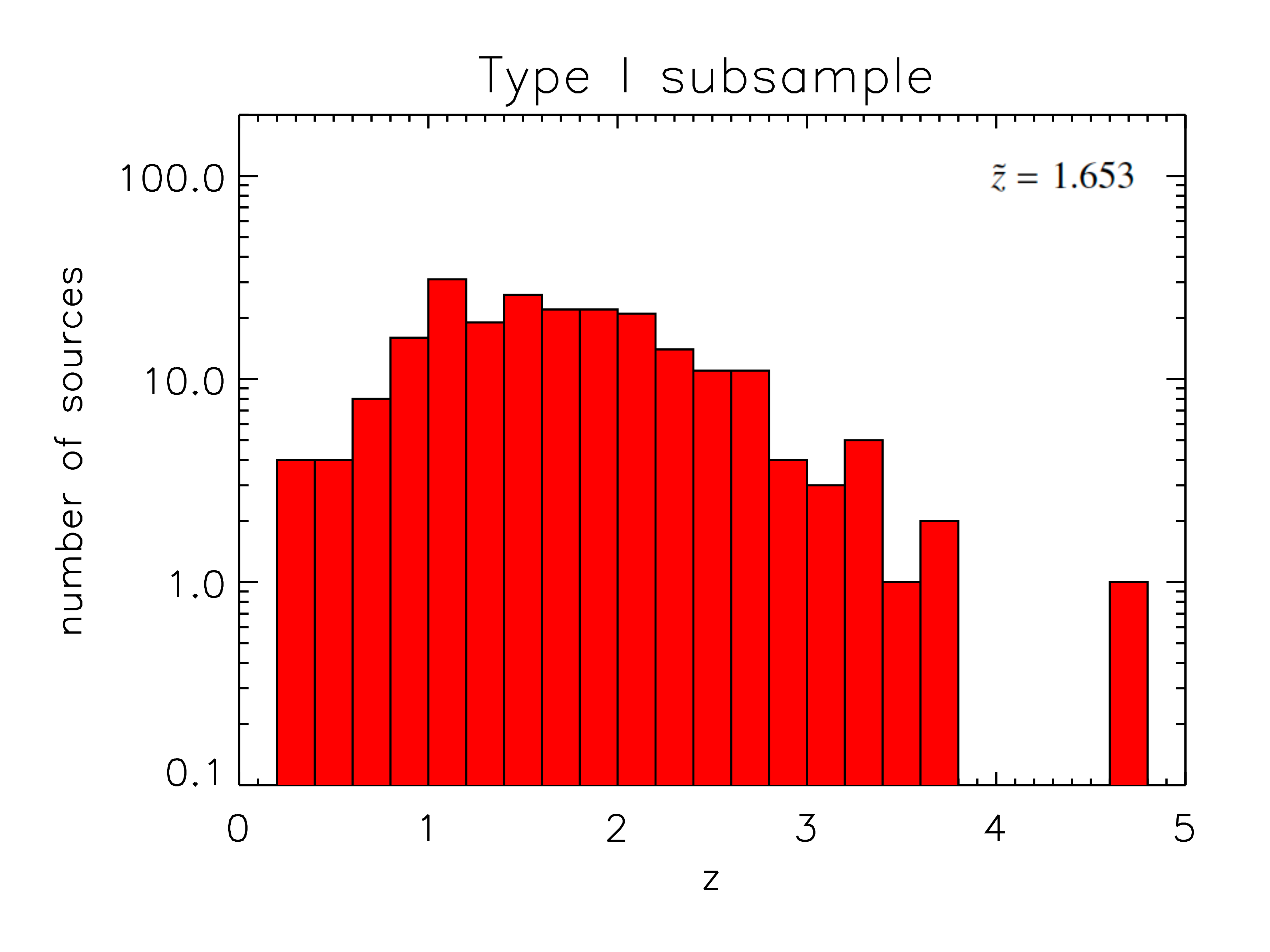}}
\subfigure
            {\includegraphics[width=6cm]{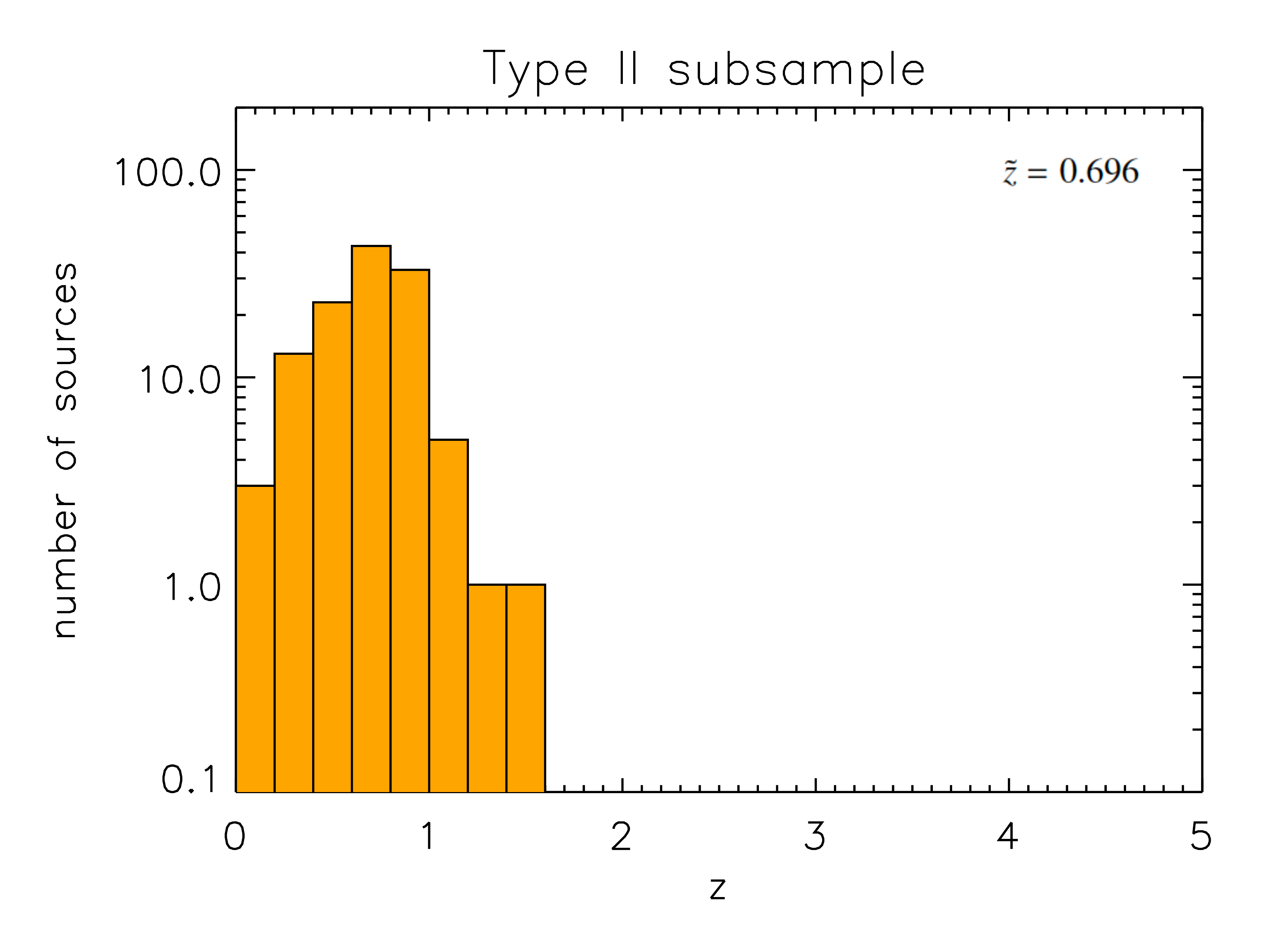}}
   \caption{Redshift distribution for the \emph{main} sample of 677 AGN (\emph{top left}), and for the five subsamples of AGN selected by means of different techniques: AGN selected via their X-ray properties (\emph{top center}), AGN selected via their MIR properties (\emph{top right}), AGN selected via optical variability (\emph{bottom left}), Type I and Type II AGN confirmed by spectroscopy (\emph{bottom center} and \emph{bottom right}, respectively). Each panel reports the median redshift value for the corresponding sample. Details about the selection criteria are reported in Section \ref{section:samples}. }\label{fig:z_hists}
   \end{figure*}

In Table \ref{tab:samples} we report the size of the various samples of AGN analyzed in this work and the corresponding median redshift value, together with the number of sources in each sample for which $M_{BH}$ and $\lambda_E$ estimates are available. It is apparent that the number of sources in the X-ray subsample with available estimates of $M_{BH}$ and $\lambda_E$ are 260, hence only four sources less than the ones in the \emph{main} sample with respect to the estimates of these quantities. As a consequence, we choose not to analyze $M_{BH}$ and $\lambda_E$ dependencies for the X-ray subsample separately as it would be redundant. We also report the redshift distribution for each of the six samples in Fig. \ref{fig:z_hists}, together with the median value for each sample. It is apparent that all the samples roughly extend to the same value ($z\approx 4$), except for the subsample of Type II AGN, which is limited to lower redshift values (up to $z\approx 1.6$).

\begin{table*}[t]
\renewcommand\arraystretch{1.4}
\caption{Number of AGN included in the \emph{main} sample and in the various subsamples selected following different criteria and used in this study.}\label{tab:samples}
\begin{center}
\begin{tabular}{l c c c c c c}
\toprule
\ & \textbf{\emph{main} sample} & \textbf{X-ray} & \textbf{MIR} & \textbf{opt. var.} & \textbf{spec. Type I} & \textbf{spec. Type II}\\
\midrule
\ total number & 677 & 605 & 225 & 301 & 225 & 122\\
\ $\tilde{z}$ & 0.876 & 0.893 & 1.214 & 1.441 & 1.653 & 0.696\\ 
\ $M_{BH}$ and $\lambda_E$ estimate & 264 & 260 & 117 & 183 & 152 & 85\\
\bottomrule
\end{tabular}
\end{center}
\footnotesize{\textbf{Notes.} The columns in the table indicate the number of AGN selected via (\emph{left to right}): $X/O$ ratio, MIR selection, optical variability selection, spectroscopic properties (returning Type I and Type II subsamples). For each sample we also report the median redshift $\tilde{z}$ and the number of sources for which $M_{BH}$ and $\lambda_E$ estimates are available.}
\end{table*}

\section{Structure function analysis}
\label{section:sf_analysis}
In what follows we discuss the definition that we adopt for the SF and present the results obtained for the various samples of AGN investigated in this work.
\subsection{Ensemble SF definition}
\label{section:sf_def}
Several definitions for the SF have been proposed in the literature (see Section \ref{section:intro}). Essentially, to each pair of visits at times $t_i$ and $t_j$ correspond two magnitude measurements mag($t_i$) and mag($t_j$); given a time lag $\Delta t$, we average all the squares of the magnitude differences $\mbox{mag($t_j$)}-\mbox{mag($t_i$)}$ associated to times such that $t_j - t_i \leq \Delta t$. As no astronomical measurement comes without noise, its contribution is usually subtracted in order to measure only intrinsic variations. 

\begin{figure}[ht]
 \centering
            {\includegraphics[width=\columnwidth]{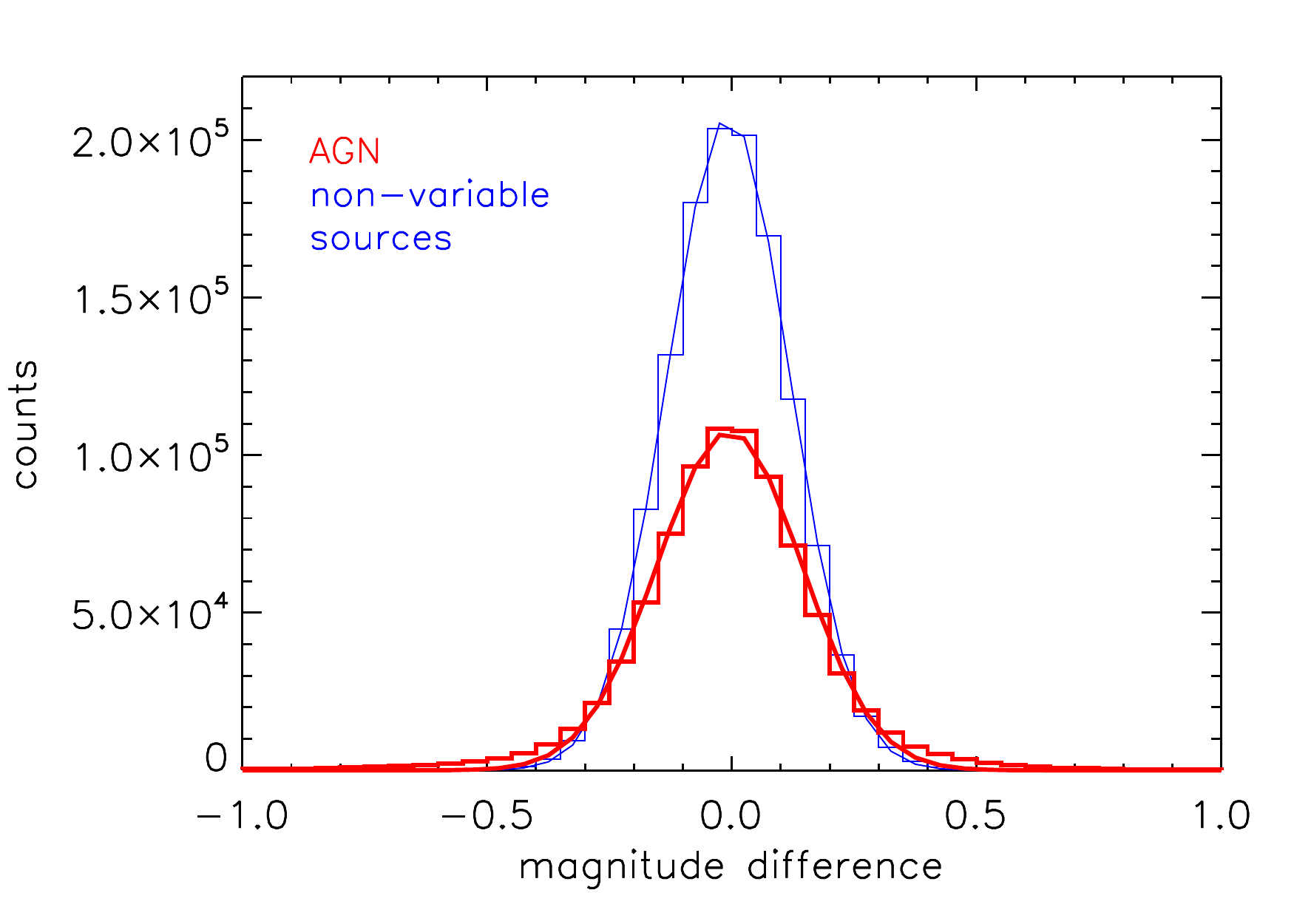}}
   \caption{Distribution of the magnitude difference for our \emph{main} sample (thick red) and for the sample of non-variable sources selected to estimate the noise contribution (thin blue). Each solid line represents the Gaussian fit for the corresponding distribution, computed via non-linear least squares.}\label{fig:delta_mag}
   \end{figure}

In this work we refer to the definition of the SF by \citet{diClemente}:
\begin{equation}
SF=\sqrt{\frac{\pi}{2}\langle |m(t_j)-m(t_i)|\rangle^2-\langle\sigma_{noise}^2\rangle}\mbox{   ,}
\label{eqn:sf}
\end{equation}
where we note that, in the first term, the square is performed after the average in order to reduce the influence of possible outliers \citep{Hook}; $\sigma_{noise}$ is the contribution due to the (uncorrelated) noise of the data. \citet{kozlowski} points out that in the original definition, using the instrumental noise, there is a missing factor of 2, leading to an underestimate of the SF slope. However here we directly measure this term on the magnitude difference of non-variable sources, and thus it represents the correct noise contribution. The overall noise contribution to subtract is defined as $\langle\sigma_{noise}^2\rangle$, and can be estimated as the average value of the squared magnitude difference of non-variable sources over bins having the size of $\Delta t$. The factor $\pi/2$ is introduced under the assumption that both the intrinsic variability and the noise can be represented via a Gaussian distribution. Fig. \ref{fig:delta_mag} reports these two distributions, showing that this is indeed the case, as both are very close to a Gaussian.

Throughout this work we choose bins of the same size on a logarithmic scale. There is no specific rule to follow in the choice of the size, except it should ensure a large enough number of points per bin; we tested different sizes, obtaining fairly consistent SFs, and in the end we settled on 0.30 dex, which allows us to have $\approx 1,500$ points in the bin corresponding to the shortest time lag. We compute error bars via error propagation on $\log(SF)$ as in what follows we use the logarithm of the SF; we weight errors for the number of points included in each bin, according to the following equation: 
\begin{equation}
err_{\log(SF)}=\frac{1}{SF}\cdot\log e \cdot err_{SF}\mbox{   ,}
\label{eqn:log_sf_err}
\end{equation}
where $err_{SF}$ is the error on the SF and is defined as
\begin{equation}
err_{SF}=\frac{1}{2SF}\sqrt{\pi^2\cdot\langle\Delta mag\rangle^2\cdot err_{\langle\Delta mag\rangle}^2+err_{\langle\sigma_{noise}^2\rangle}^2}\mbox{   ,}
\label{eqn:sf_err}
\end{equation}
and $\langle\Delta mag\rangle$ is the average magnitude difference computed for each pair of visits falling into the bin at issue.

\subsection{The SF of our six samples of AGN}
\label{section:sf_six}
\begin{figure*}[ht]
 \centering
\subfigure
            {\includegraphics[width=5.5cm]{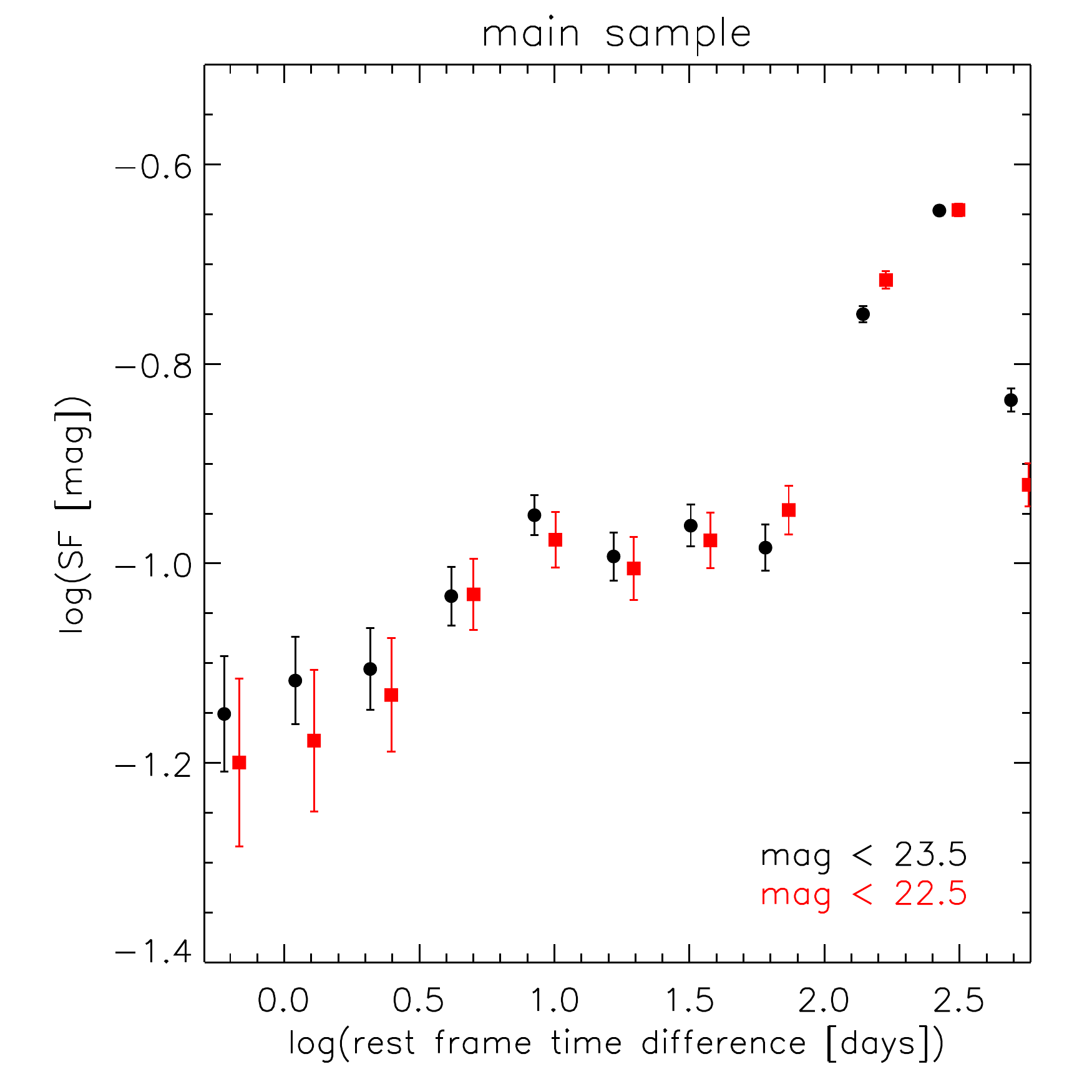}}
\subfigure
            {\includegraphics[width=5.5cm]{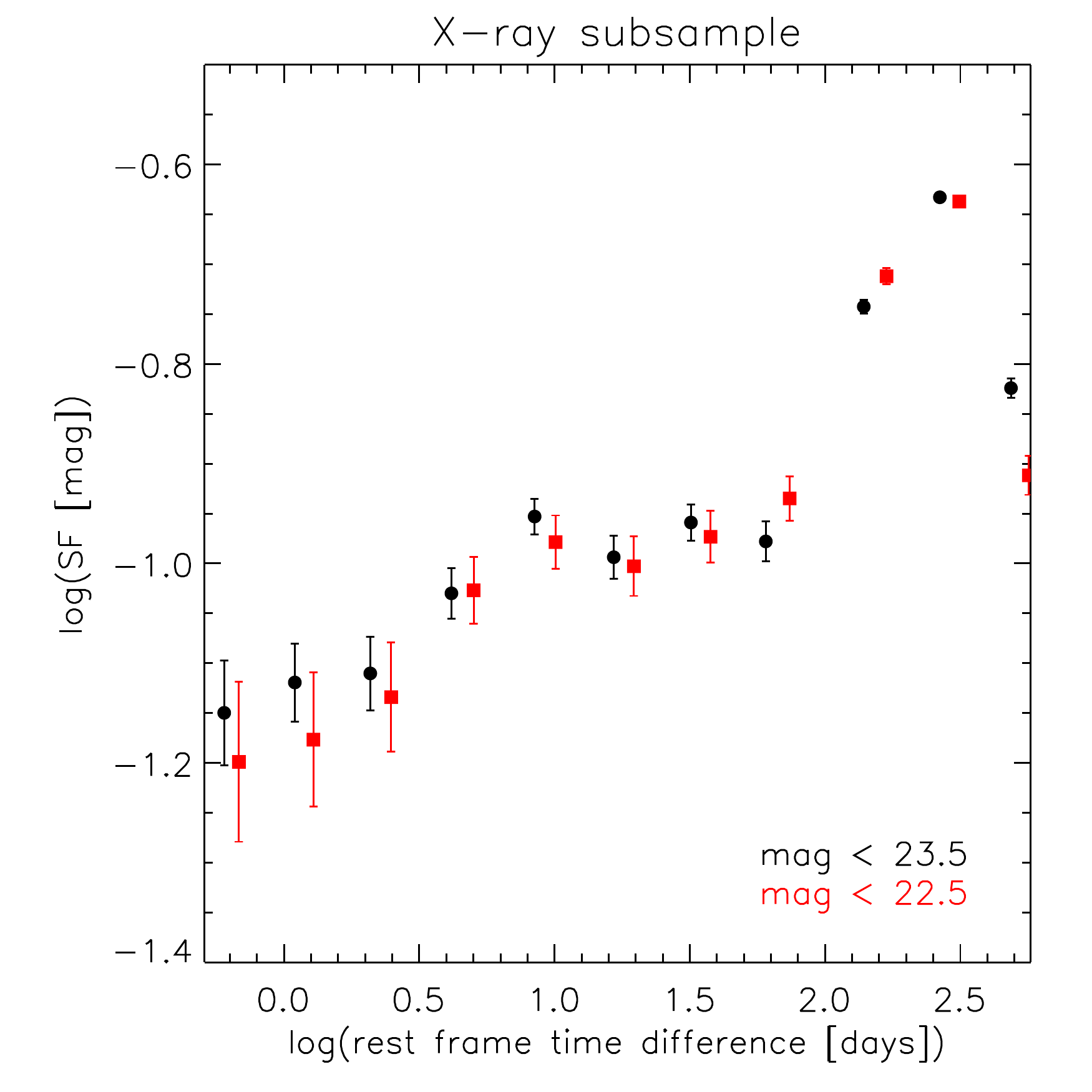}}
\subfigure
            {\includegraphics[width=5.5cm]{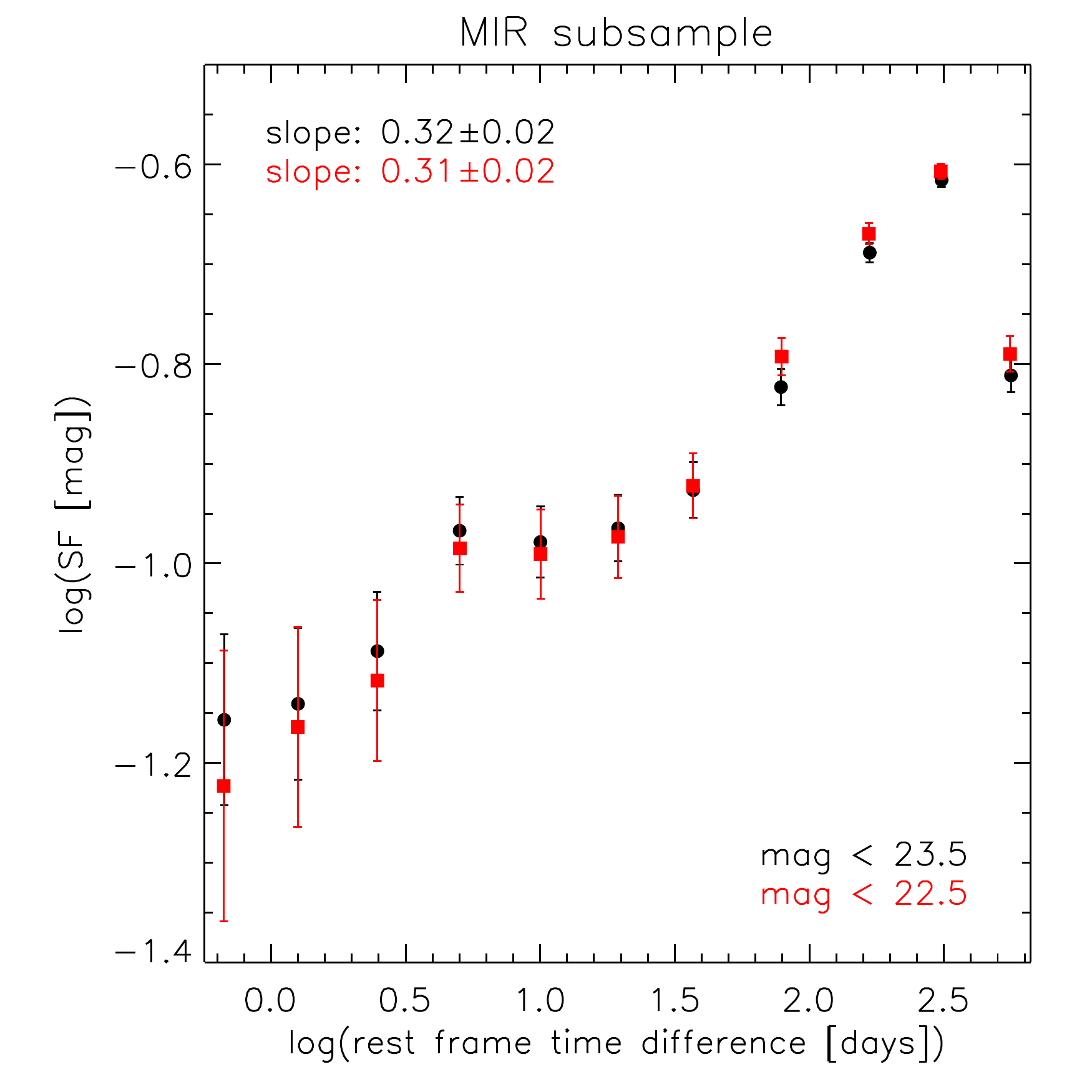}}
\subfigure
            {\includegraphics[width=5.5cm]{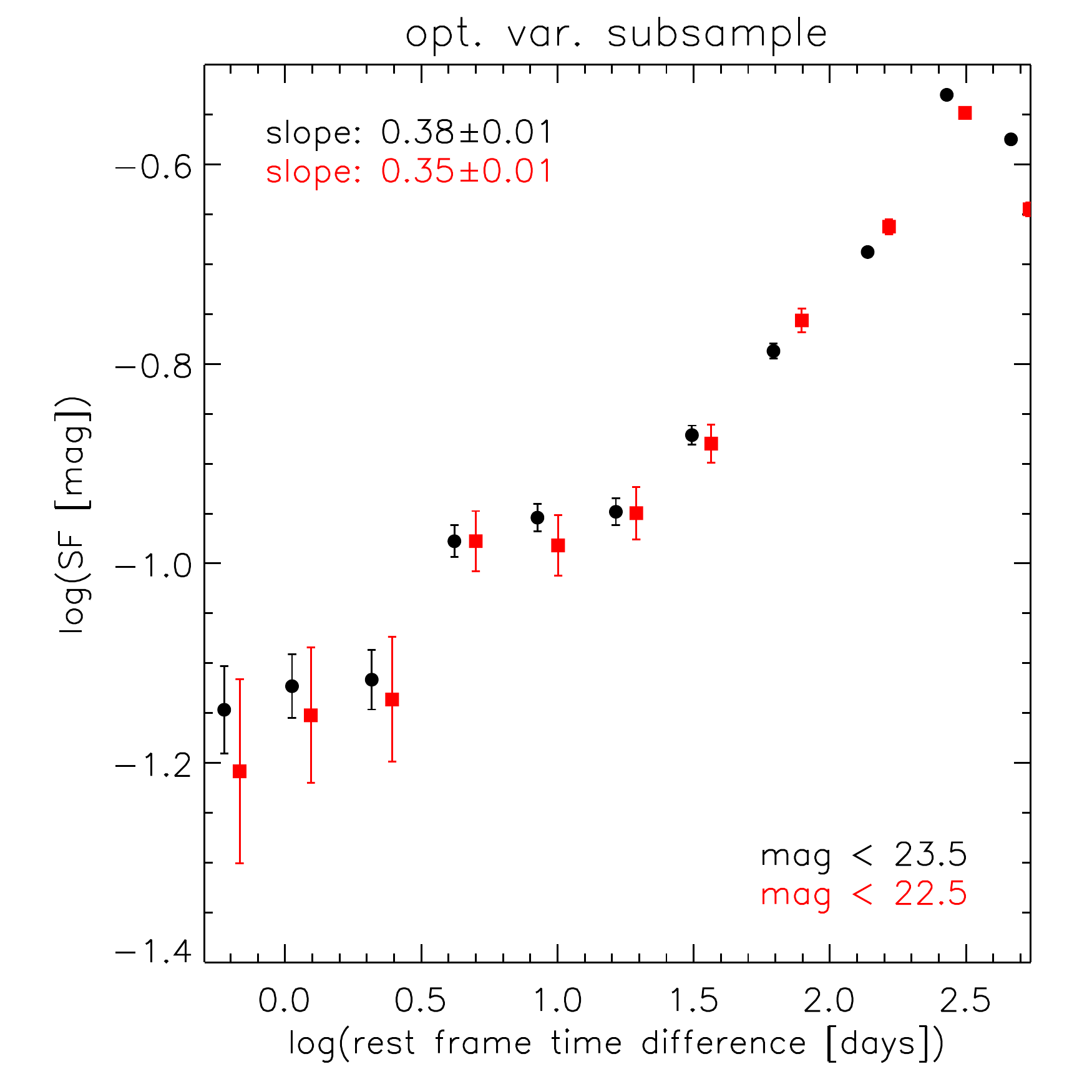}}
\subfigure
            {\includegraphics[width=5.5cm]{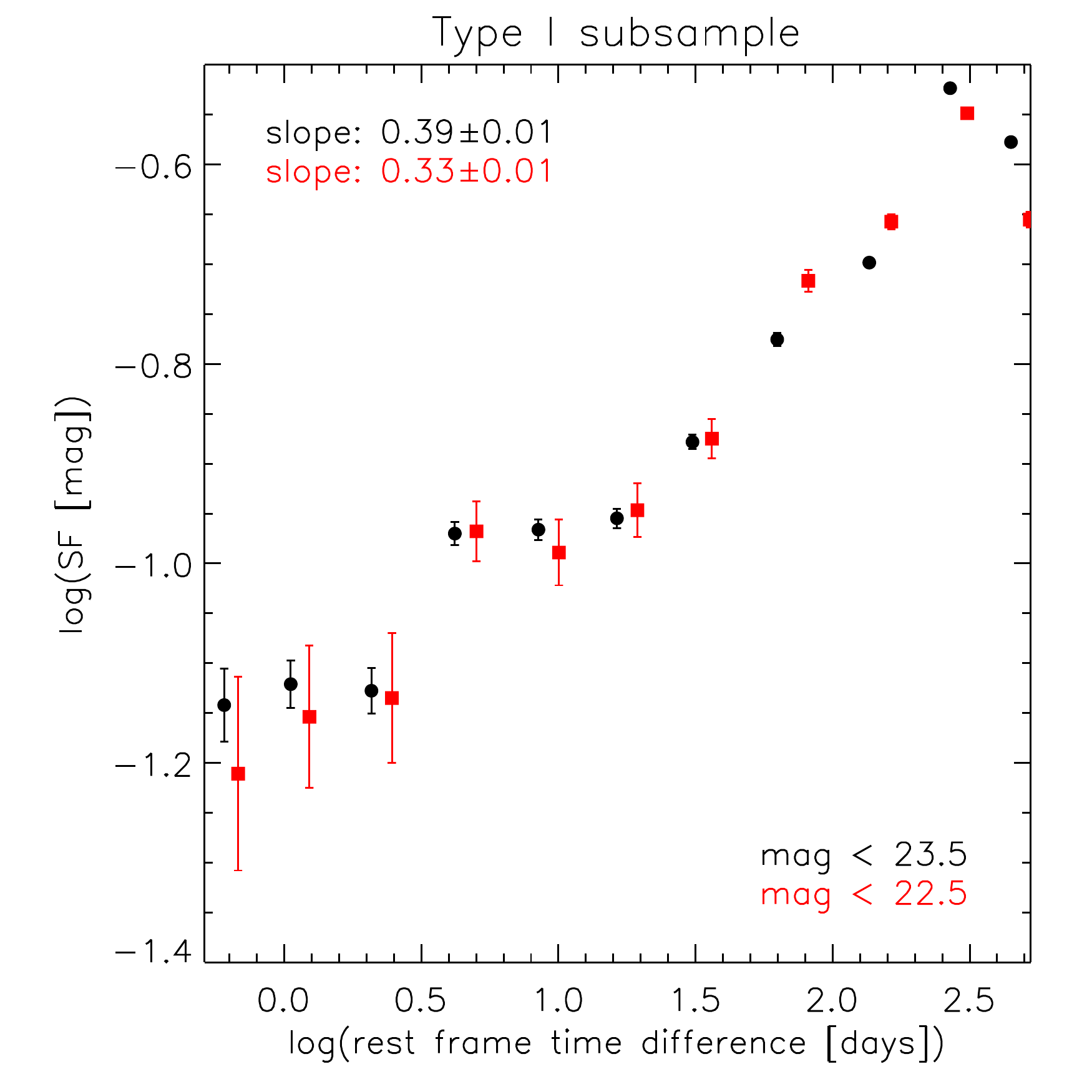}}
\subfigure
            {\includegraphics[width=5.5cm]{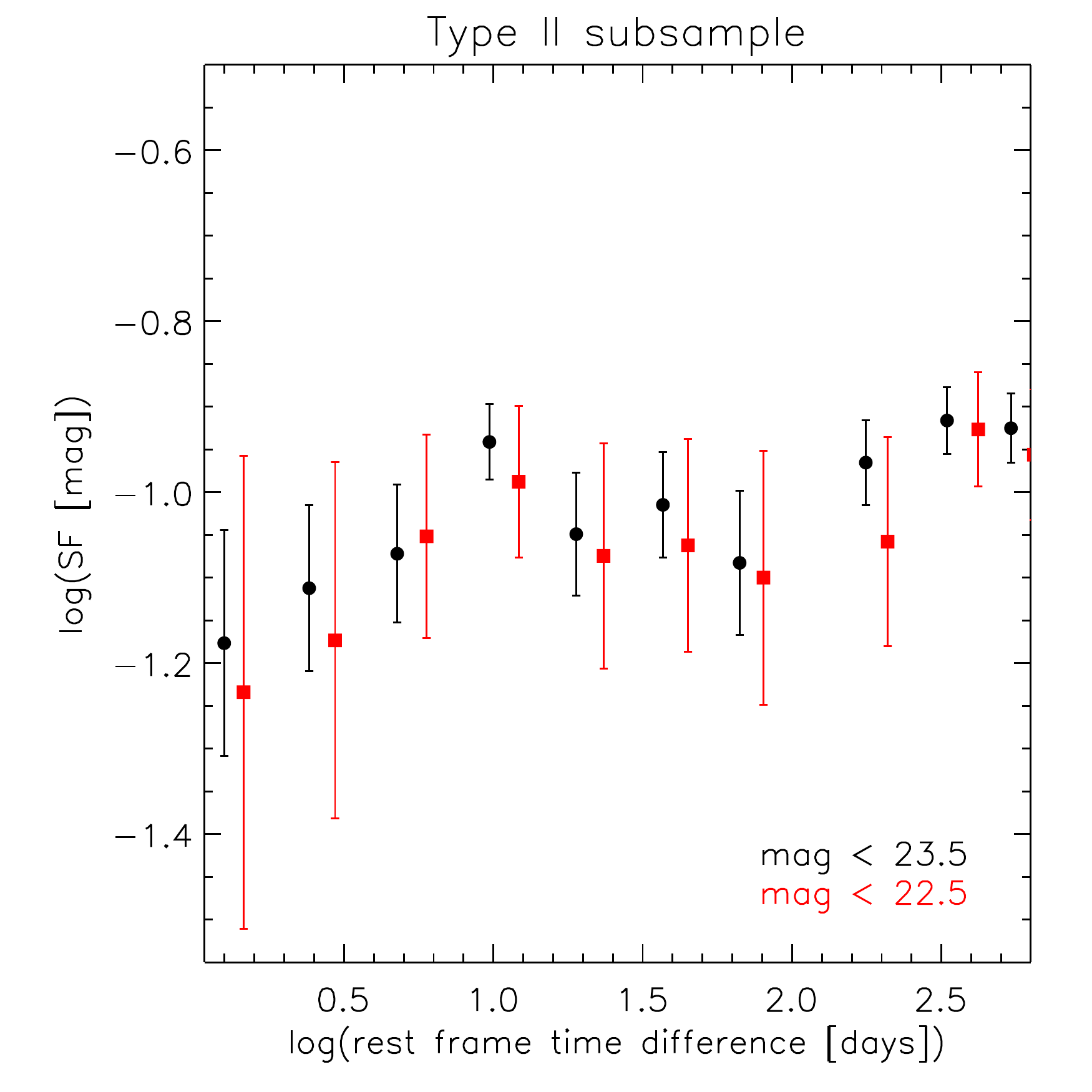}}
   \caption{SF of the six samples of AGN selected via different properties and investigated in this study: \emph{main} sample (\emph{top left}), X-ray AGN (\emph{top center}), MIR AGN (\emph{top right}), optically variable AGN (\emph{bottom left}), Type I AGN (\emph{bottom center}), and Type II AGN (\emph{bottom right}). We test two different magnitude thresholds: $r < 23.5$ mag (black dots) and $r < $ mag (red squares). The ranges on both axes are roughly the same to ease comparison among the various samples and thresholds. The three panels corresponding to MIR, optically variable, and Type I AGN, where a clear linear region in the SF can be identified in correspondence of the logarithmic range $1.0-2.6$ for the baseline, include estimates for the corresponding slopes, for each of the two adopted thresholds. The slopes are the best-fit lines computed via weighted least squares regression.}\label{fig:sf}
   \end{figure*}

In Figure \ref{fig:sf} we show the SFs obtained for each of our six samples of AGN selected by means of different properties and investigated in this study. Time differences are measured at rest frame in order to be able to study the dependence of the intrinsic variability on time. Each panel includes the SF obtained when adopting a 22.5 mag threshold, with the aim of assessing whether a different depth leads to significant differences in the SF shape.

The plot corresponding to the subsample of Type I AGN is the one representing the typical AGN structure function; the one corresponding to optically variable AGN is very similar, as that subsample is highly biased towards Type I AGN (75\% of the sources). In these two plots three distinct regions can be identified. When time lags are too short to detect variability, we can see small amplitude variations characterized by larger error bars. Starting from time lags of $\approx10$ days, the SF can be represented via a power law, and therefore on a logarithmic scale it is approximately linear; this linear trend originates from the shape of the autocorrelation function describing the stochastic process leading to variability, and therefore related to the properties of the light curves. For timescales longer than $\approx250$ days ($\log(\mbox{time}) \approx 2.4$) we observe a drop in variability amplitude. We may be tempted to believe that we reached a maximum in variability amplitude. Indeed, several works show how, as the time difference $\Delta t$ increases, a turnover is observed at a characteristic timescale $\tau$, and then the SF switches from a red-noise regime to a white-noise regime \citep[e.g.,][]{Bauer,kozlowski}. However this $\tau$ is estimated to be on the order of $10^2$ rest-frame days at optical wavelengths for SDSS quasars \citep[e.g.,][]{cp01,kelly09,macleod10}, suggesting that we need a more extended baseline in order to investigate this timescale without the current sampling limitations. Thus we believe that the observed turnover is not real but an effect of the poor sampling at large time lags, and it would not be observed if the sampling were even and if it covered longer timescales \citep[e.g.,][]{rengstorf,Bauer,emmanoulopoulos}. In particular, \citet{Bauer} investigate the windowing effects due to irregular data sampling by simulating quasar light curves with uniform cadence, and their results support the thesis that the turnover is not evidence for the existence of a maximum timescale in AGN variability, but rather an effect of inadequate sampling.

We visually identify the linear region in correspondence of the logarithmic values $1.0-2.6$ for the baseline, that is to say $\approx 10-400$ days. The measured slopes for the linear regions of the SFs of Type I AGN and optically variable AGN are obtained via a weighted least squares regression\footnote{Throughout this work, when referring to weighted least square regression, we consider as weights the error bars of each point included in the fit. As mentioned earlier in this section, each of these points is the result of a binning.} and are $0.39\pm0.01$ and $0.38\pm0.01$, respectively, for the 23.5 mag threshold, and $0.33\pm0.01$ and $0.35\pm0.01$, respectively, for the 22.5 mag threshold. These values are fairly consistent with some values reported in the literature, e.g., \citet{Bauer} and \citet{vdB}, those being $0.3607\pm0.0075$ and $0.336\pm0.033$, respectively. 

If we focus on the SF corresponding to the subsample of Type~II AGN, we see that the variations detected on short timescales are consistent with the ones detected for Type I AGN, but this does not hold for long timescales, where we clearly see that this subsample of AGN is characterized by smaller variations than Type I AGN. A line with a much shallower slope could in principle fit the whole set of points, but they are characterized in this case by much larger error bars than the previous subsamples. These ``pure'' subsamples of AGN, consisting only of Type I or Type II AGN, represent two extreme cases, returning very different SFs. When we include AGN selected via different properties in a sample, we obtain a SF that reflects this heterogeneity: we can see this from the SFs corresponding to the remaining three subsamples of sources analyzed in this work, where the linear region is less defined, while the zone dominated by the measurement noise is larger. This is particularly evident for the largest samples analyzed, i.e., the X-ray sample and the \emph{main} sample, which largely overlap, as apparent from the Venn diagram in Fig. \ref{fig:venn}. We note that the fraction of known spectroscopic Type I AGN decreases as the sample includes more sources, being 60\% in the MIR AGN sample, 37\% in the X-ray AGN sample, and 33\% in the \emph{main} sample of AGN. Conversely, the corresponding fractions of known Type II AGN roughly increase, as they are 11\%, 20\%, and 18\%, respectively. Assuming that this two trends hold for the fractions of sources lacking a spectroscopic classification in each of the three subsamples (29\% for MIR AGN, 43\% for X-ray AGN, and 49\% for the \emph{main} sample), all this would suggest that the inclusion of less variable AGN damps the average SF slope, as the inclusion of AGN seen along obscured lines of sight damps the intrinsic variability properties on an individual basis. Therefore, all of the SF may be underestimating the intrinsic variability properties. Because of these characteristics, the subsample of Type II AGN turns out to be unsuitable for our study based on the optical variability of AGN. We note that we can still identify a clear linear region in the SF of the MIR subsample, its slope being $0.32\pm0.02$ for the 23.5 mag threshold and $0.31\pm0.02$ for the  mag threshold. These values are slightly lower than the ones obtained for the subsamples of Type I AGN and optically variable AGN, supporting the thesis that the inclusion of Type II AGN in the MIR subsample is flattening the SF.

The noise contribution for each sample was estimated in two different ways. As mentioned above, the noise is what we measure on timescales too short to detect significant flux variations. Based on this, for each sample of sources we can obtain a noise estimate as the average of the first two points of the SF, corresponding to rest-frame timescales shorter than three days. The second way is more accurate and makes use of a sample of non-variable sources. For such objects, we expect the magnitude difference between two visits to be due only to stochastic fluctuations and show no dependence on timescale (i.e., white noise). For this estimate we take advantage of a sample of 1,000 sources used in \citet{decicco21} as a labeled set of ``inactive'' galaxies. Based on all the COSMOS catalogs we consulted, these galaxies show no sign of nuclear activity and can therefore be assumed to be non-variable. In this case, consistent with the noise definition in the adopted SF expression (see Eq. \ref{eqn:sf}), we compute the square of the magnitude difference for this sample of sources and adopt its average value (hence corresponding to $\langle\sigma_{noise}^2\rangle)$ as our error estimate. Both methods return values consistent down to the third decimal digit.

\subsection{The overlap issue}
\label{section:overlap}
In Section \ref{section:samples} we mentioned that the various subsamples of AGN selected for this work are not disjointed; on the contrary, their overlap is in some cases considerable, as is apparent from Table \ref{tab:samples} and Fig. \ref{fig:venn}. This is a consequence of the multiple properties that AGN typically exhibit and of the different sensitivities of the different methods to AGN activity (as a function of various properties like obscuration, orientation, radio-loudness, etc.), which allow selection via more than one diagnostic or technique. If two or more subsamples of sources largely overlap, we expect them to be characterized by similar SFs in terms of shape, amplitude and steepness. Nonetheless, the partitions of objects that only belong to one subsample could in principle have peculiar properties -- possibly explaining why they were not detected via other methods -- or share the same properties of the rest of their parent subsample. It is therefore interesting to inspect these partitions, in this case 222 X-ray selected AGN, 50 MIR-selected AGN, and 19 optical variability-selected AGN.

\begin{figure*}[ht]
 \centering
\subfigure
            {\includegraphics[width=5.5cm]{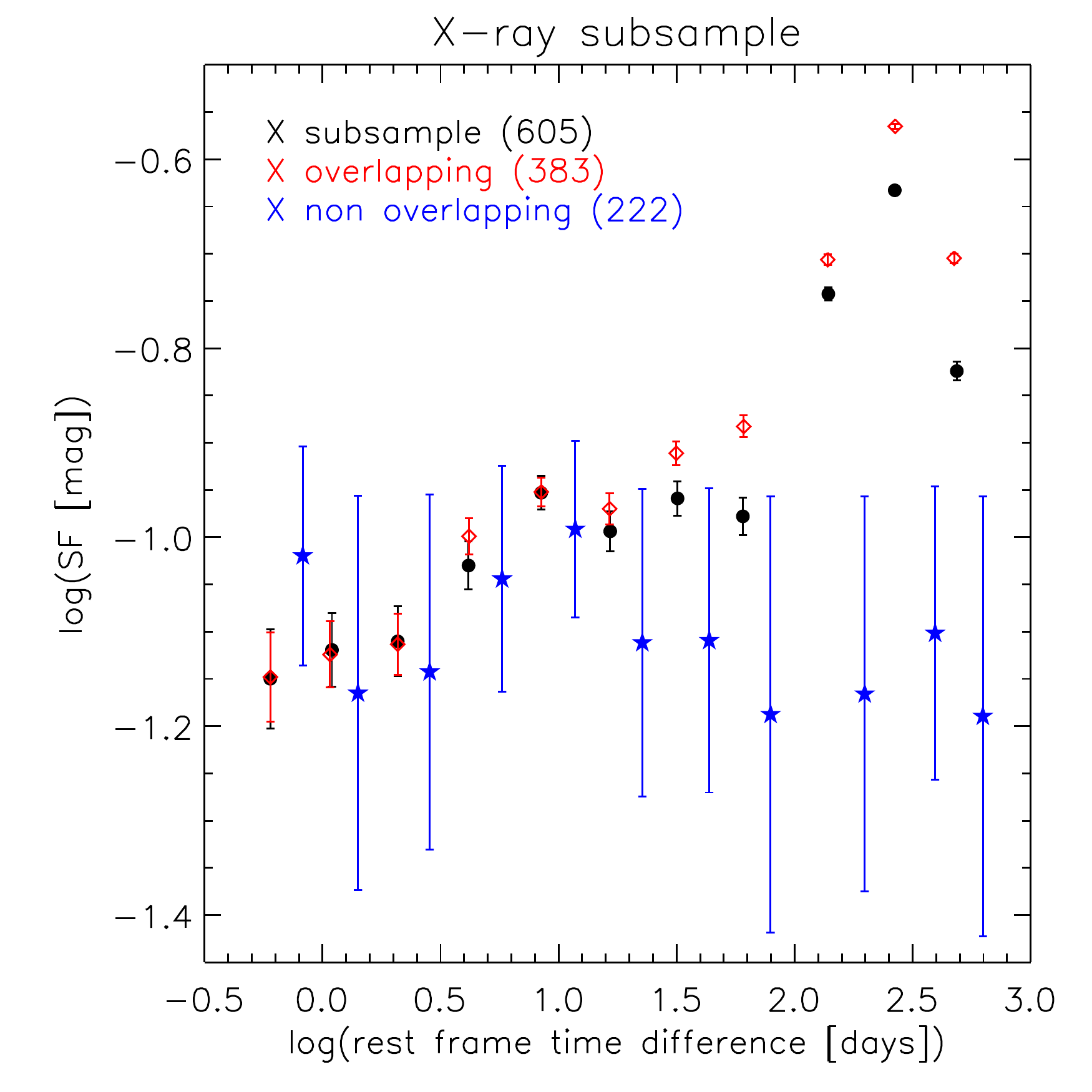}}
\subfigure
            {\includegraphics[width=5.5cm]{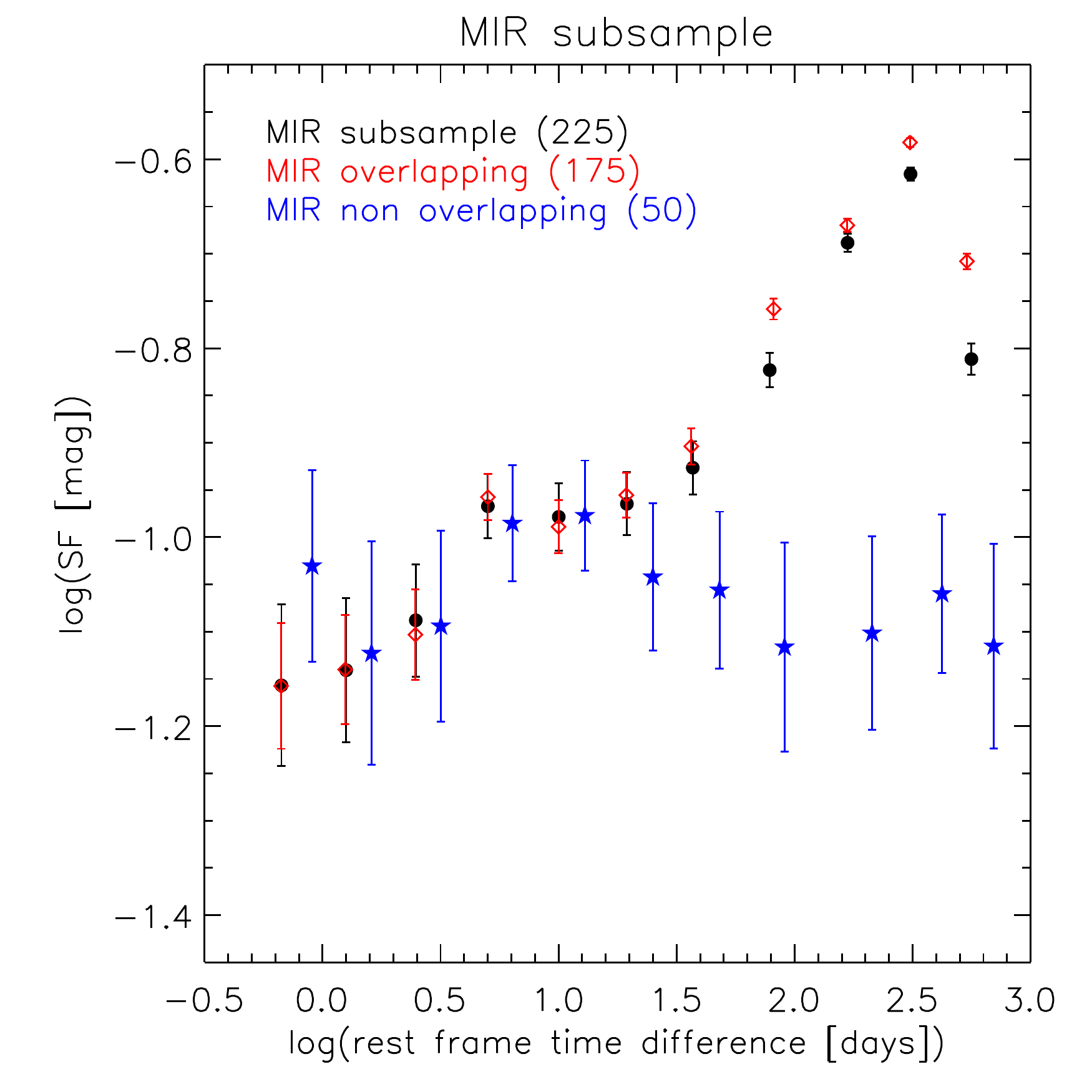}}
\subfigure
            {\includegraphics[width=5.5cm]{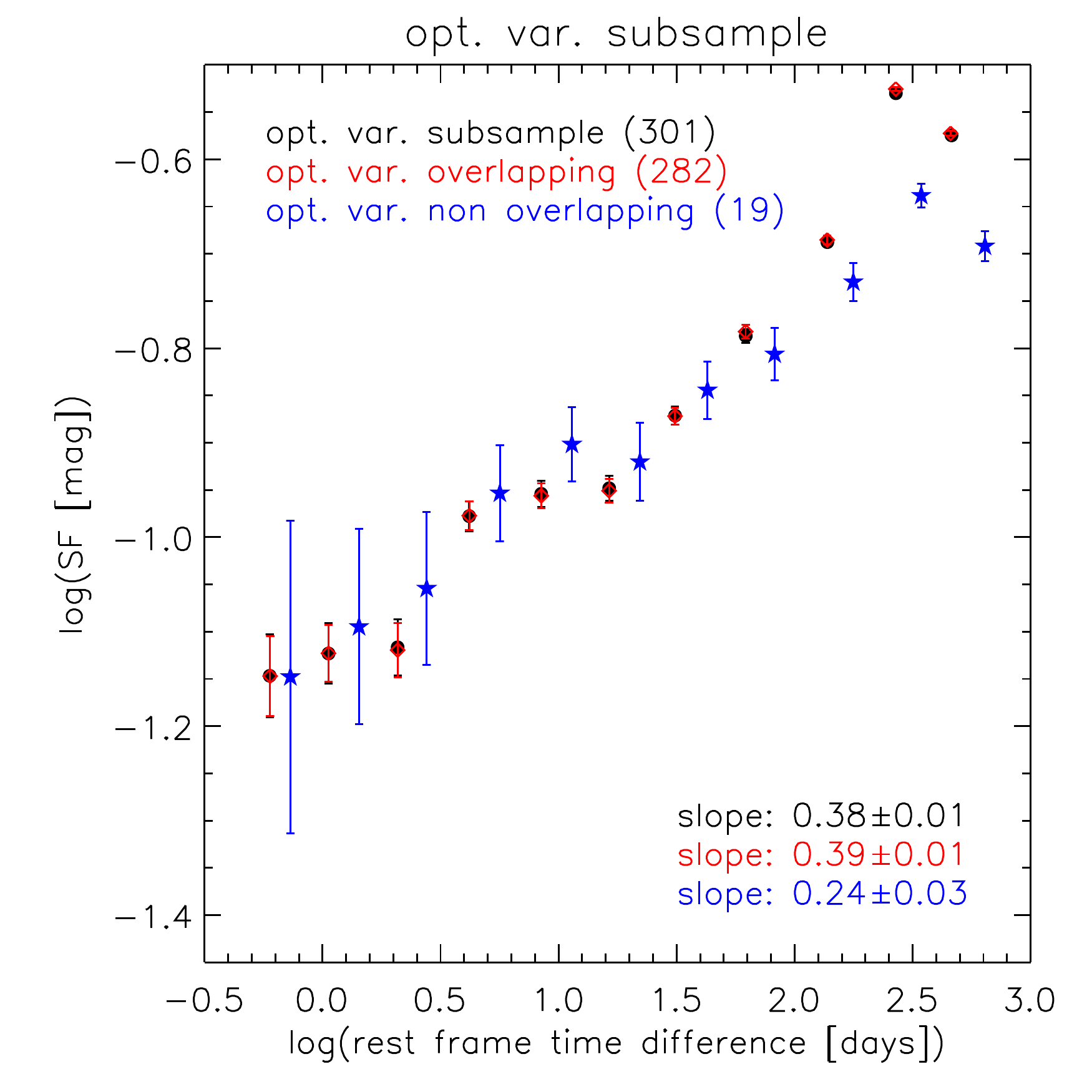}}
   \caption{SF of the three subsamples of AGN including a partition of sources not overlapping any other AGN subsamples: X-ray AGN (\emph{left}), MIR AGN (\emph{center}), optically variable AGN (\emph{right}). Each subsample consists of two complementary subsets: one overlapping one or more AGN subsamples, and the other not overlapping any other AGN subsample. For each subsample we show the SF obtained from all of its sources (black dots), the SF of the overlapping subset (red diamonds), and the SF of the non-overlapping subset (blue stars). The panel corresponding to optically variable AGN includes estimates for the slope of the linear region of each SF, which can be identified in correspondence of the logarithmic range $1.0-2.6$ for the baseline. The slopes are the best-fit lines computed via weighted least squares regression.}\label{fig:non_overlap}
   \end{figure*}

In order to do so we divide the X-ray, the MIR, and the optically-variable subsamples in two complementary subsets each: the one overlapping one or more other AGN subsamples (hereafter, overlapping subset), and the one not overlapping any other AGN subsamples (hereafter, non-overlapping subset). We then compute the SF for each pair of subsets, and show our results in Figure \ref{fig:non_overlap}, where we also include the SF of the whole subsample at issue to ease comparison. The figure shows that the two complementary subsets of AGN selected via optical variability have similar SFs as for shape; the linear region is less steep for the non-overlapping subset, which is also characterized by slightly lower amplitude values in that region. We include estimates for the slope of each linear region. The amplitude of the global SF of this subsample is dominated by the overlapping subset, which constitutes the majority of the subsample (282 overlapping sources vs. 19 non overlapping). All this is not surprising, since the SF represents the variability of the investigated sources, and the sample at issue consists of AGN selected on the basis of their variability properties. These sources do not have a counterpart in the X-ray catalogs used in this work, while they do have a MIR infrared counterpart, but are not classified as AGN based on that. 

Results are very different for the other two subsamples of X-ray and MIR AGN: in both cases the non-overlapping subset is characterized by a flat SF, while the SF of the overlapping subset is steeper and includes a linear region. As a consequence, the global SF of each subsample -- dominated by the overlapping sources -- is slightly less steep than the SF for the corresponding non-overlapping subset. Since optical variability is not a selection criterion for the X-ray and MIR AGN subsamples, these sources are not necessarily optically variable and, in particular, the two non-overlapping subsets can include Type II AGN, for which identification via optical variability is typically hard. A spectroscopic follow-up, at least for the brightest sources in these three non-overlapping subsets, would help shed some light on their nature, and we therefore plan to apply for observing time to pursue this goal. We note that, at present, only 16 of the sources in the three non-overlapping subsets (13 in the X-ray subsample and three in the MIR subsample) have BH mass and Eddington ratio estimates available that were used for this study.

\section{Analysis of the main physical properties}
\label{section:phys}
One of the main goals of this work is the investigation of possible connections between the optical variability of our AGN samples and some physical properties, namely $M_{BH}$, $\lambda_E$, $L_{bol}$, redshift, and absorption. 
\subsection{Black hole mass, accretion rate, and bolometric luminosity dependence}
\label{section:bel}
We mentioned (see Section \ref{section:samples}) that $M_{BH}$ and $\lambda_E$ estimates are available for 264 AGN in the \emph{main} sample. In what follows, we do not take into account the subsample of Type II AGN as, based on Fig. \ref{fig:sf} and on the arguments presented in Section \ref{section:sf_six}, they do not vary significantly over the investigated timescale. 

\begin{figure*}[th]
 \centering
\subfigure
            {\includegraphics[width=6cm]{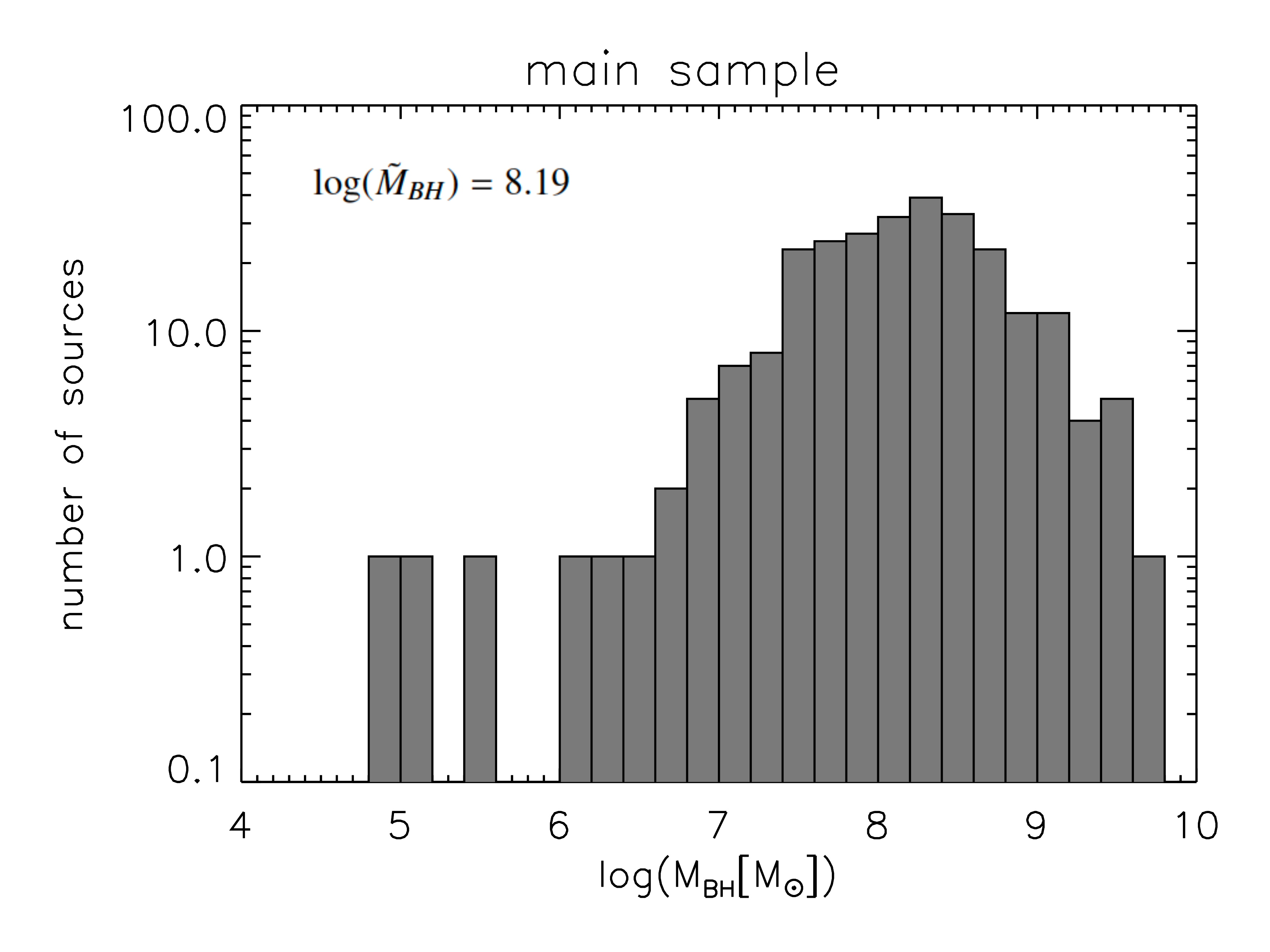}}
\subfigure
            {\includegraphics[width=6cm]{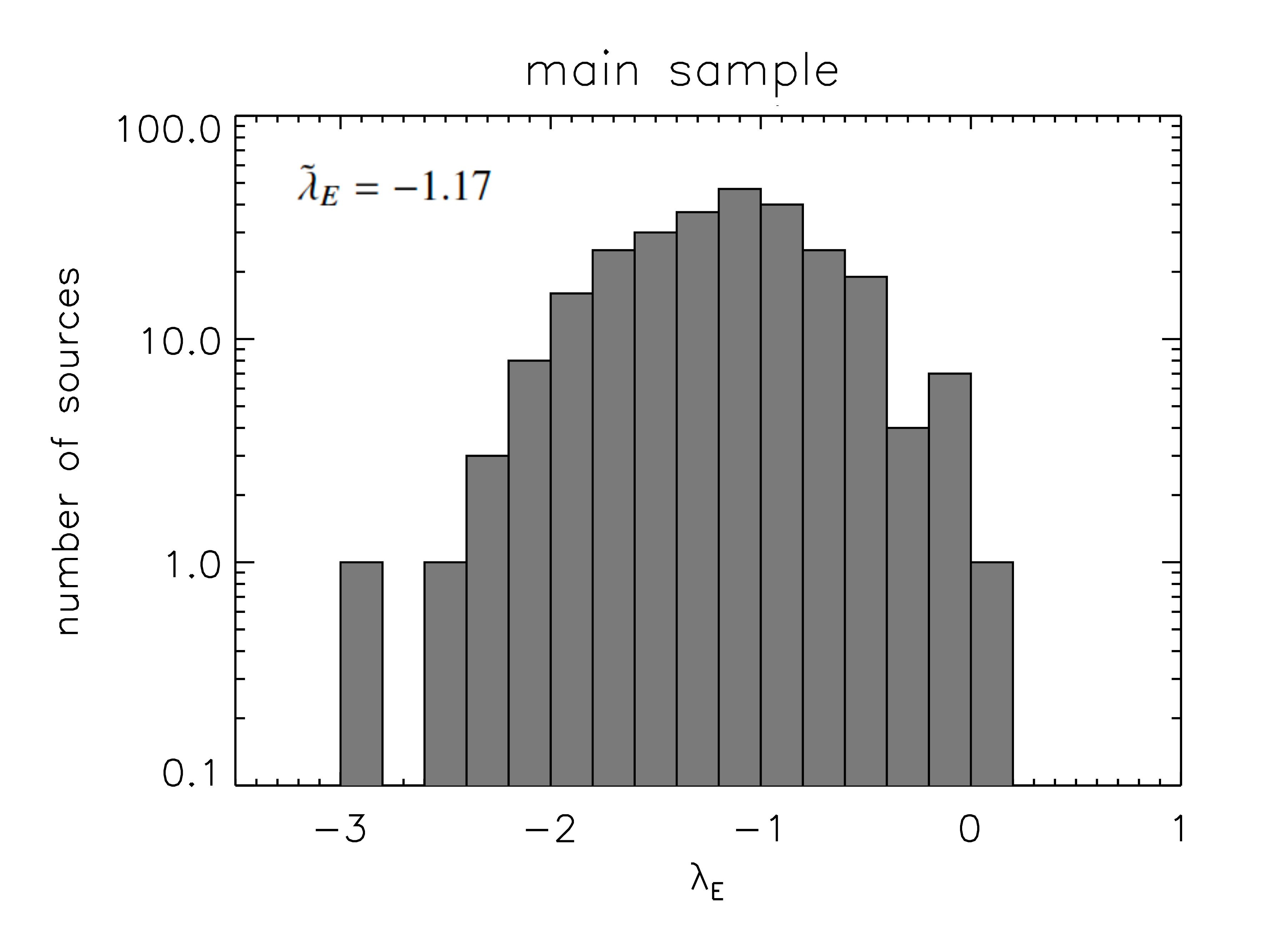}}
\subfigure
            {\includegraphics[width=6cm]{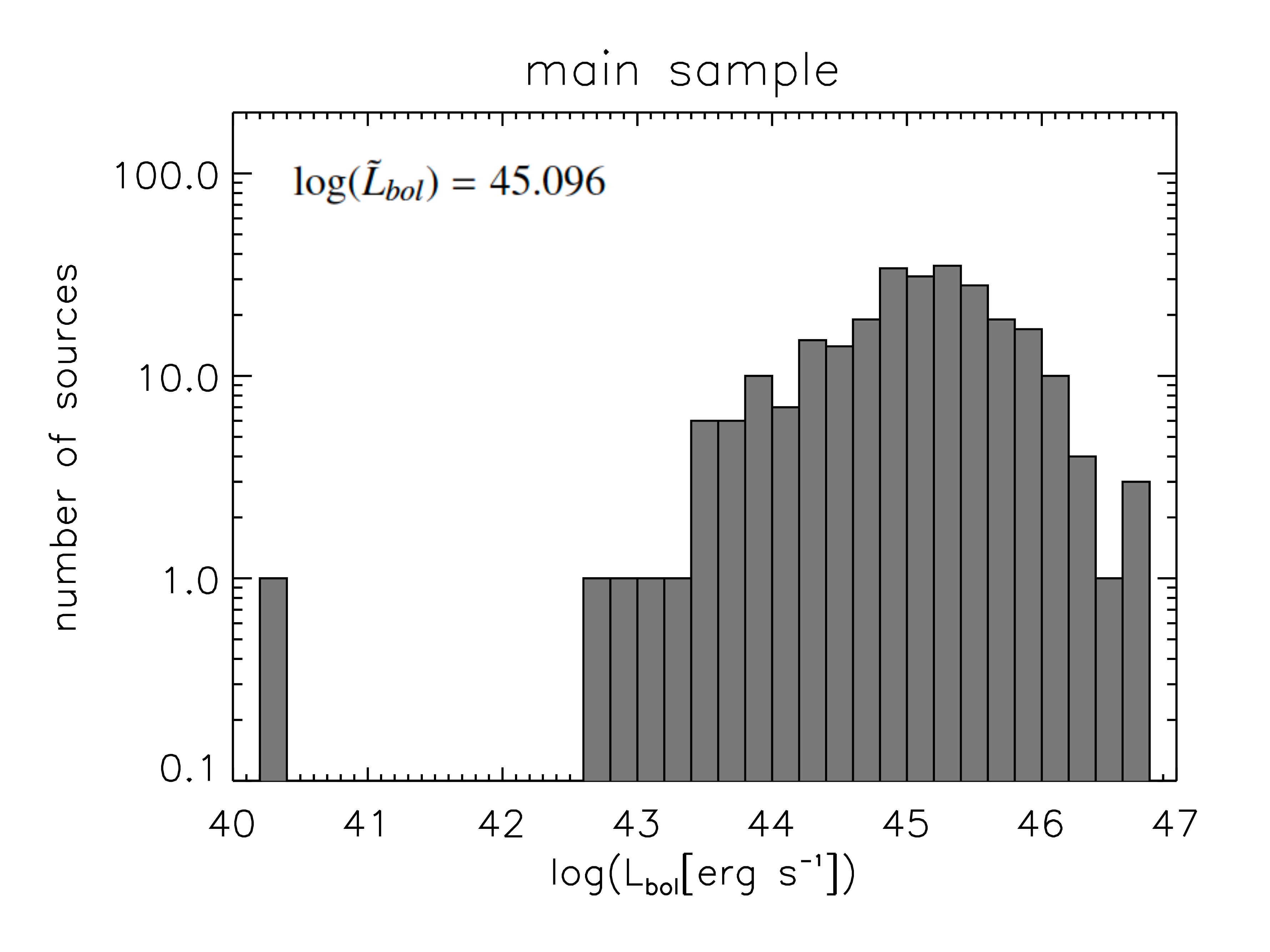}}\\
\subfigure
            {\includegraphics[width=6cm]{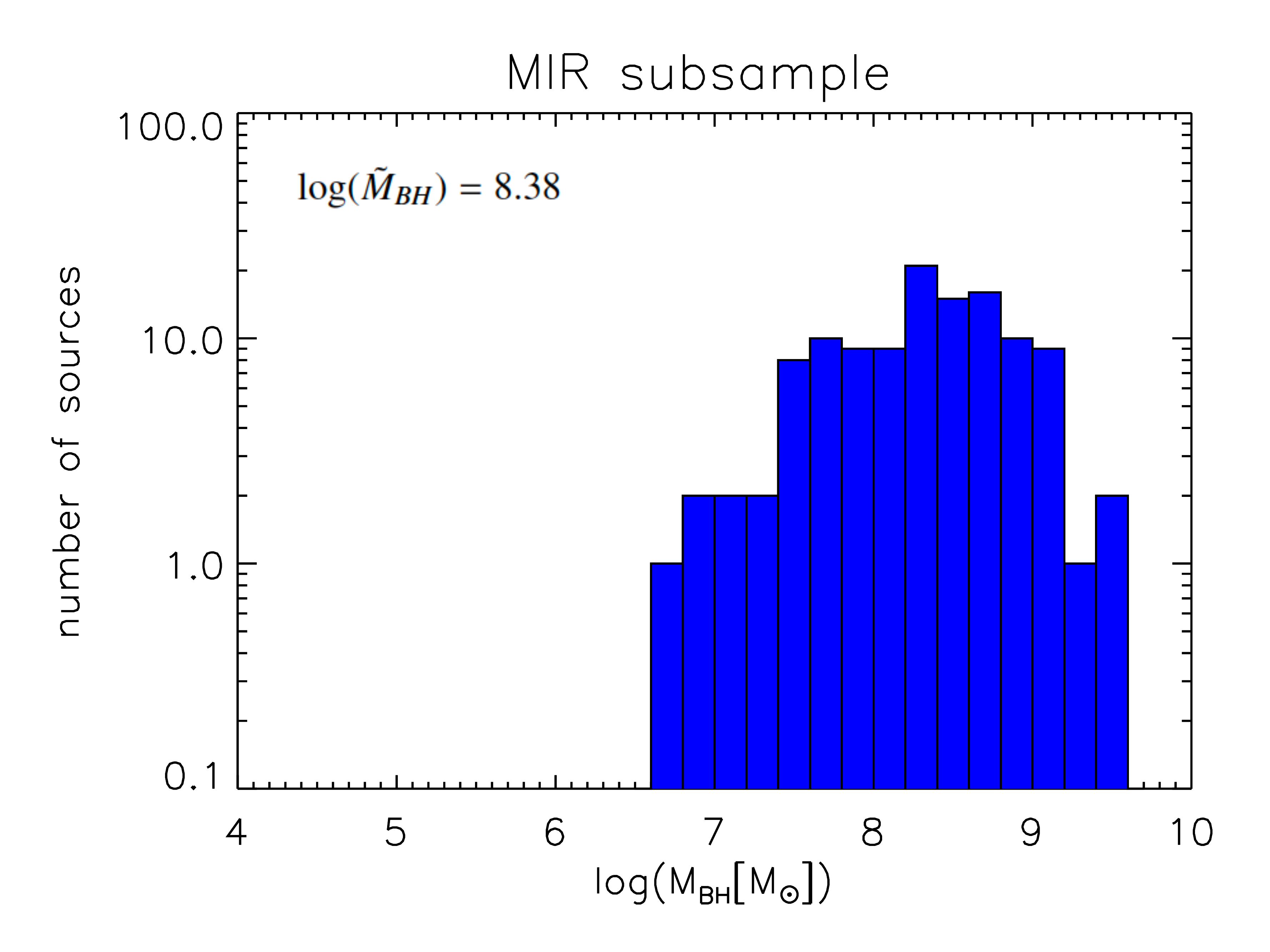}}
\subfigure
            {\includegraphics[width=6cm]{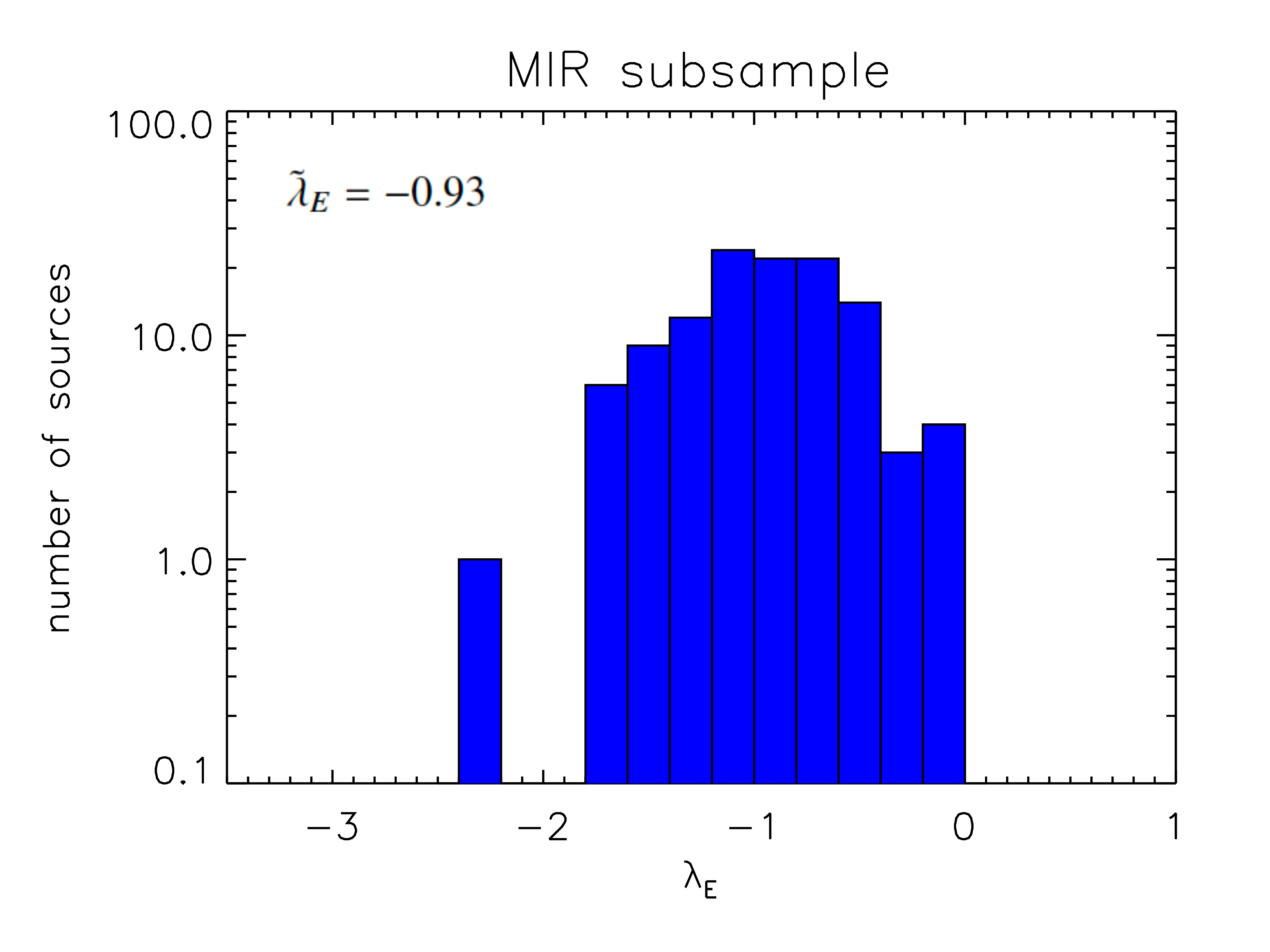}}
\subfigure
            {\includegraphics[width=6cm]{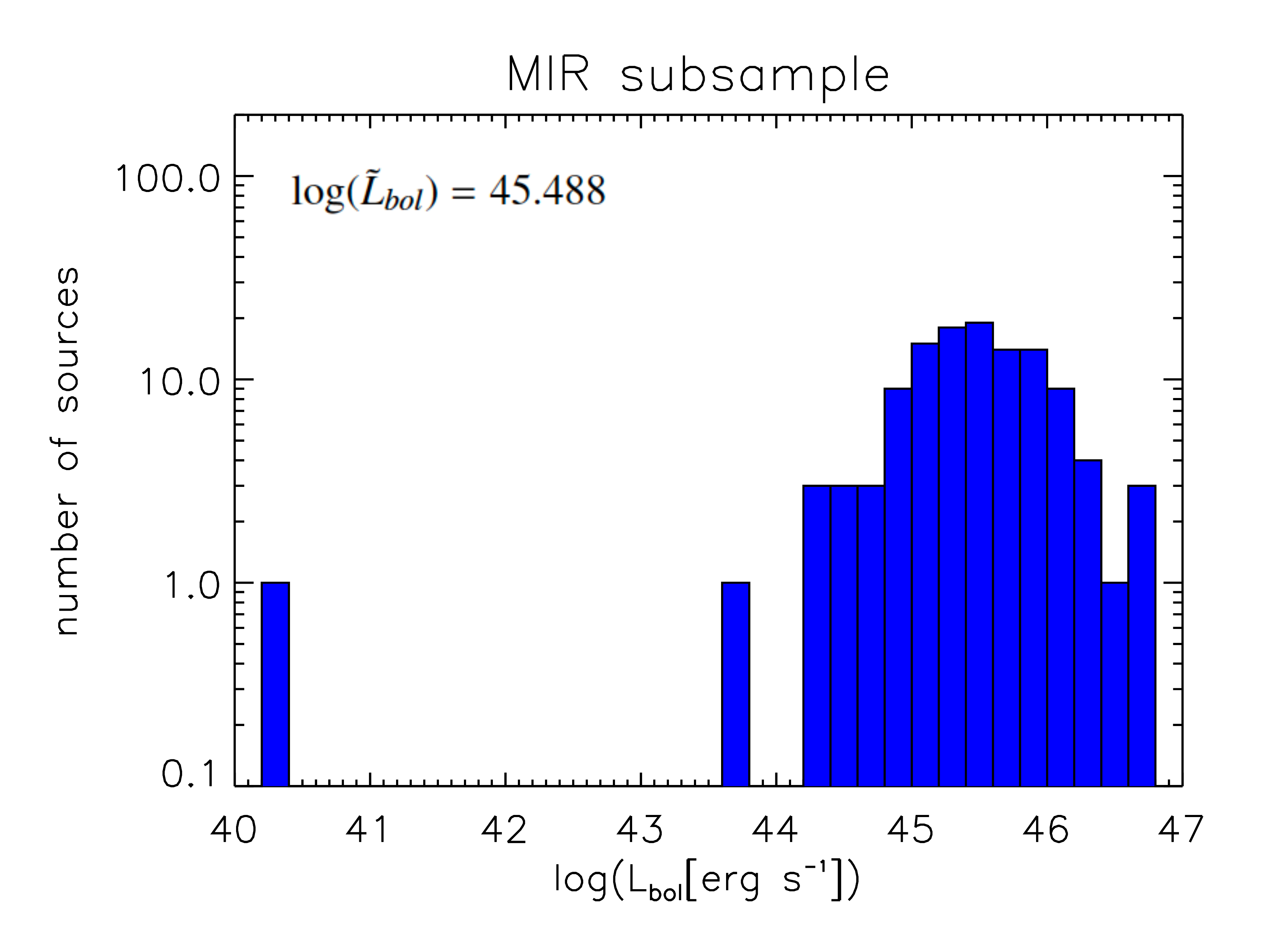}}\\
\subfigure
            {\includegraphics[width=6cm]{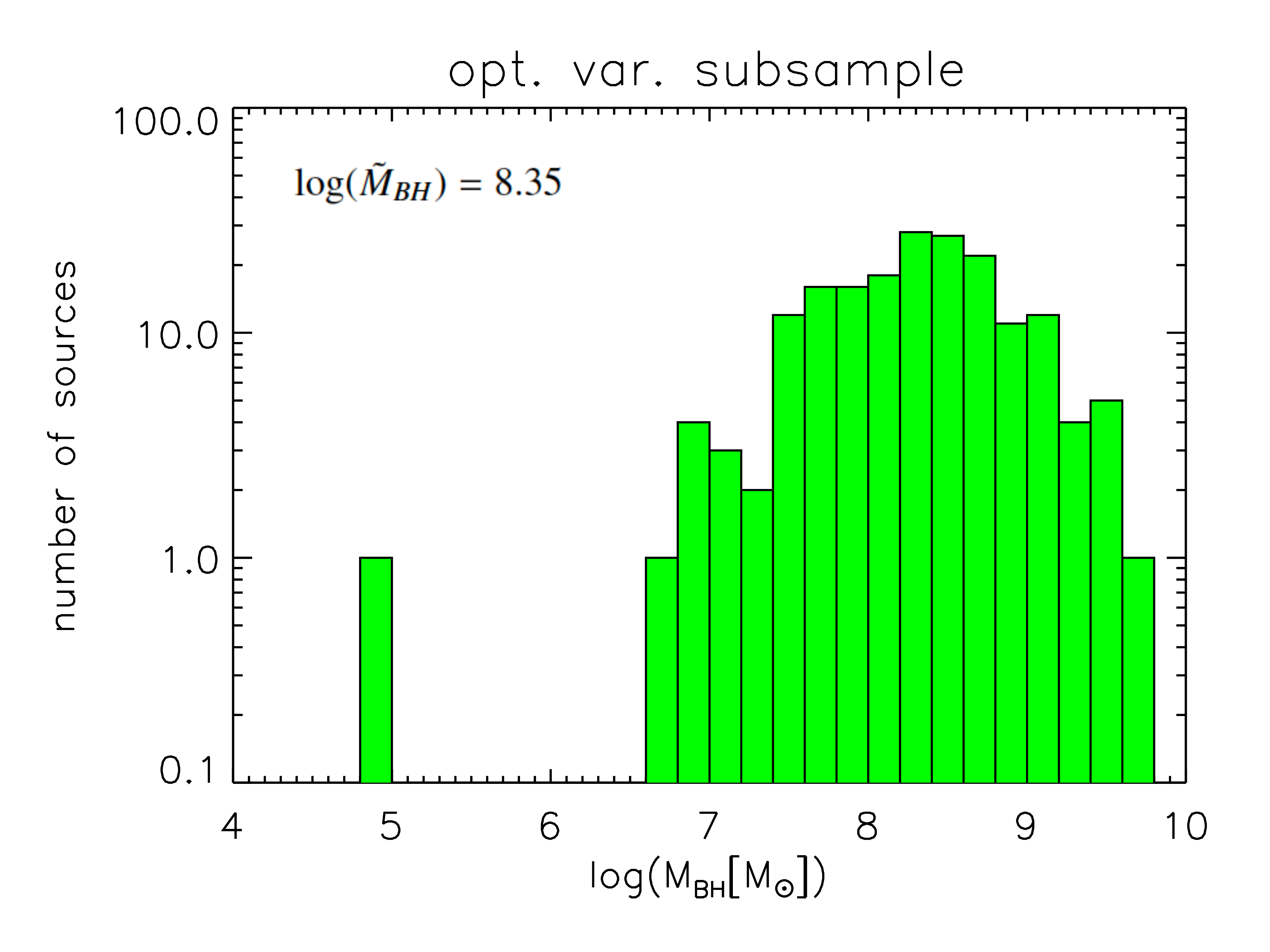}}
\subfigure
            {\includegraphics[width=6cm]{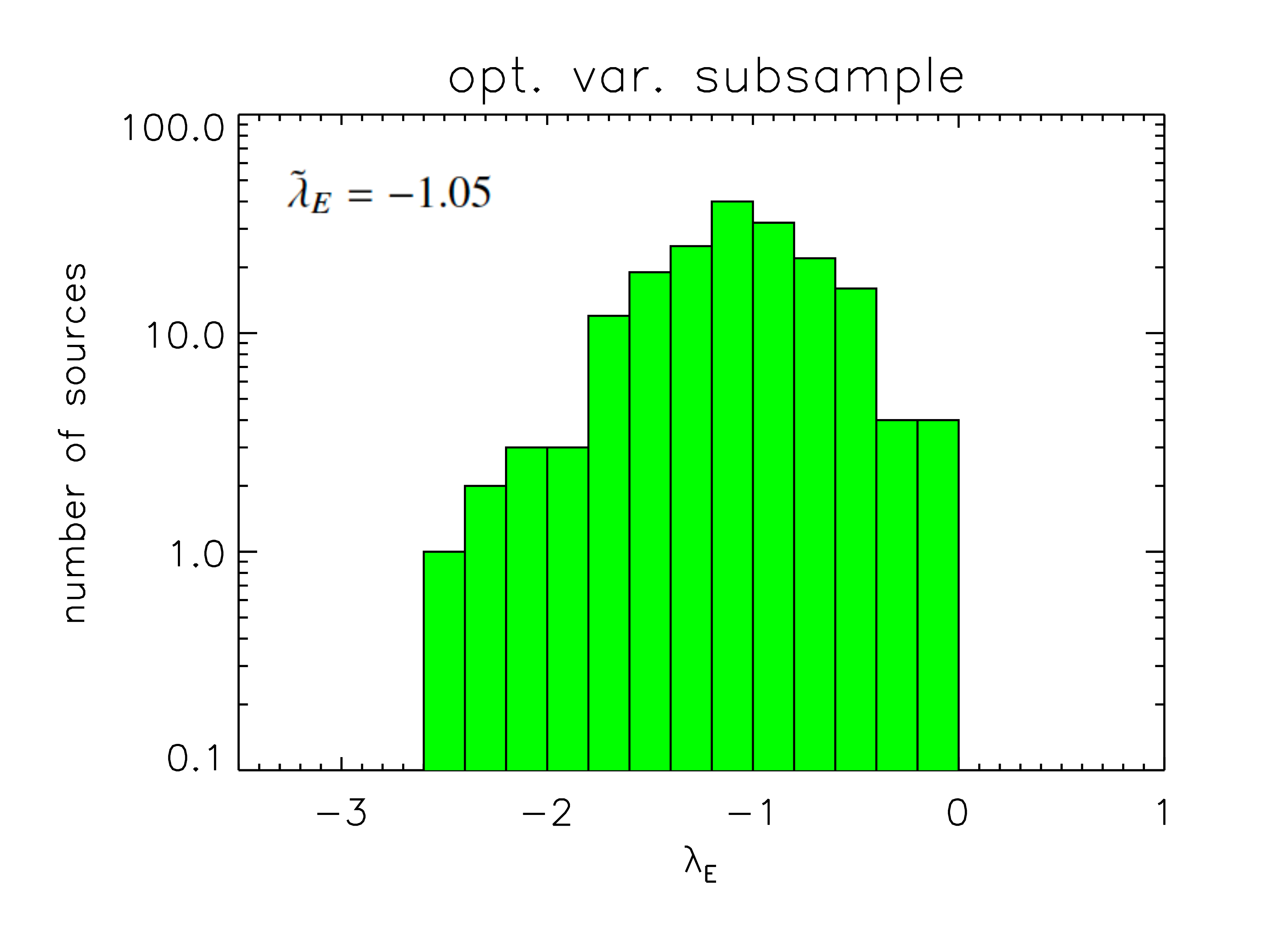}}
\subfigure
            {\includegraphics[width=6cm]{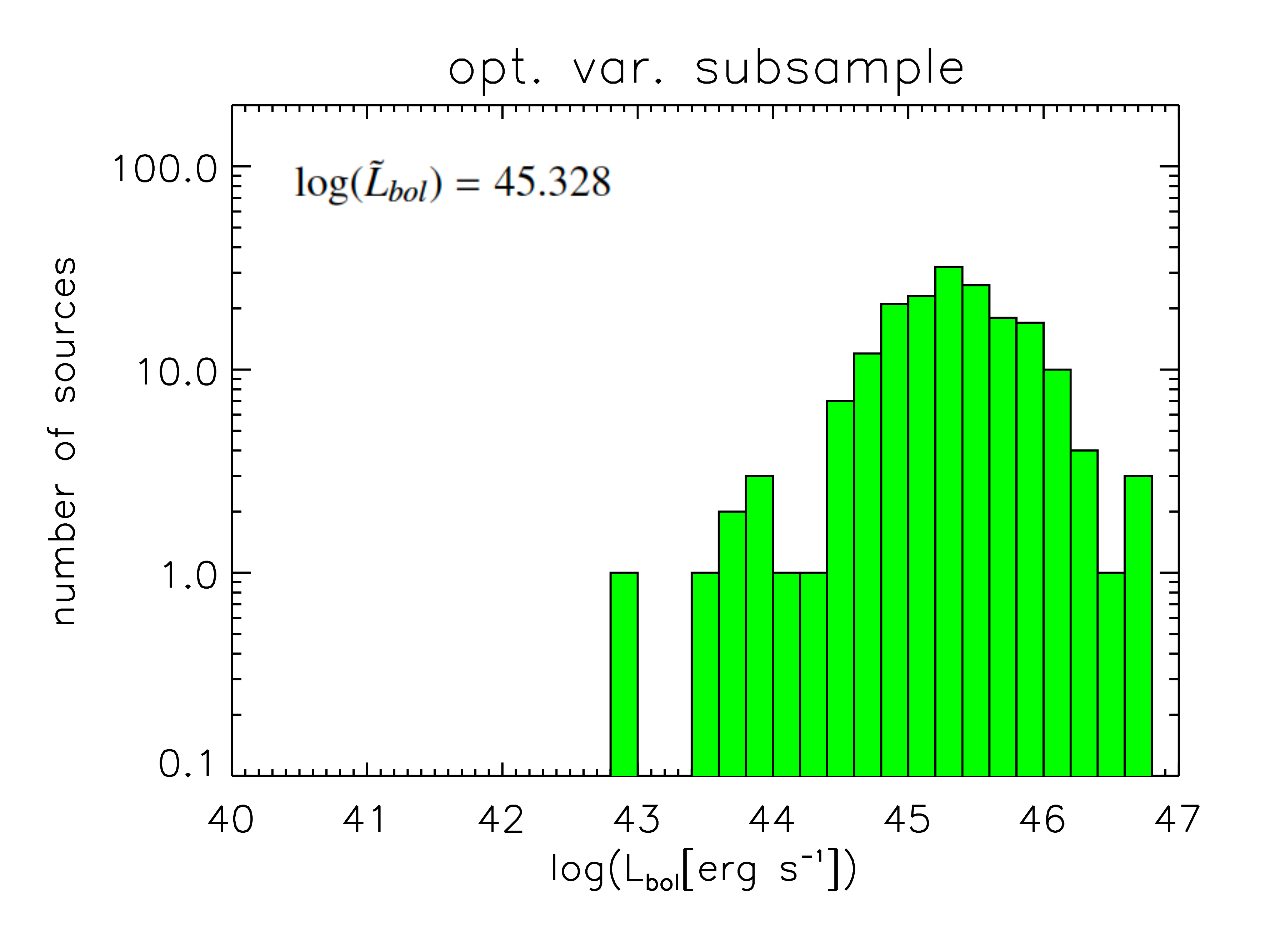}}\\
\subfigure
            {\includegraphics[width=6cm]{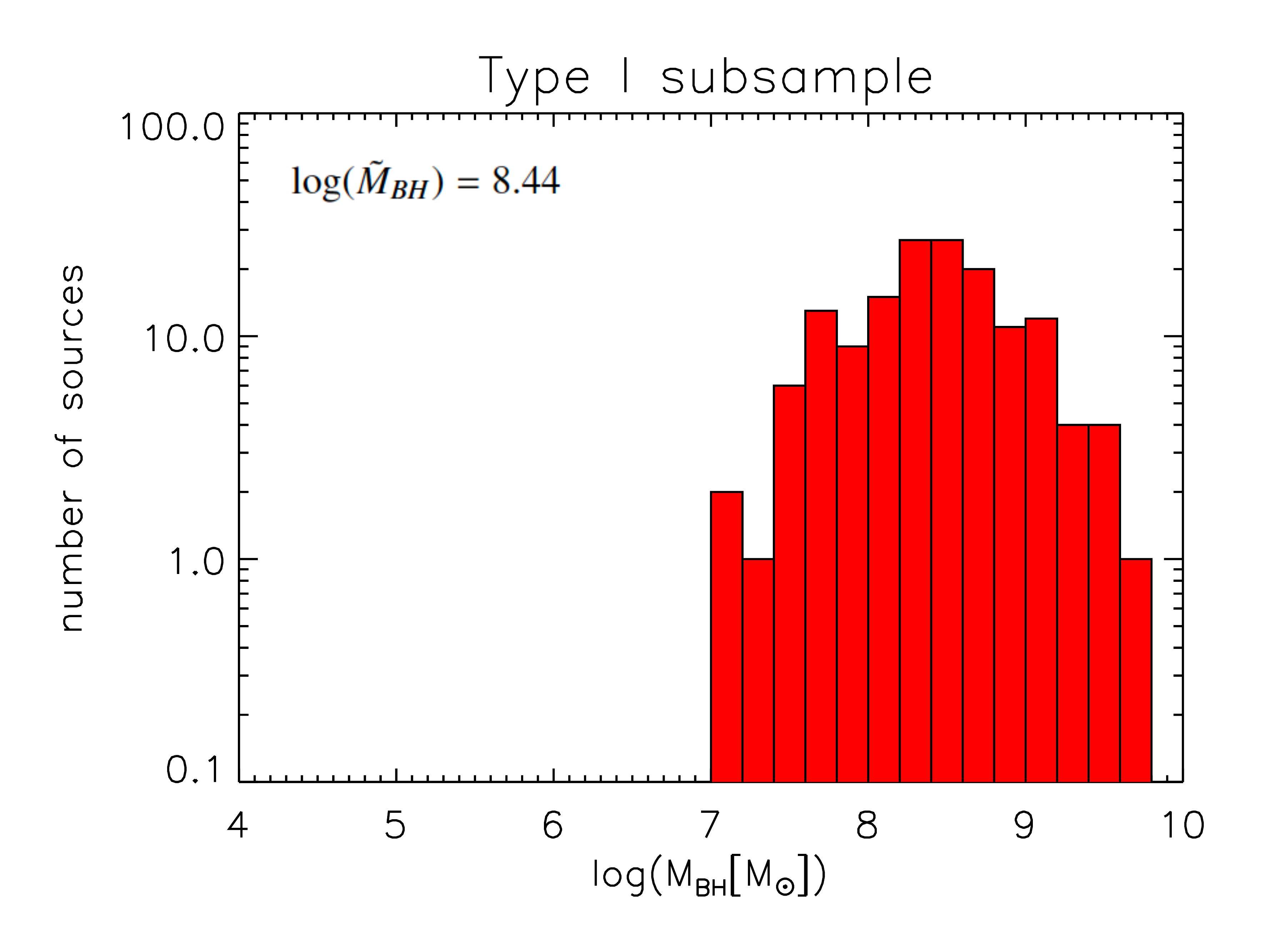}}
\subfigure
            {\includegraphics[width=6cm]{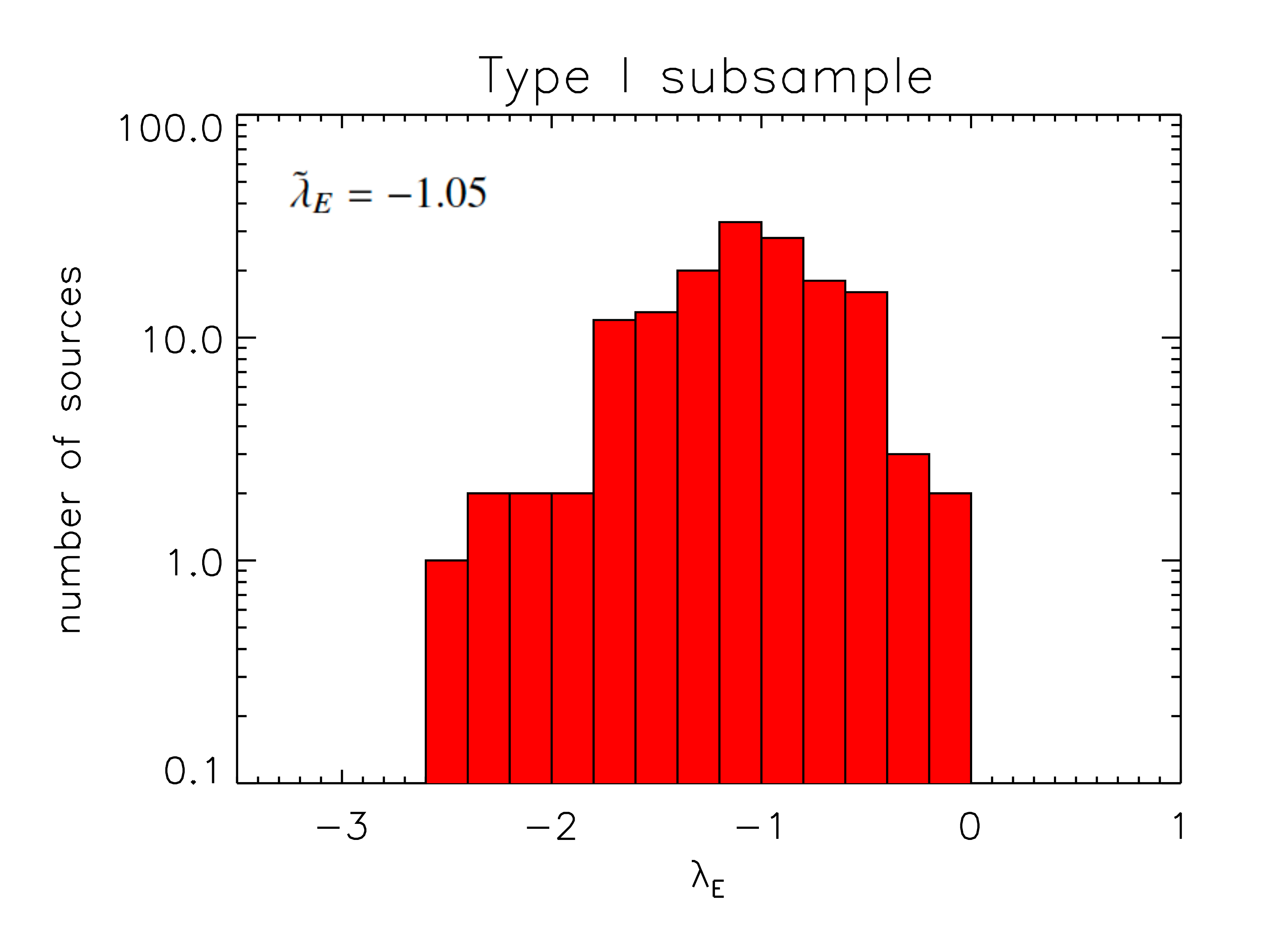}}
\subfigure
            {\includegraphics[width=6cm]{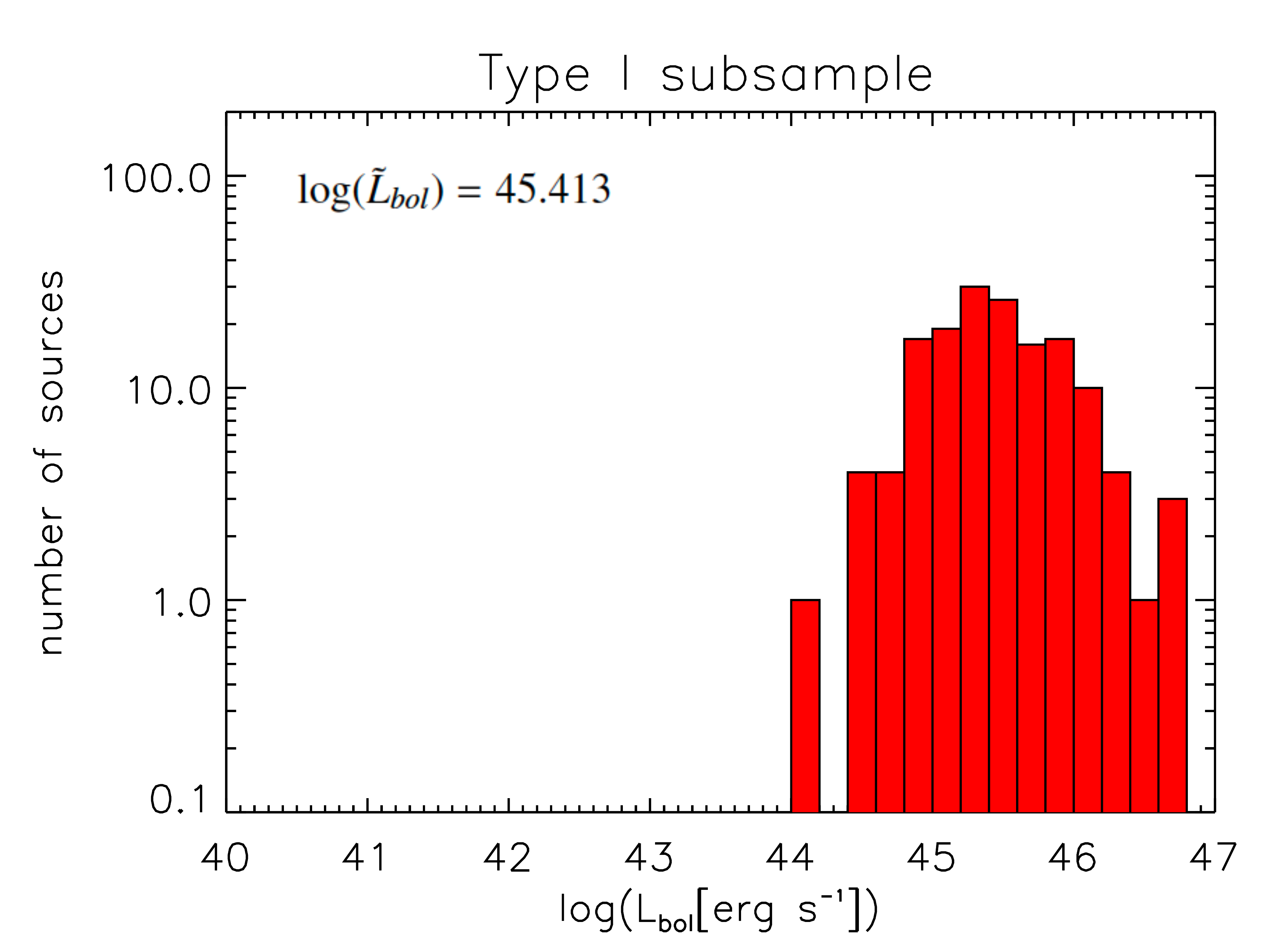}}\\
   \caption{Black-hole mass (\emph{left column}), $\lambda_E$ (\emph{middle column}), and $L_{bol}$ (\emph{right column}) distributions for the \emph{main} sample of 677 AGN (\emph{top line}), and for the three subsamples of AGN selected to investigate dependence on these physical properties: AGN selected via their MIR properties, AGN selected via optical variability, and Type I AGN confirmed by spectroscopy (\emph{second to bottom lines}). Each panel reports the (logarithm of the) median value of the analyzed physical quantity for the corresponding sample; $M_{BH}$ are in solar mass units and bolometric luminosities are in erg s$^{-1}$. Details about the selection criteria are reported in Section \ref{section:samples}.}\label{fig:hists}
   \end{figure*}

Here we analyze the possible dependence of the SF on the $M_{BH}$, $\lambda_E$, and $L_{bol}$ of our AGN. For the luminosity we limit the analysis to the sample of sources for which $M_{BH}$ and $\lambda_E$ estimates are available, in order to focus on the same sets of objects for each test. Figure \ref{fig:hists}, shows the distributions of $M_{BH}$, $\lambda_E$, and $L_{bol}$ values, respectively, for each of the analyzed samples. We note that both $M_{BH}$ and $\lambda_E$ distributions cover larger ranges and, in particular, include lower values than several past works such as, e.g., \citet{Bauer,simm}.

\begin{figure*}[th]
 \centering
\subfigure
            {\includegraphics[width=5.3cm]{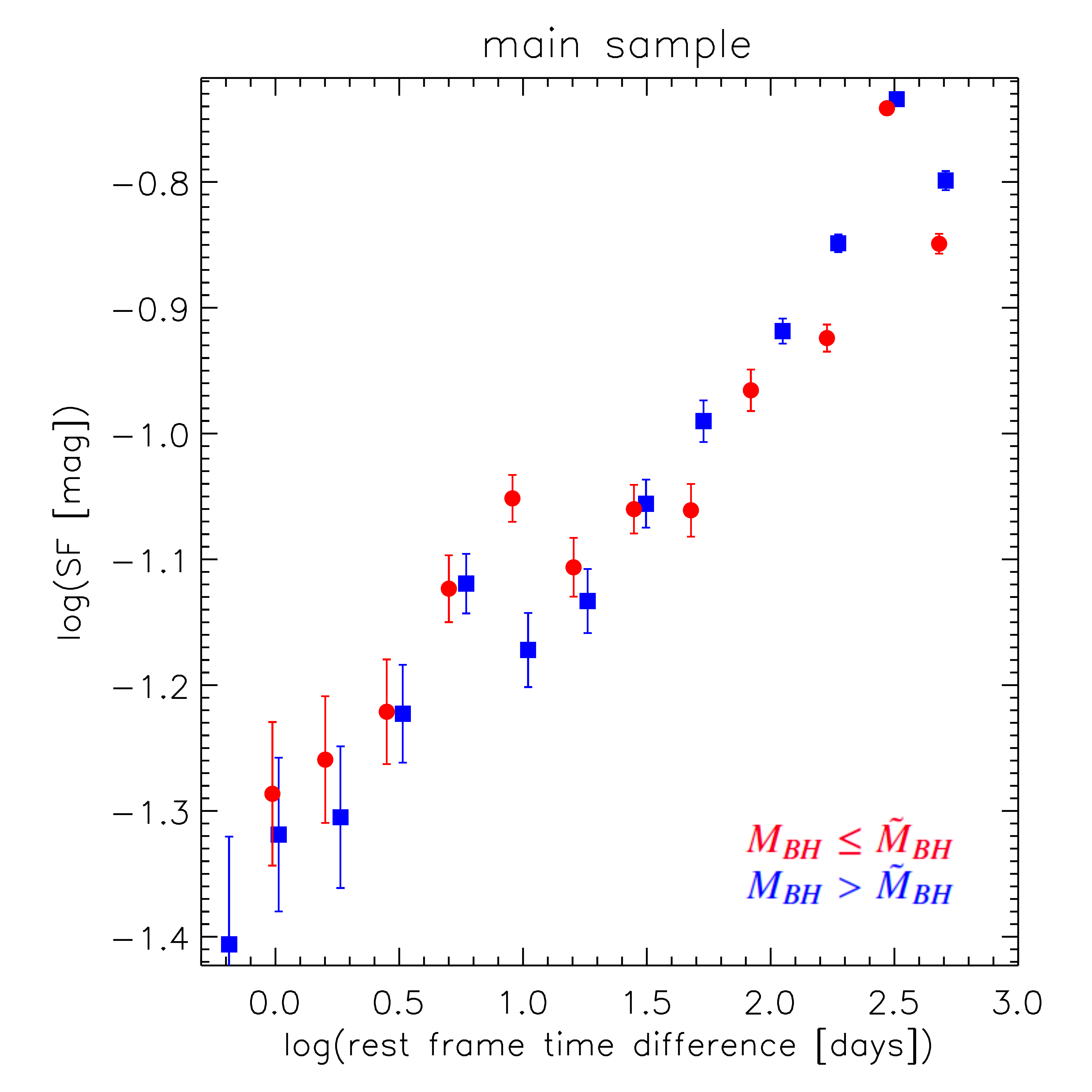}}
\subfigure
            {\includegraphics[width=5.3cm]{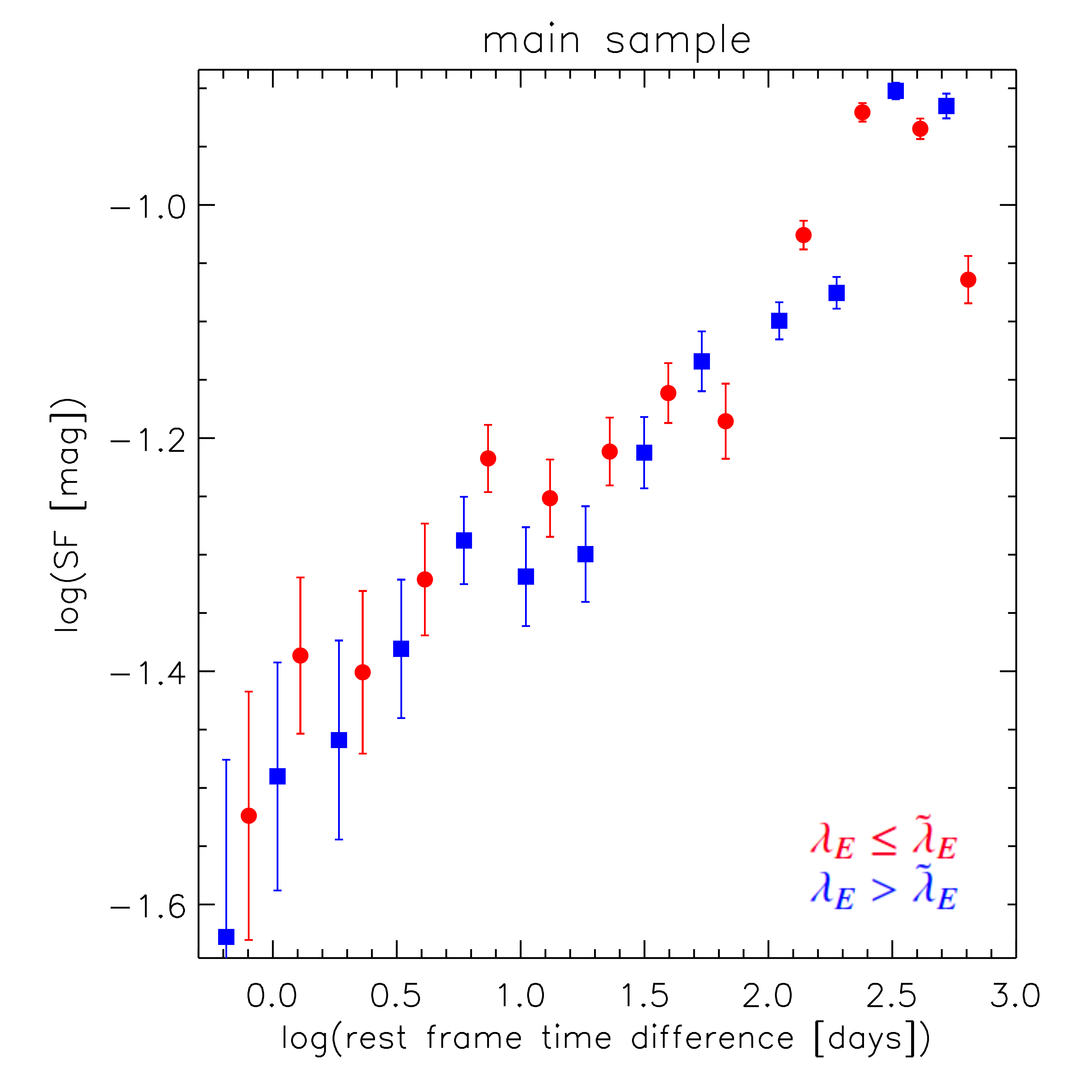}}
\subfigure
            {\includegraphics[width=5.3cm]{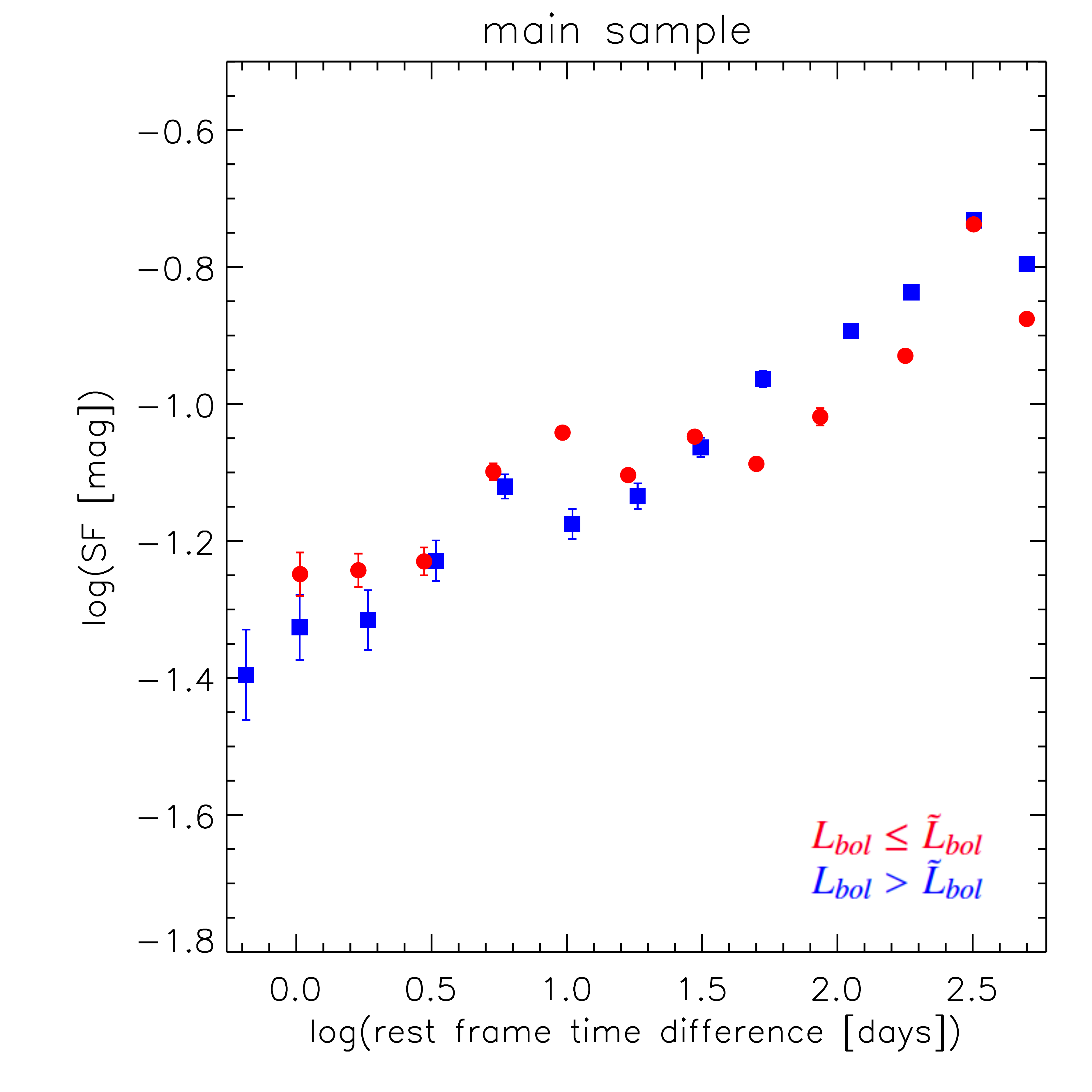}}\\
\subfigure
            {\includegraphics[width=5.3cm]{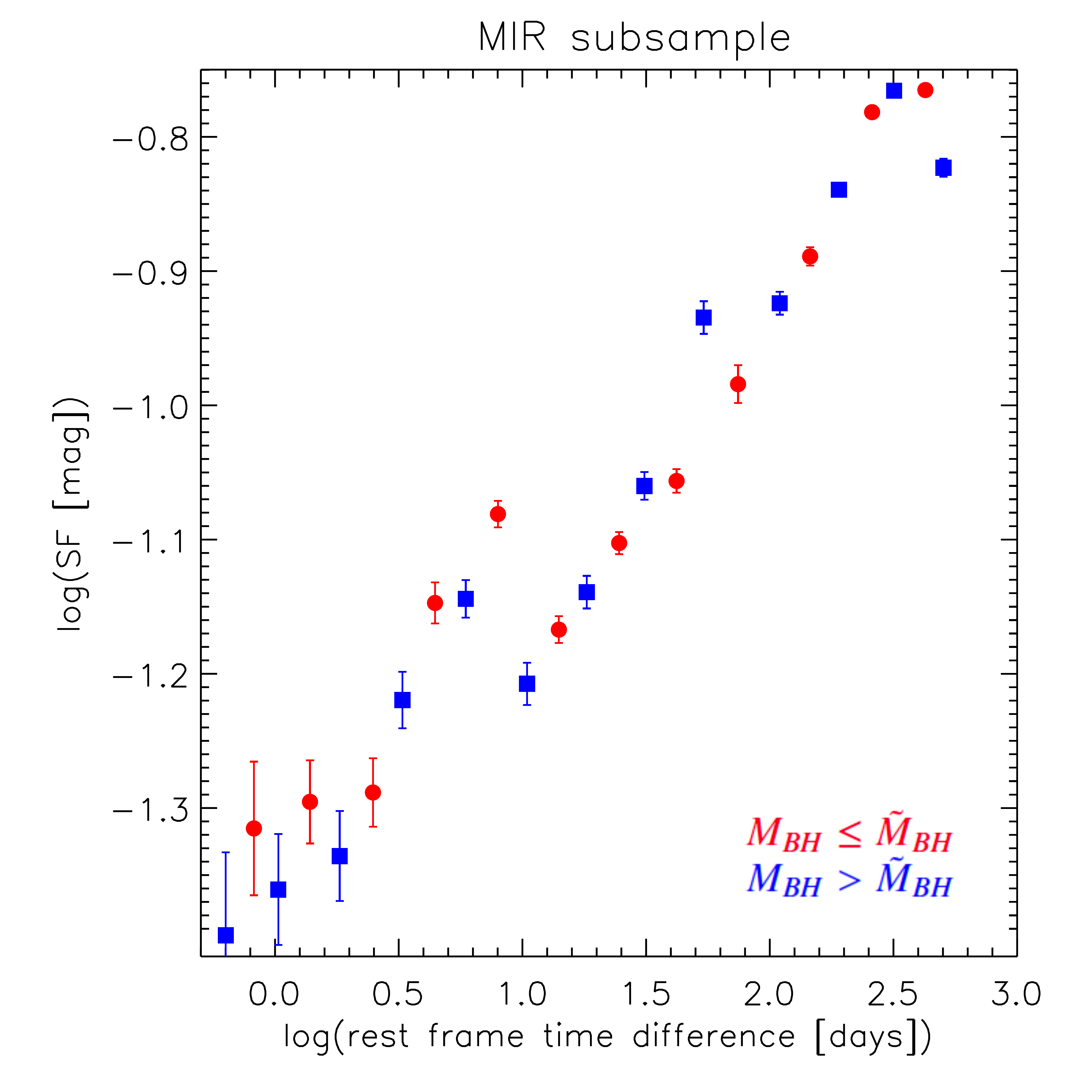}}
\subfigure
            {\includegraphics[width=5.3cm]{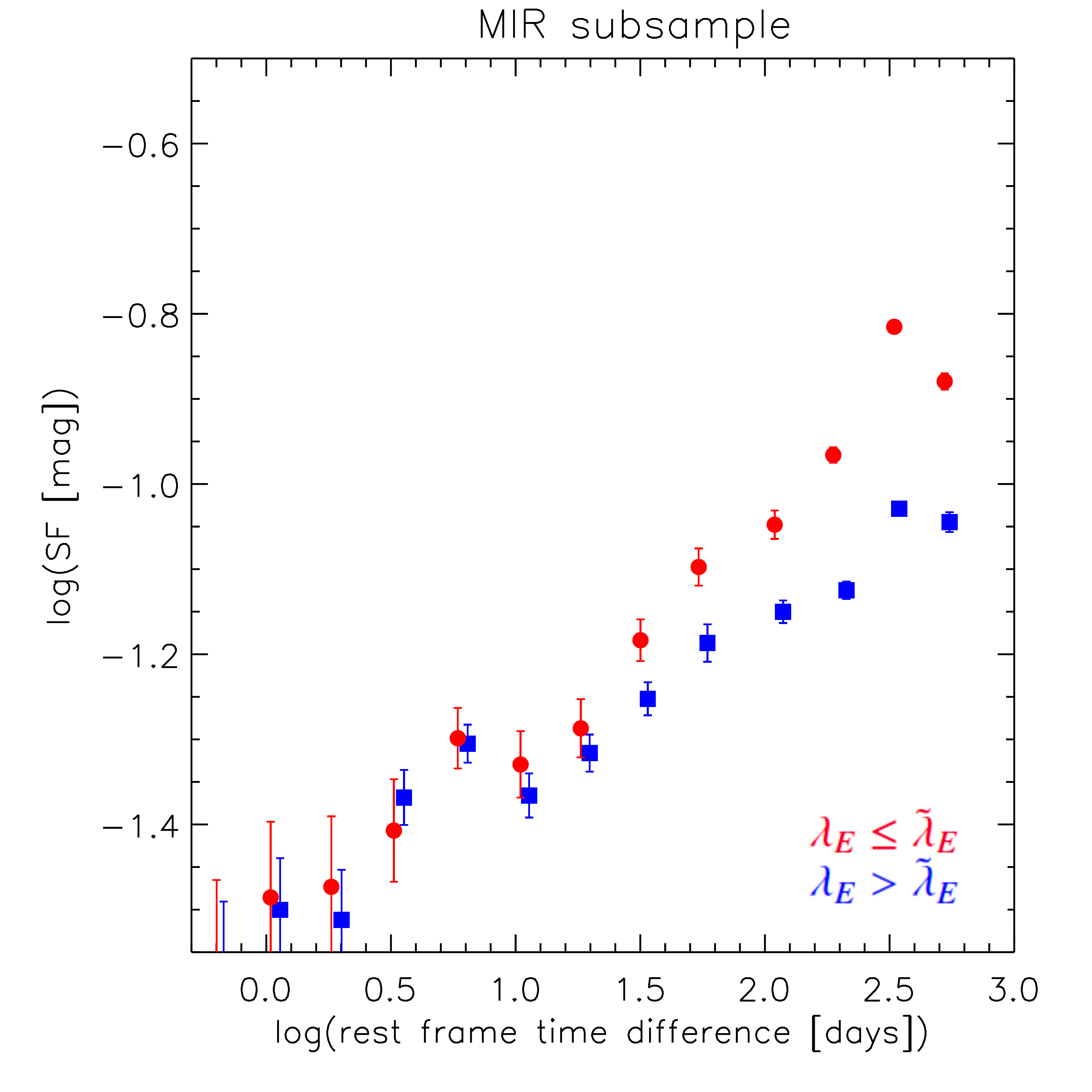}}
\subfigure
            {\includegraphics[width=5.3cm]{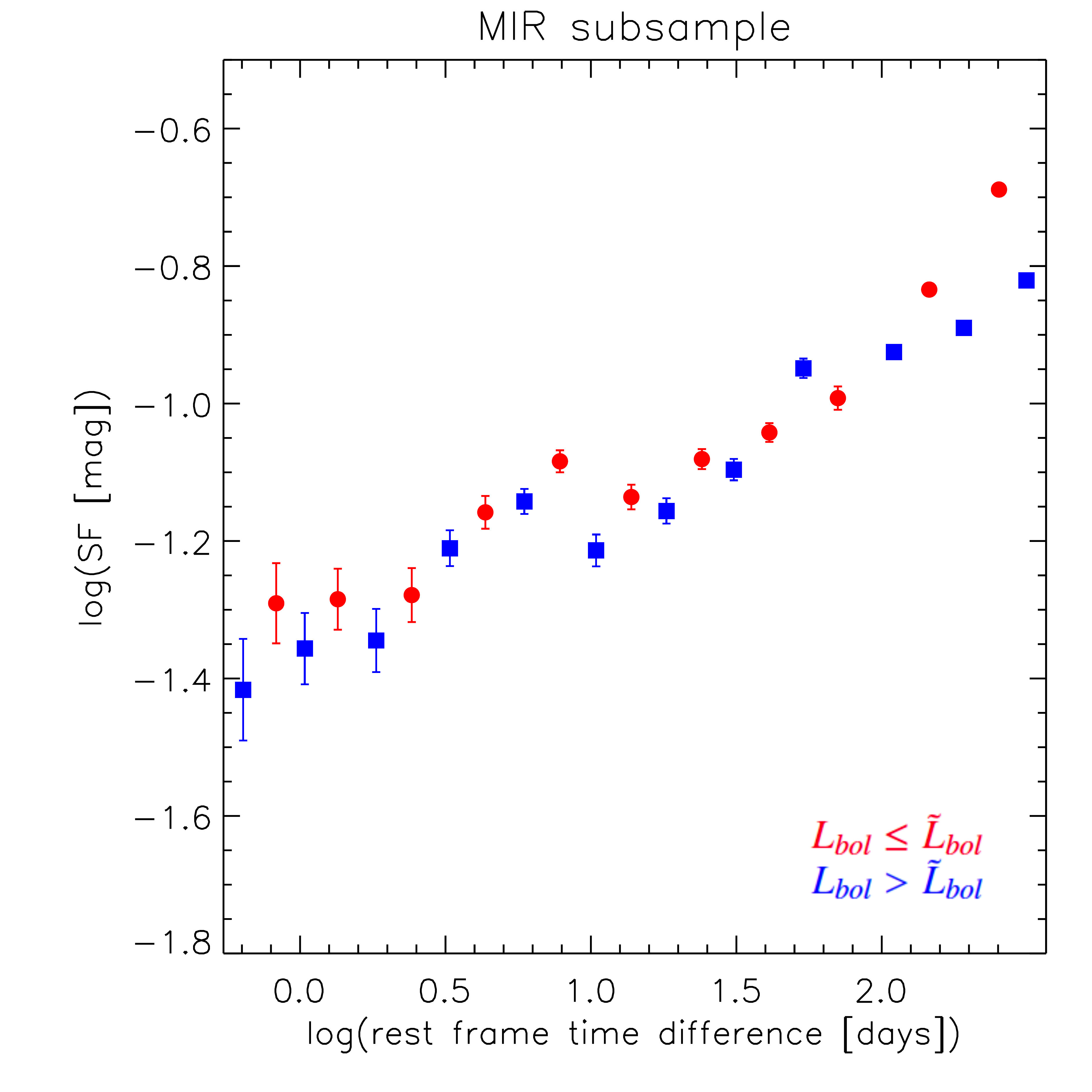}}\\
\subfigure
            {\includegraphics[width=5.3cm]{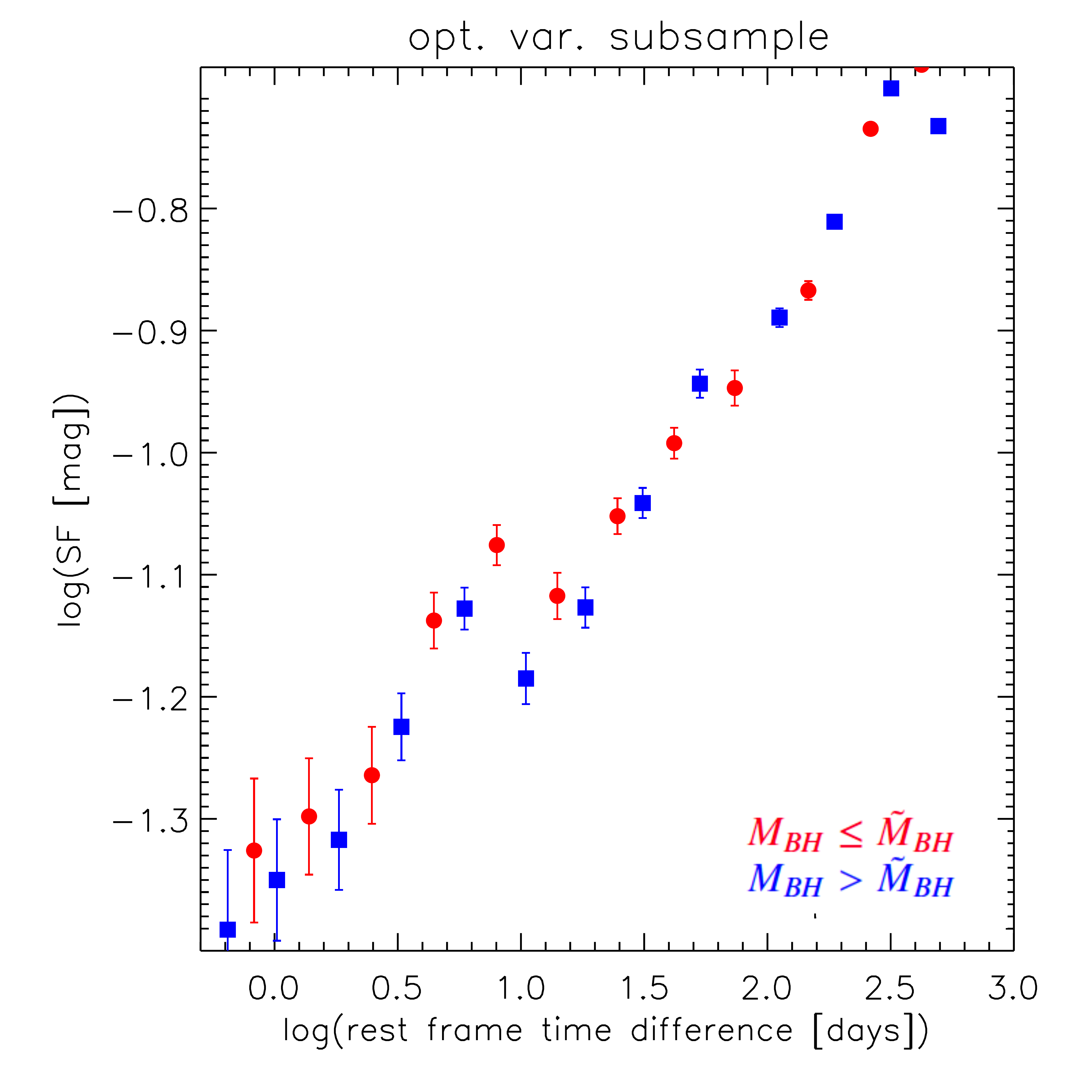}}
\subfigure
            {\includegraphics[width=5.3cm]{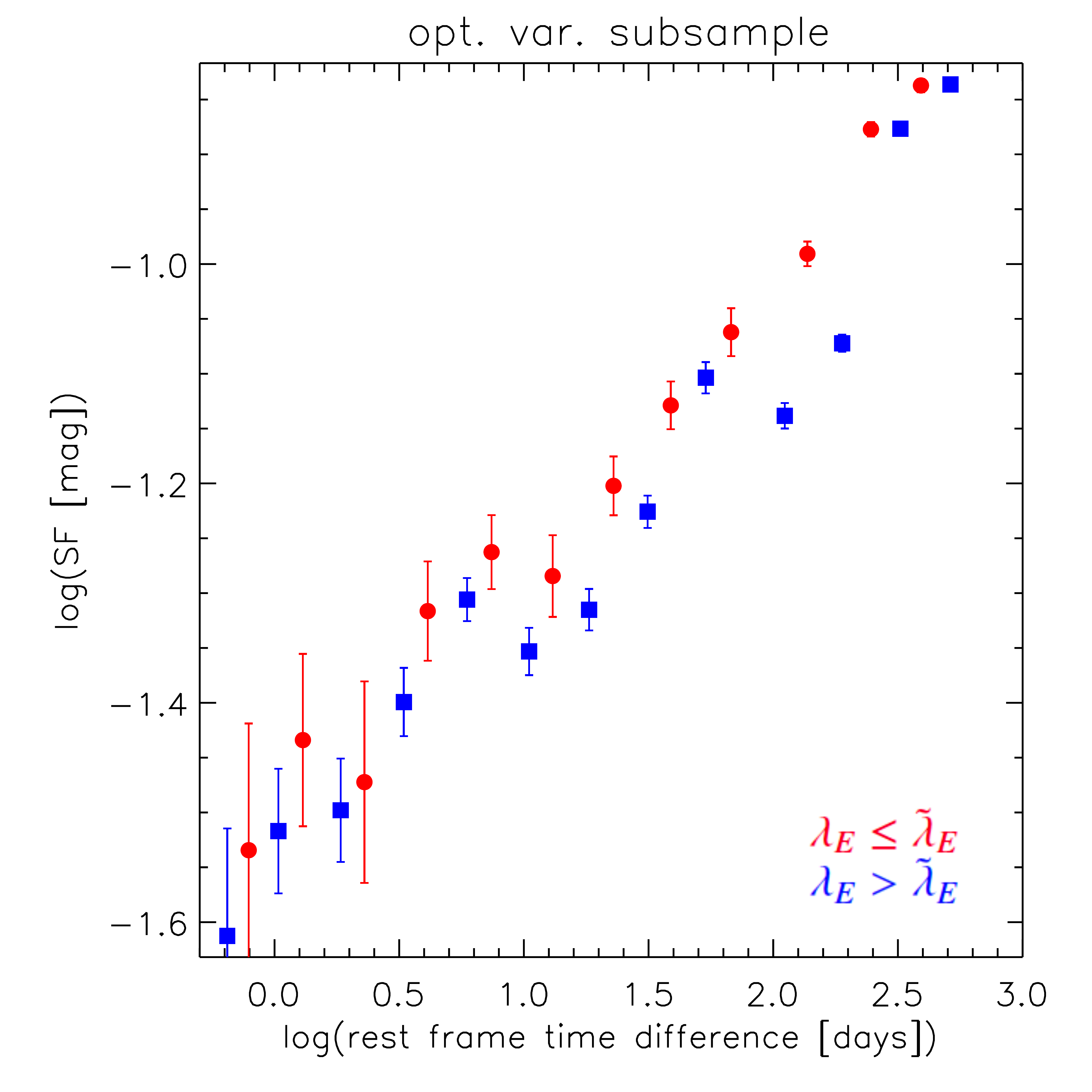}}
\subfigure
            {\includegraphics[width=5.3cm]{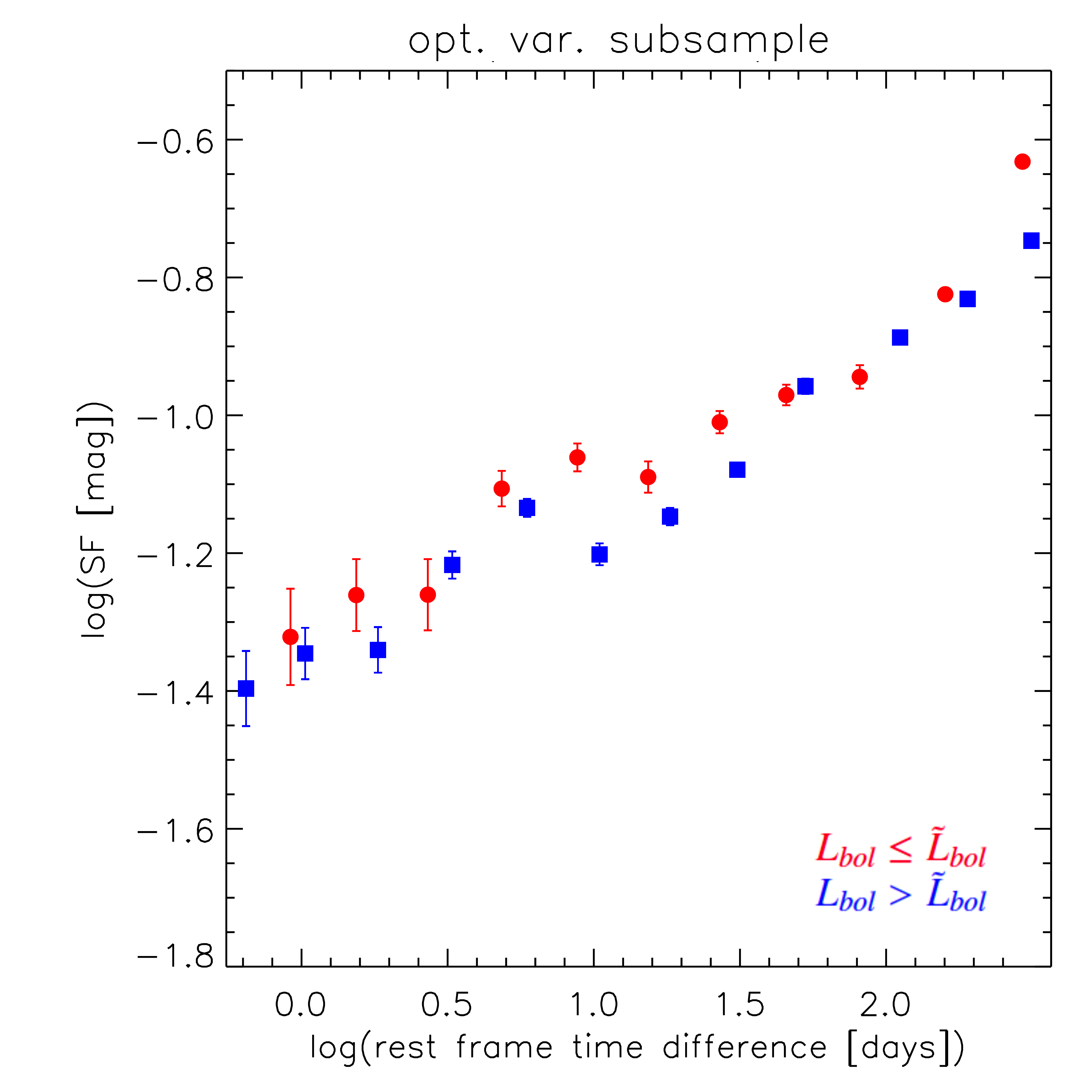}}\\
\subfigure
            {\includegraphics[width=5.3cm]{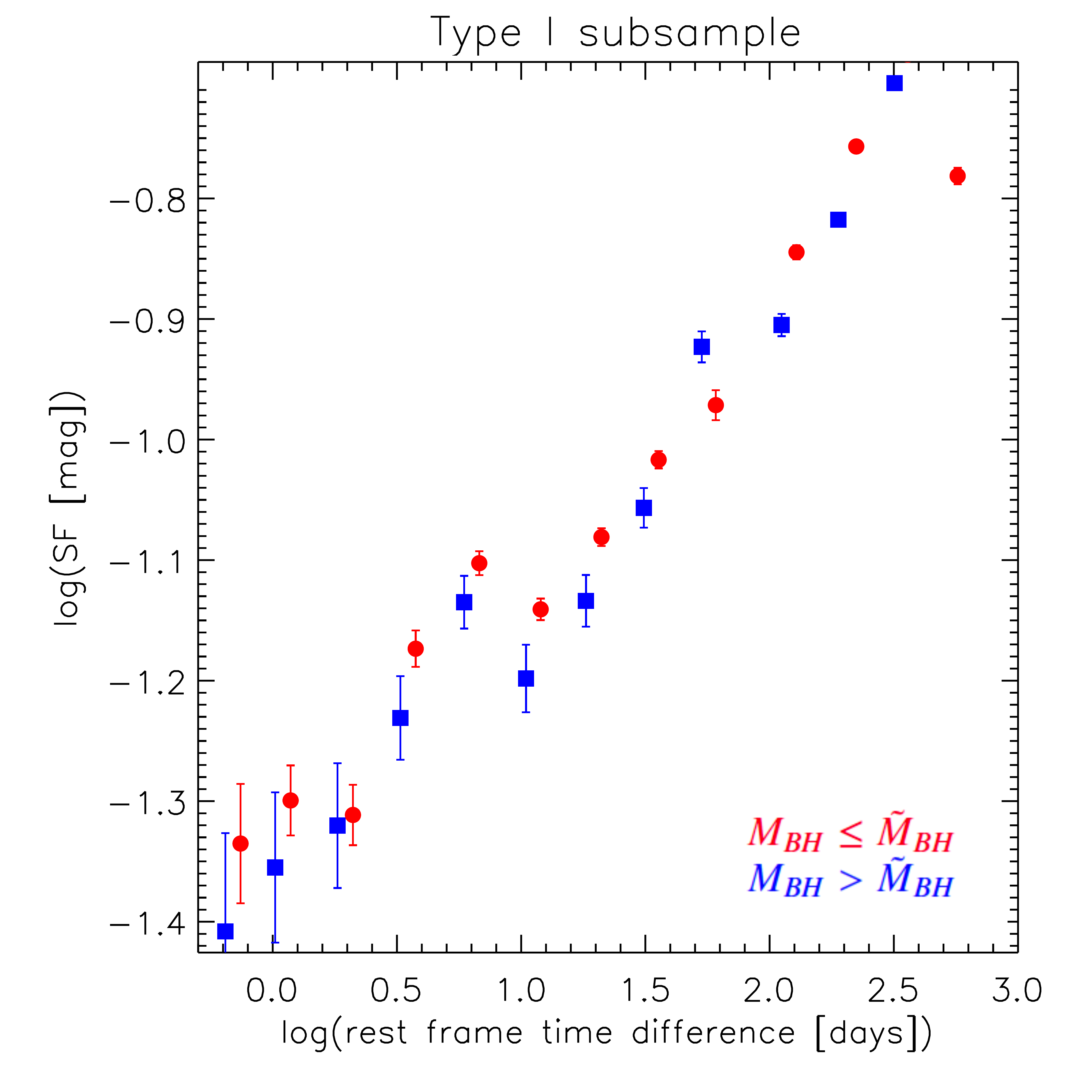}}
\subfigure
            {\includegraphics[width=5.3cm]{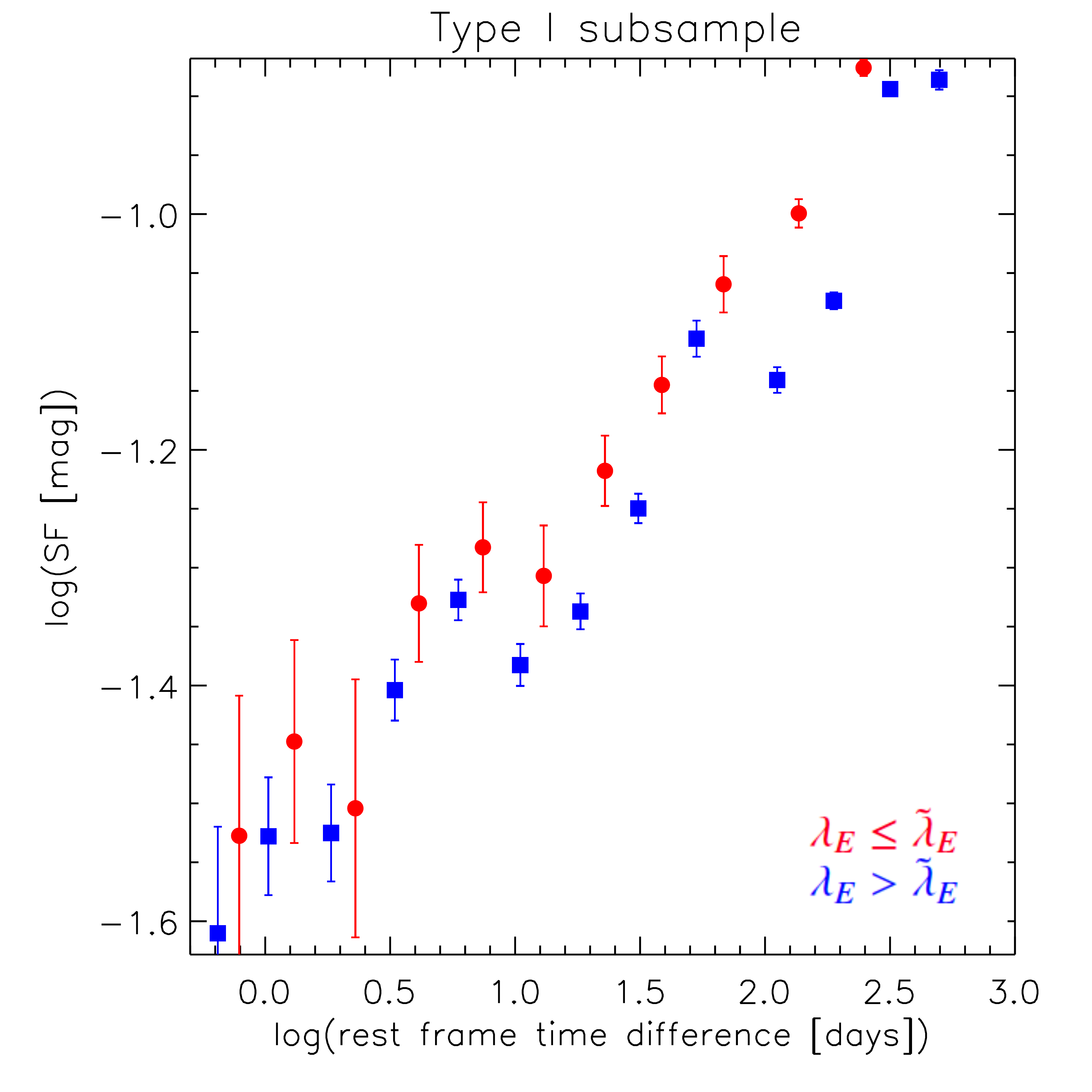}}
\subfigure
            {\includegraphics[width=5.3cm]{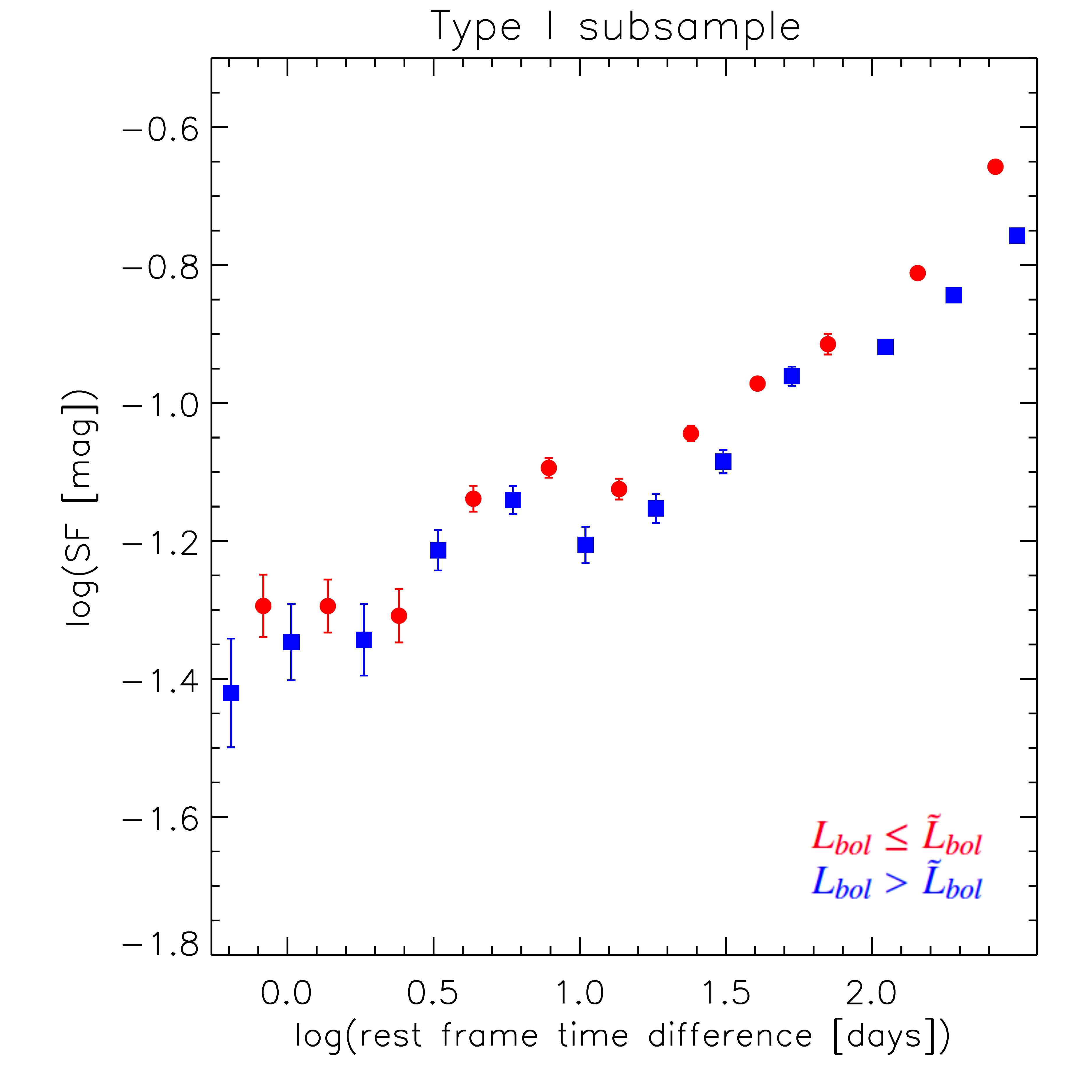}}\\
  \caption{SF for the four samples of AGN for which we investigate possible dependencies on $M_{BH}$ (\emph{left column}), $\lambda_E$ (\emph{middle column}), and $L_{bol}$ (\emph{right column}). The various panels show results for the \emph{main} sample, MIR AGN, optically variable AGN, and Type I AGN (\emph{top to bottom lines}). In each panel, the sample was divided into two subsets based on the (logarithm of the) median value of the physical property of interest: red dots indicate the samples of sources with lower $M_{BH}$/$\lambda_E$/$L_{bol}$ values and blue squares the sources with higher values. 
  }\label{fig:bel_analysis}
   \end{figure*}

\begin{figure*}[th]
 \centering
\subfigure
            {\includegraphics[width=5.3cm]{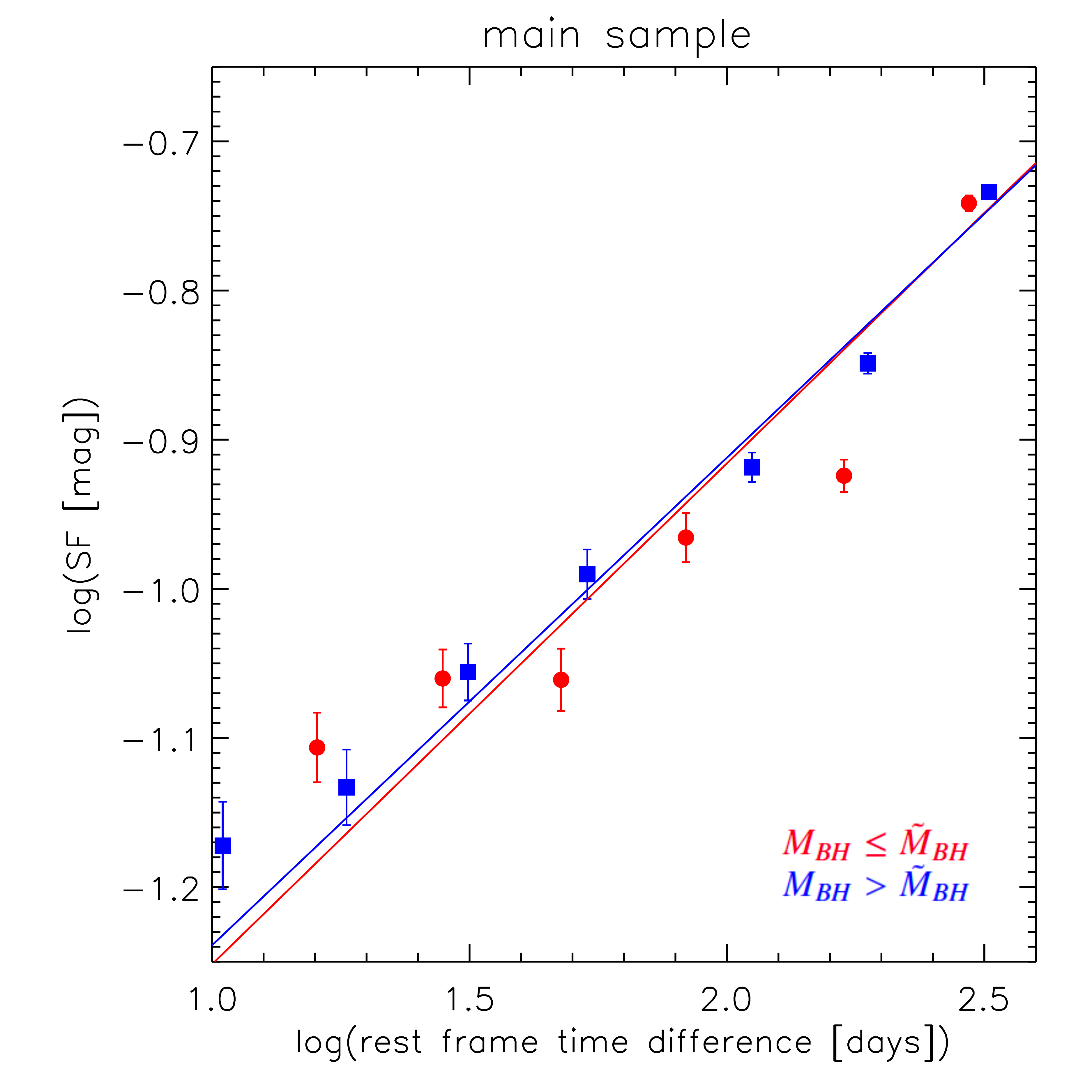}}
\subfigure
            {\includegraphics[width=5.3cm]{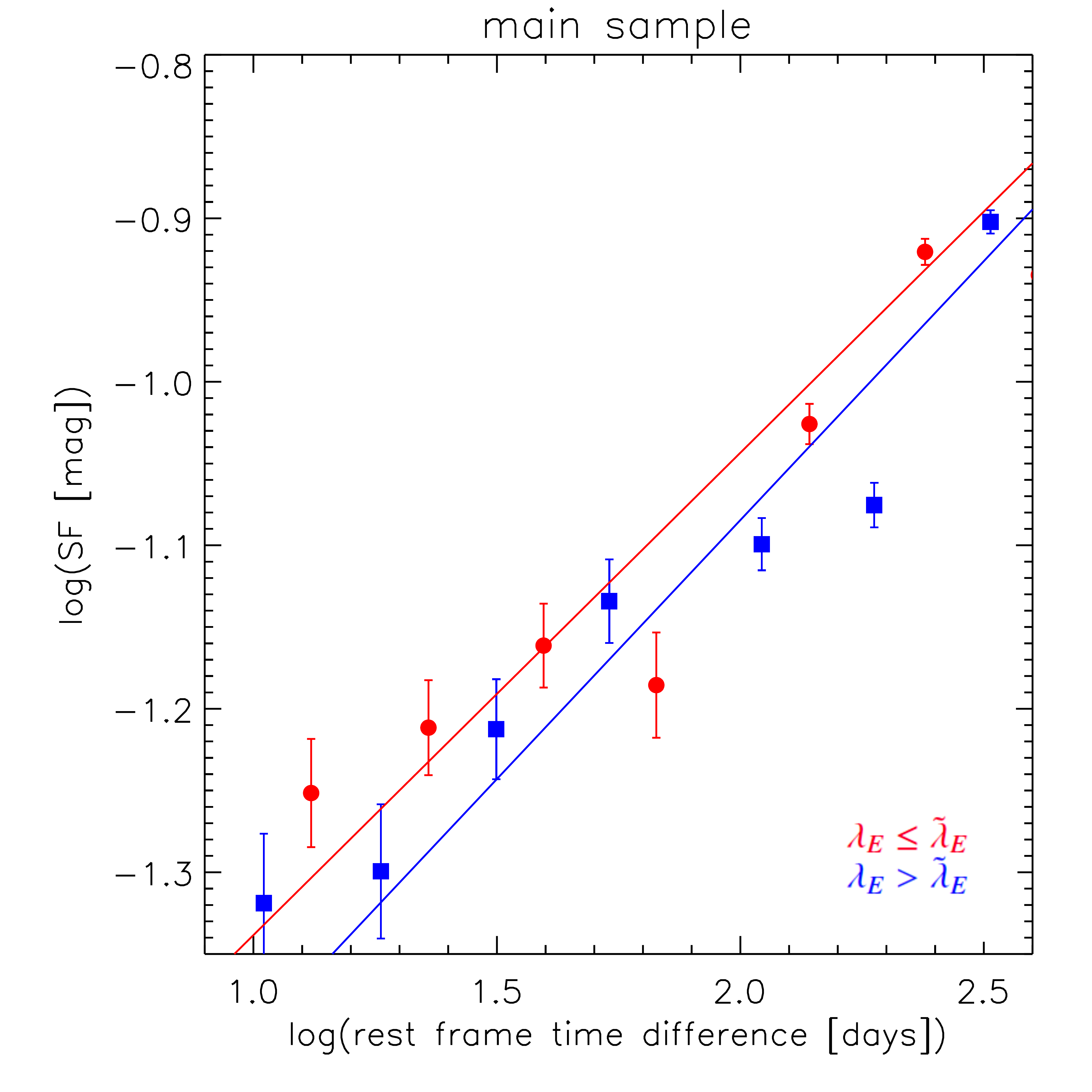}}
\subfigure
            {\includegraphics[width=5.3cm]{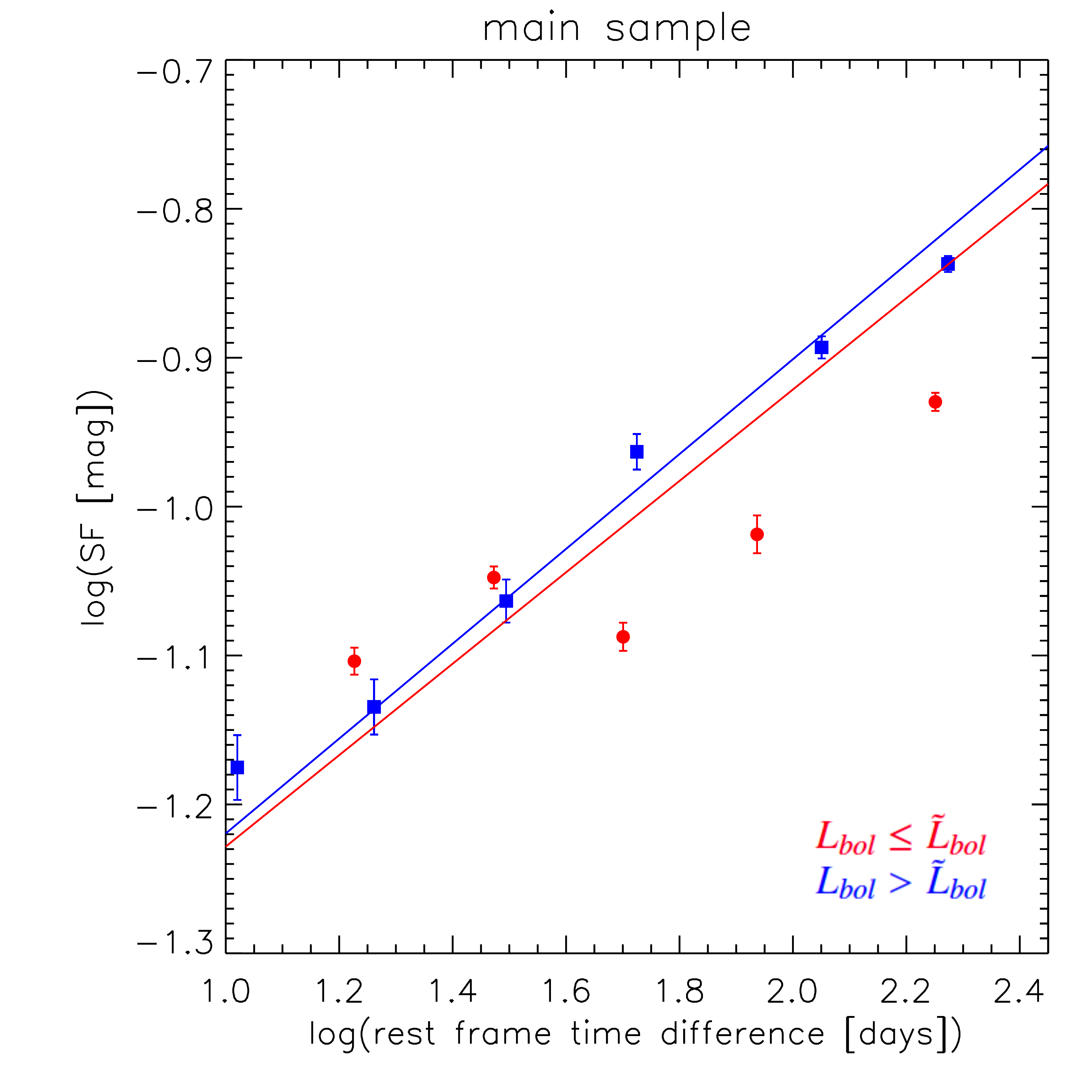}}\\
\subfigure
            {\includegraphics[width=5.3cm]{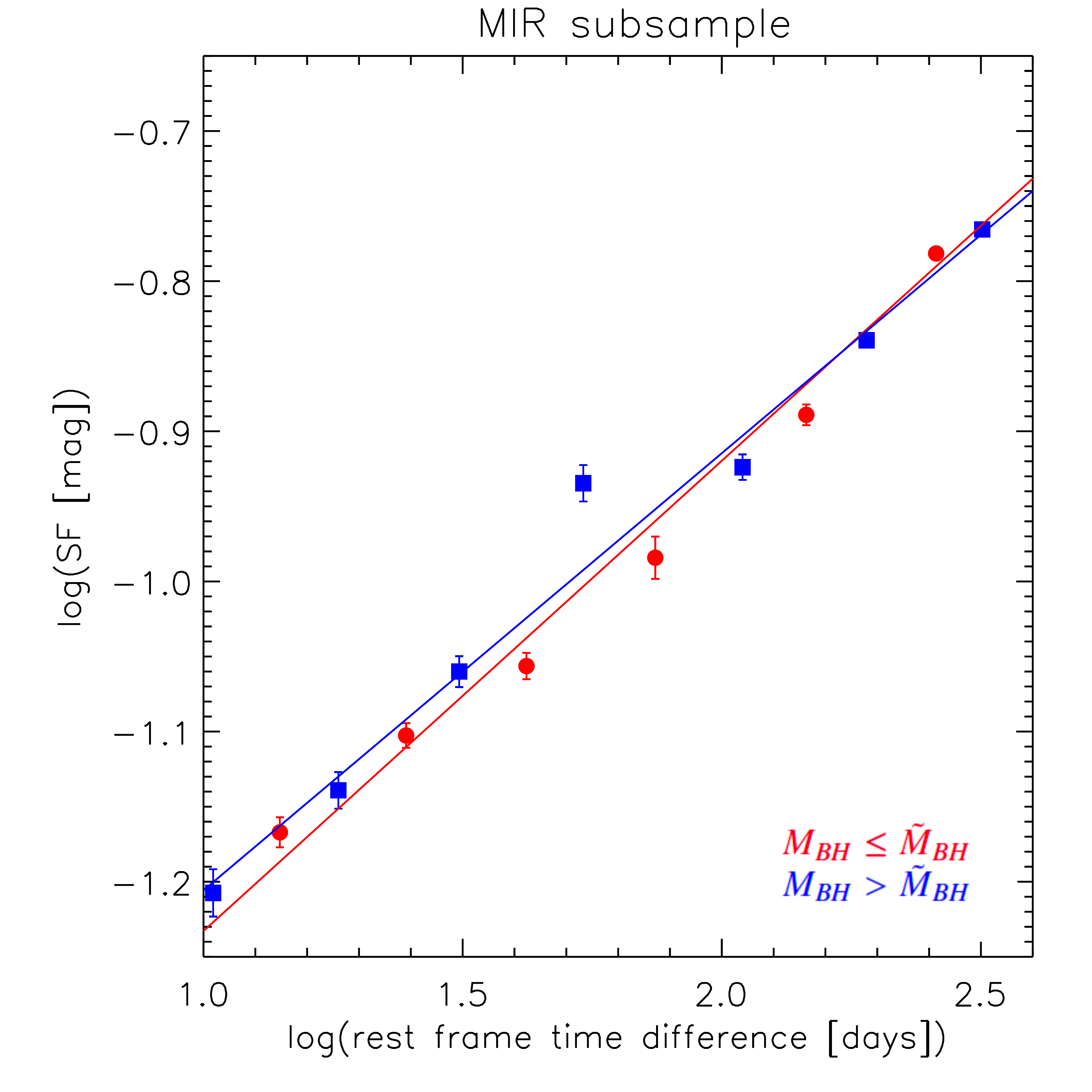}}
\subfigure
            {\includegraphics[width=5.3cm]{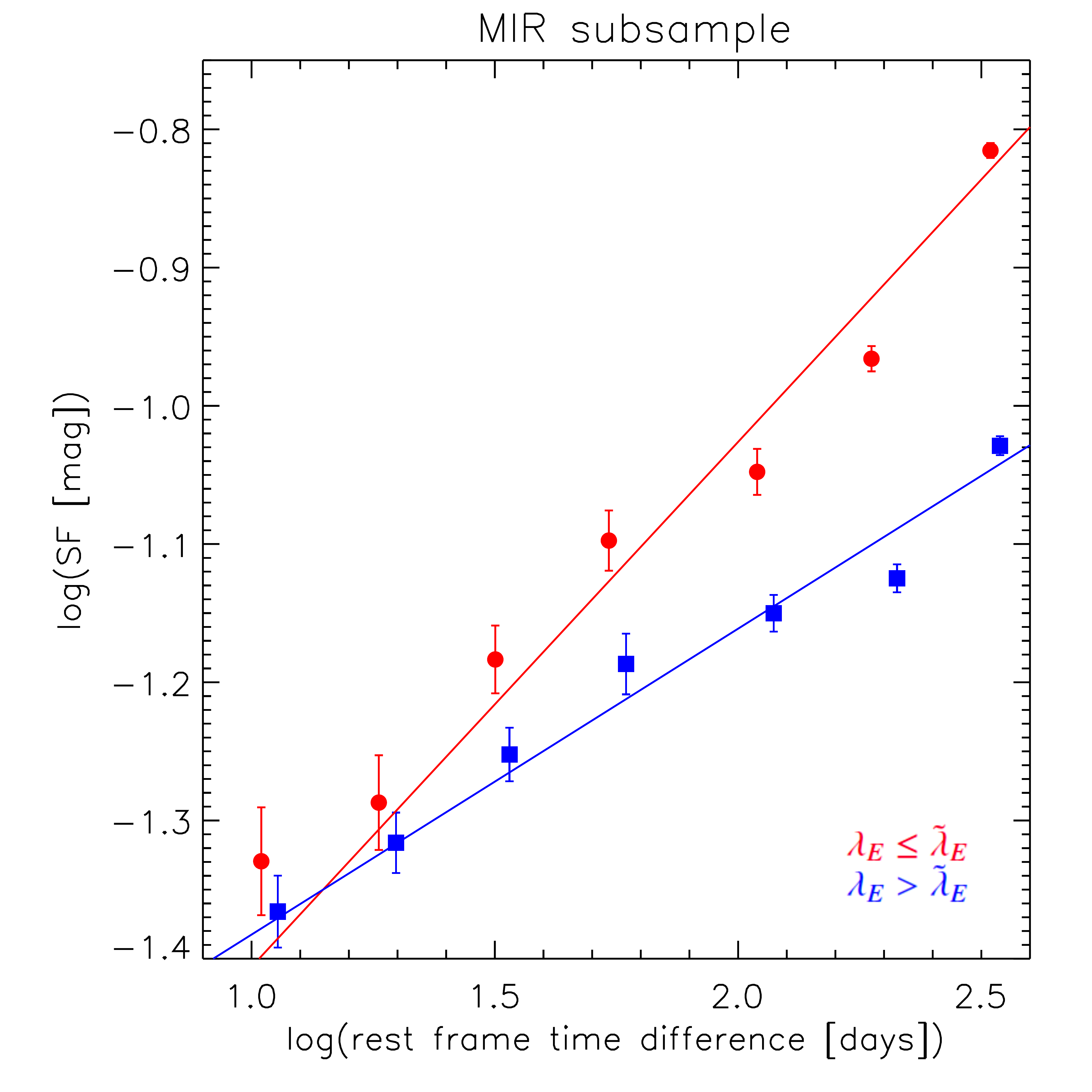}}
\subfigure
            {\includegraphics[width=5.3cm]{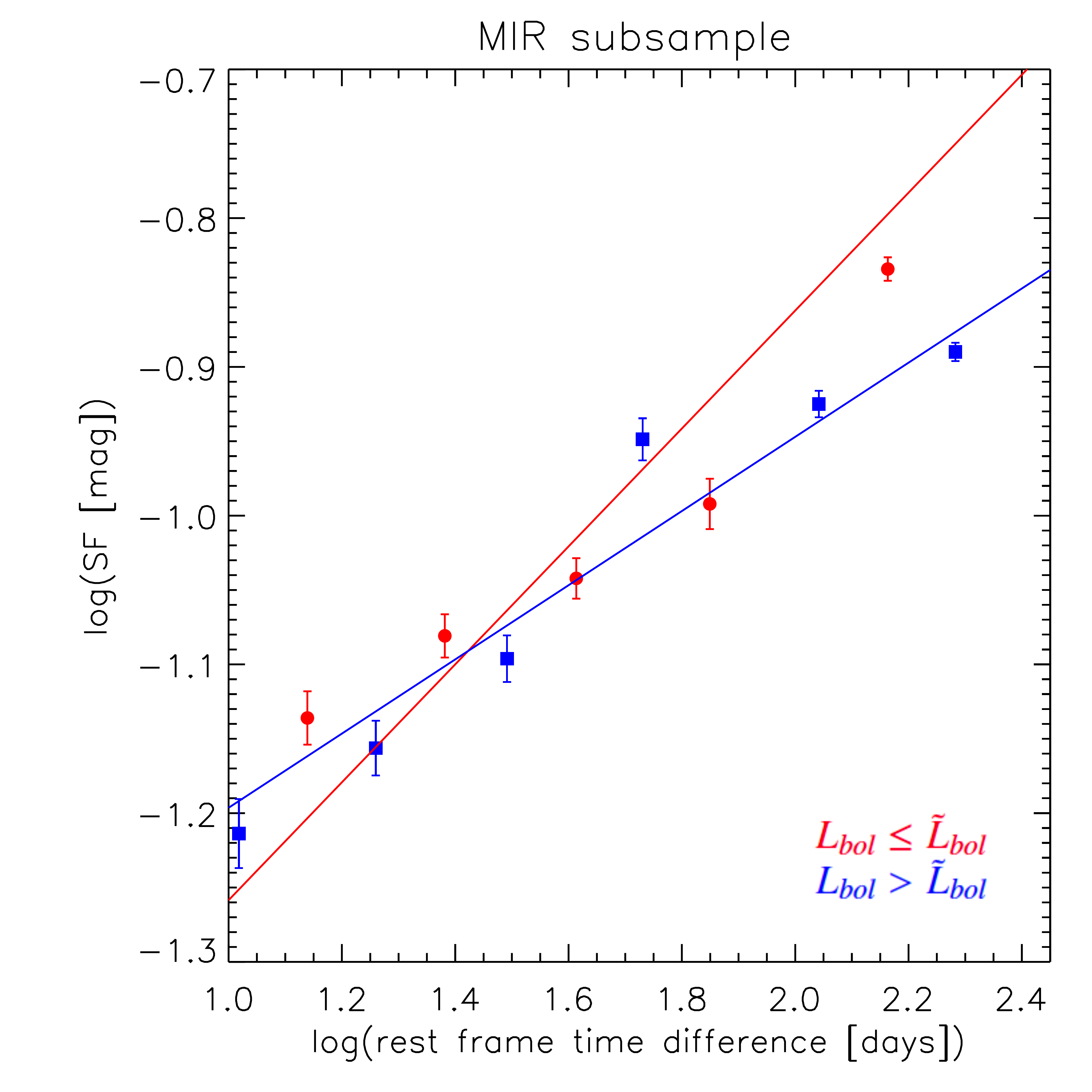}}\\
\subfigure
            {\includegraphics[width=5.3cm]{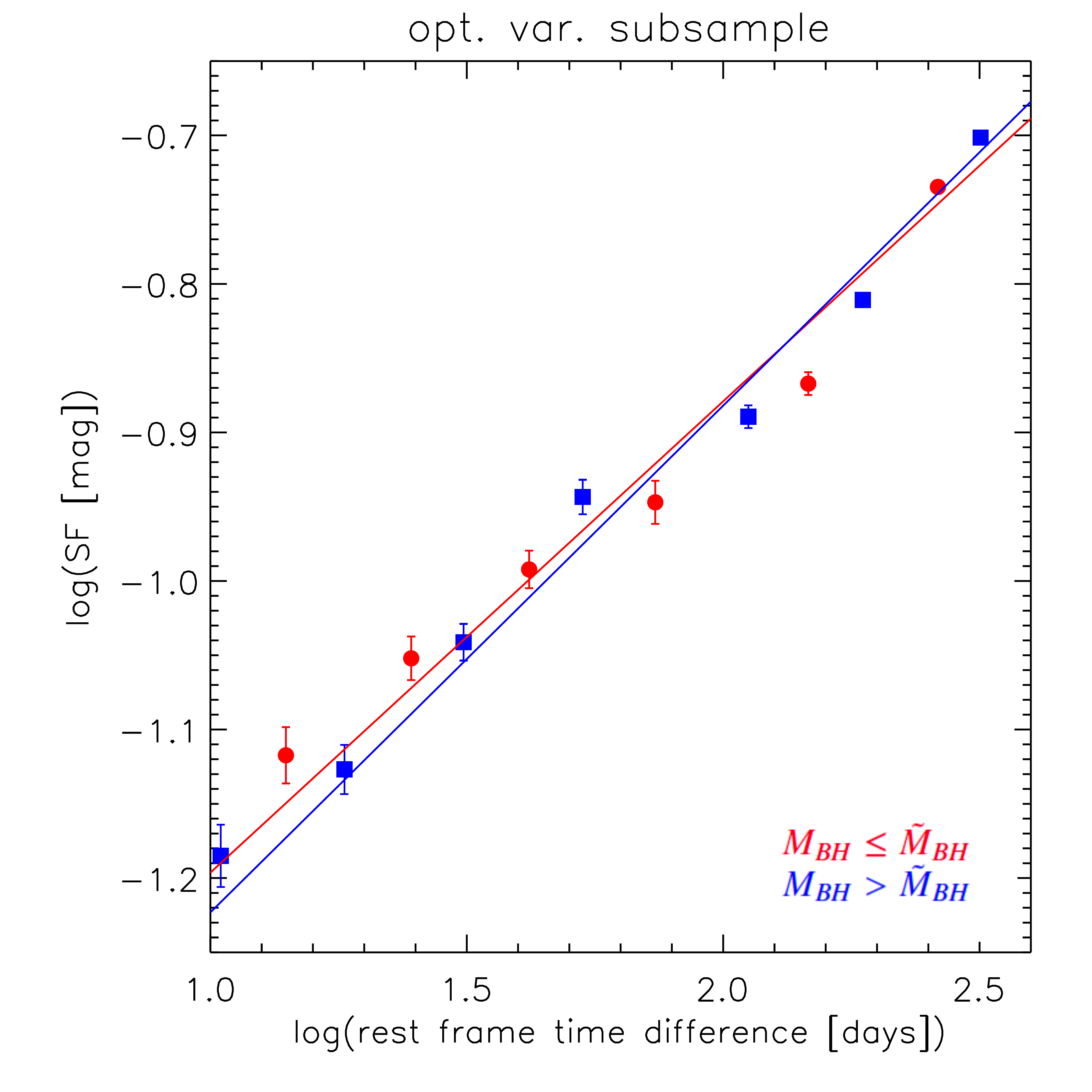}}
\subfigure
            {\includegraphics[width=5.3cm]{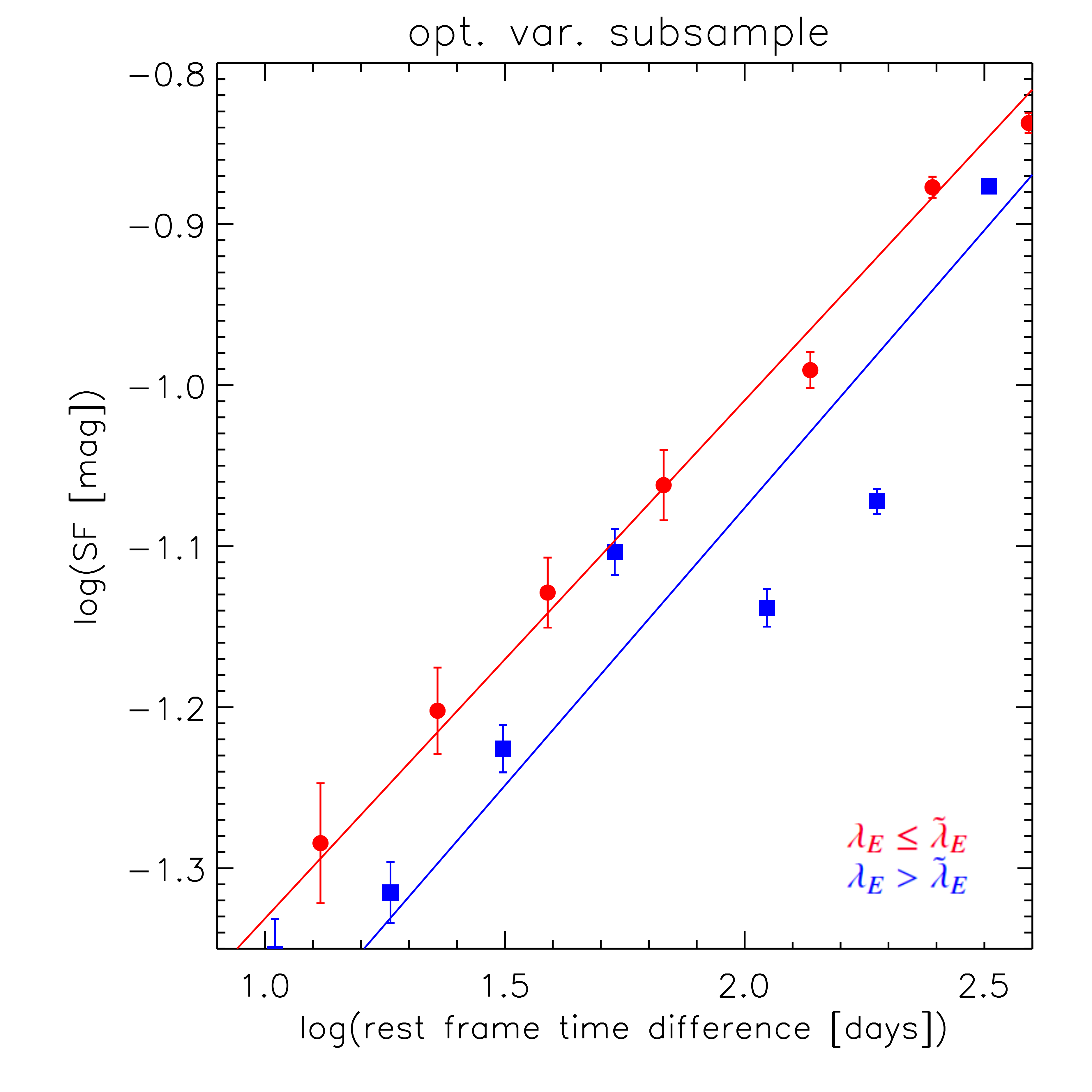}}
\subfigure
            {\includegraphics[width=5.3cm]{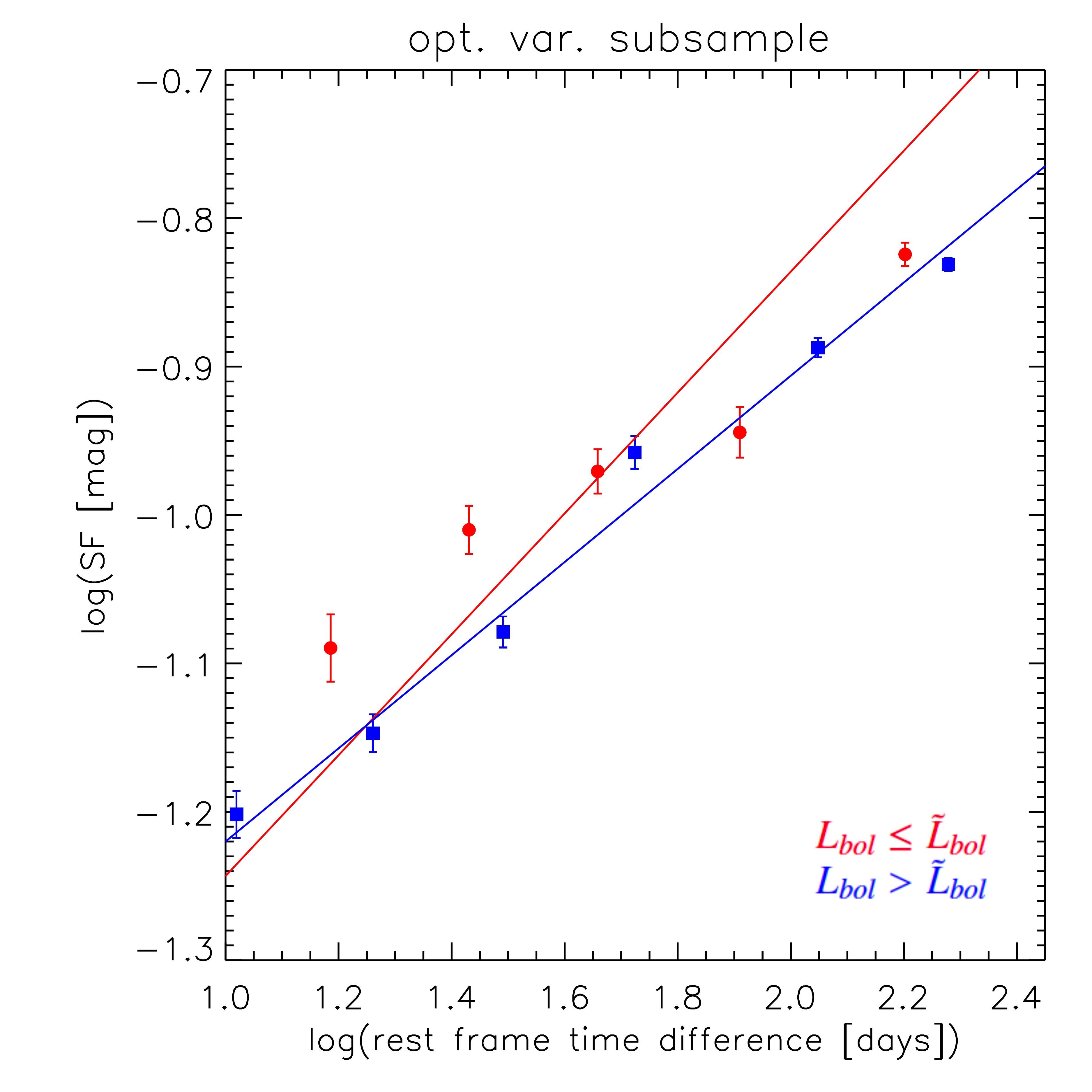}}\\
\subfigure
            {\includegraphics[width=5.3cm]{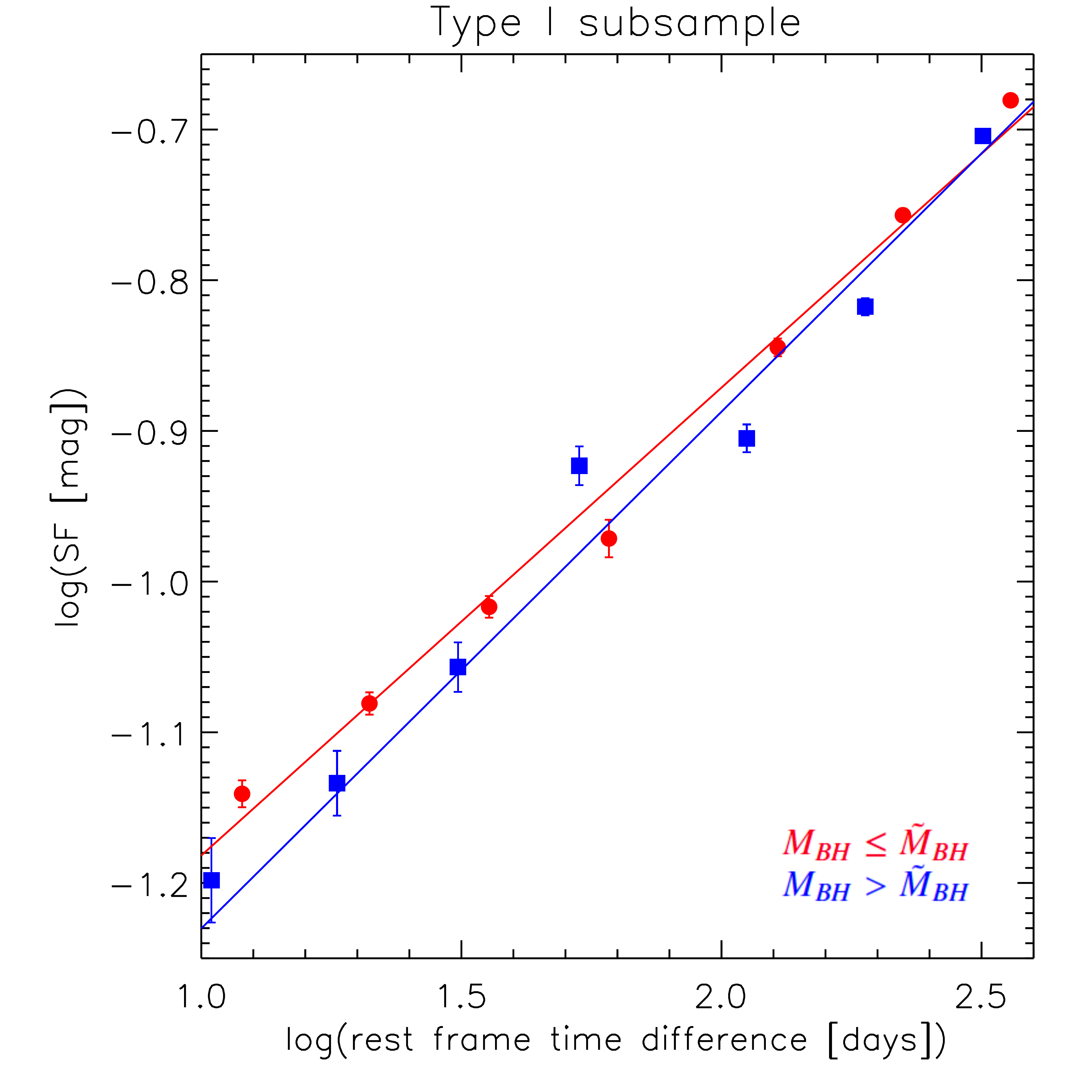}}
\subfigure
            {\includegraphics[width=5.3cm]{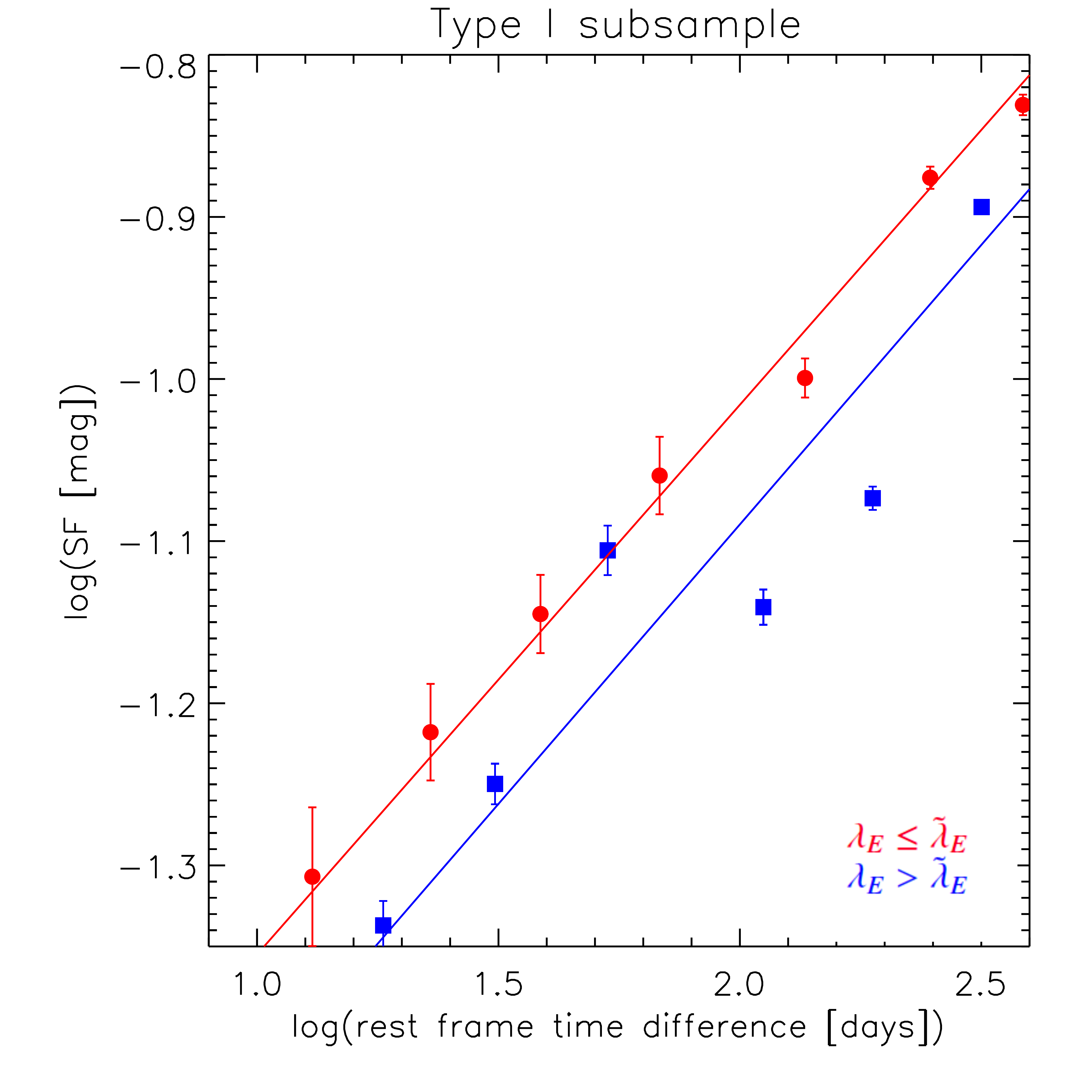}}
\subfigure
            {\includegraphics[width=5.3cm]{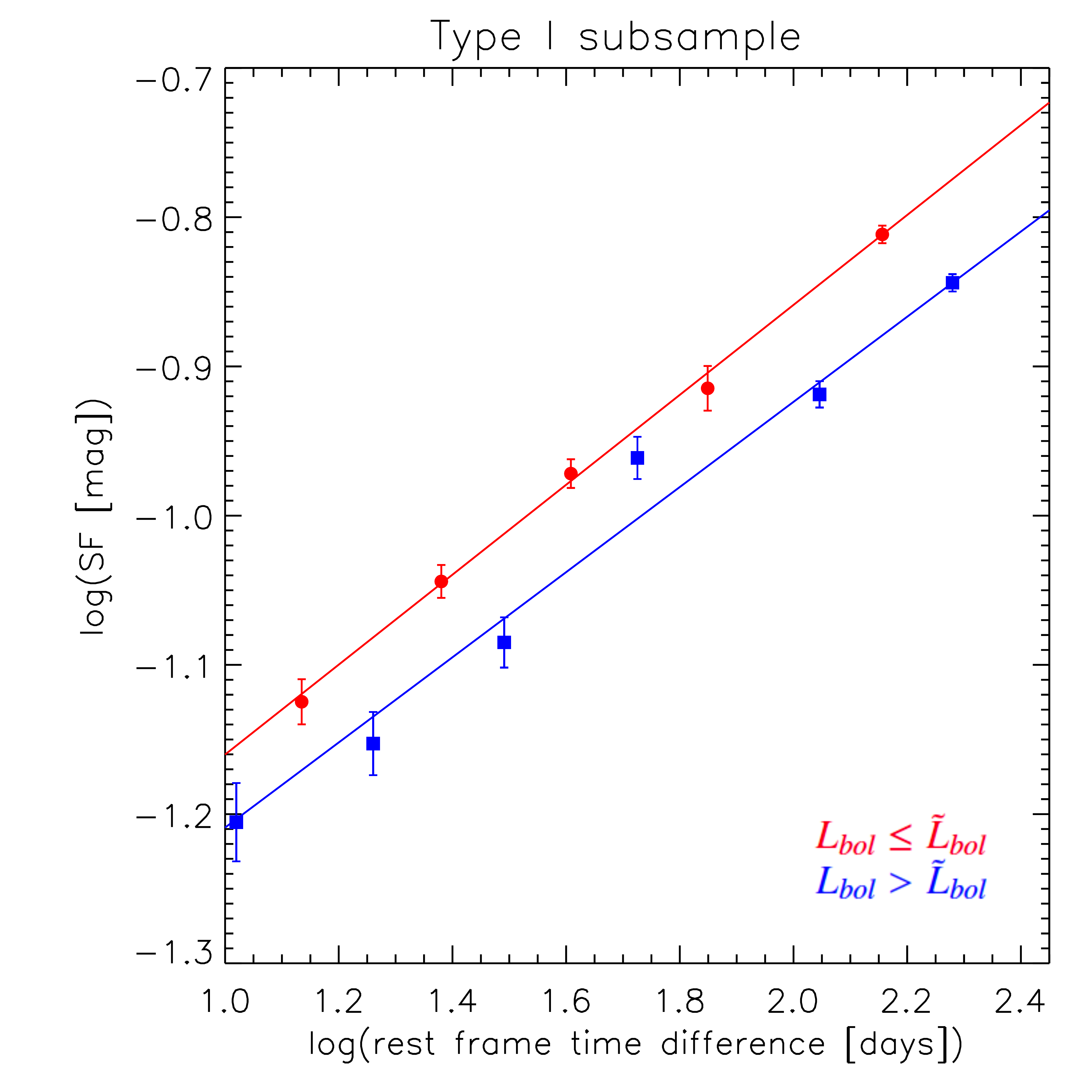}}\\
  \caption{Zoom in on the linear region of the SF for the four samples of AGN selected to investigate possible dependencies on $M_{BH}$ (\emph{left column}), $\lambda_E$ (\emph{middle column}), and $L_{bol}$ (\emph{right column}). The various panels show results for the \emph{main} sample, MIR AGN, optically variable AGN, and Type I AGN (\emph{top to bottom lines}). As in Fig. \ref{fig:bel_analysis}, in each panel we show the two subsets of sources delimited by the (logarithm of the) median value of the physical property of interest: red dots indicate the samples of sources with lower $M_{BH}$/$\lambda_E$/$L_{bol}$ values and blue squares indicate the sources with higher values. The represented rest-frame baseline corresponds to $\approx 10-400$ days.
  }\label{fig:bel_lin}
   \end{figure*}

For each test, we divide each of the four analyzed samples of sources into two bins based on the median value (always indicated by a superimposed tilde) of the investigated physical property in each sample. 
Figures \ref{fig:bel_analysis} presents the SFs obtained from the analysis of $M_{BH}$, $\lambda_E$, and $L_{bol}$ dependence for each sample of sources (\emph{main} sample, Type I AGN, MIR AGN, and optically variable AGN). For each pair of SFs reported in each panel we also identify a region of linearity based on the SFs shown in Fig. \ref{fig:sf}. This roughly corresponds to the logarithmic values $1.0-2.6$ for the baseline, that is to say $\approx 10-400$ days; we show the zoomed-in linear regions in Fig. \ref{fig:bel_lin}. For each linear region we estimate the best-fit line via weighted least squares regression. These estimates are gathered together in Table \ref{tab:bel}. The table also reports the median redshift value for each subset of sources.  

\begin{table*}[htb]
\renewcommand\arraystretch{1.4}
\caption{Results from the investigation of possible dependencies of the SF on $M_{BH}$ (\emph{top section}), $\lambda_E$ (\emph{middle section}), and $L_{bol}$ (\emph{bottom section}) for our four samples of AGN (listed in the top row) selected via different properties and used for this part of the analysis.}
\label{tab:bel}
\begin{center}
\footnotesize
\begin{tabular}{c c c c c}
\toprule
\ & \textbf{\emph{main} sample} & \textbf{MIR} & \textbf{opt. var.} & \textbf{spec. Type I}\\
\midrule
\ & & $M_{BH}$ dependence & & \\
\midrule
\ \makecell{$\log({M}_{BH})$ range \\ ($M_{BH}$[$M_\odot$])} & 4.95 to 9.65 & 6.68 to 9.47 & 4.95 to 9.65 & 7.02 to 9.65\\
\midrule
\vspace{1mm}
\ 1st quartile & 7.73 & 7.89 & 7.89 & 8.07\\
\vspace{1mm}
\ median ($\log(\tilde{M}_{BH})$) & 8.19 & 8.38 & 8.35 & 8.44\\
\ 3rd quartile & 8.54 & 8.70 & 8.69 & 8.72\\
\midrule
\ $M_{BH} < \tilde{M}_{BH}$ slope & $0.34\pm0.01$ & $0.31\pm0.01$ & $0.32\pm0.01$ & $0.31\pm0.01$\\
\ $\tilde{z}$ of this subset & 0.801 & 1.061 & 1.122 & 1.298\\
\midrule
\ $M_{BH} > \tilde{M}_{BH}$ slope & $0.33\pm0.01$ & $0.29\pm0.01$ & $0.34\pm0.01$ & $0.34\pm0.01$\\
\ $\tilde{z}$ of this subset & 1.450 & 1.560 & 1.531 & 1.597\\
\midrule
\ & & $\lambda_E$ dependence & & \\
\midrule
\ $\lambda_E$ range  & -2.85 to 0.09 & -2.21 to -0.09 & -2.40 to -0.14 & -2.40 to -0.14\\
\midrule
\vspace{1mm}
\ 1st quartile  & -1.55 & -1.18 & -1.35 & -1.32\\
\vspace{1mm}
\ median ($\tilde{\lambda}_E$)  & -1.17 & -0.93 & -1.05 & -1.05\\
\ 3rd quartile  & -0.85 & -0.70 & -0.81 & -0.79\\
\midrule
\ $\lambda_E < \tilde{\lambda}_E$ slope & $0.30\pm0.02$ & $0.38\pm0.01$ & $0.32\pm0.02$ & $0.34\pm0.02$\\
\ $\tilde{z}$ of this subset & 0.938 & 1.195 & 1.306 & 1.345\\
\midrule
\ $\lambda_E > \tilde{\lambda}_E$ slope & $0.32\pm0.02$ & $0.22\pm0.01$ & $0.34\pm0.01$ & $0.34\pm0.01$\\
\ $\tilde{z}$ of this subset & 1.258 & 1.531 & 1.444 & 1.550\\
\midrule
\ & & $L_{bol}$ dependence & & \\
\midrule
\ \makecell{$\log(\tilde{L}_{bol})$ range \\ $L_{bol}$ in erg s$^{-1}$)} & 40.276 to 46.676 & 45.299 to 46.676 & 42.865 to 46.676 & 44.023 to 46.676\\
\midrule
\vspace{1mm}
\ 1st quartile & 44.634 & 45.149 & 44.995 & 45.149\\
\vspace{1mm}
\ median ($\log(\tilde{L}_{bol})$) & 45.096 & 45.488 & 45.328 & 45.413\\
\ 3rd quartile & 45.513 & 45.815 & 45.665 & 45.76s\\
\midrule
\ $L_{bol} < \tilde{L}_{bol}$ slope & $0.31\pm0.01$ & $0.40\pm0.01$ & $0.41\pm0.01$ & $0.31\pm0.01$\\
\ $\tilde{z}$ of this subset & 0.762 & 0.970 & 1.033 & 1.279\\
\midrule
\ $L_{bol} > \tilde{L}_{bol}$ slope & $0.32\pm0.01$ & $0.25\pm0.01$ & $0.31\pm0.01$ & $0.29\pm0.01$\\
\ $\tilde{z}$ of this subset & 1.527 & 1.729 & 1.646 & 1.681\\
\bottomrule

\end{tabular}
\end{center}
\footnotesize{\textbf{Notes.} For each physical quantity we report the (logarithm of the) corresponding range boundaries, as well as the 1st quartile, median, and 3rd quartile values. Each sample is divided in two subsets on the basis of the (logarithm of the) median value of the physical property of interest in that sample. Each section reports, for each subset, the median redshift value and the estimated slope value for the best-fit line approximating the corresponding set of points in the linear region of the SF, with its error. We show the linear plots for each of the investigated samples of sources in Fig. \ref{fig:bel_lin}, comparing results for each of the performed tests.}
\end{table*}

From this point on, all the SFs computed will include the so-called V-correction, after \citet{vagnetti16}. This allows to take into account the dependence of variability on wavelength and, essentially, consists of a factor that is generally subtracted from the usual SF definition, producing an upwards shift of the SF, without affecting its slope. The V-correction depends on a parameter $\beta$, quantifying the variations of the spectral index in correspondence with monochromatic flux variations in the band of interest. Following \citet{laurenti}, we resort to their same estimate for this corrective factor, which is based on extrapolations from \citet{morganson14}, and thus obtain the following corrective factor:
\begin{equation}
\mbox{V-corr} \equiv \delta \log\mbox{SF} \simeq -\log e \cdot \beta \log(1+z)\mbox{   ,}
\label{eqN:Vcorr}
\end{equation}
with $\beta\simeq1$ (see \citealt{vagnetti16} and \citealt{laurenti} for details).

The figures and the information in the section in Table \ref{tab:bel} corresponding to the analysis of $M_{BH}$ dependence show that the slopes in each pair of subsets (low vs. high $M_{BH}$) analyzed are consistent within the errors. In the beginning, before applying the V-correction by \citet{vagnetti16}, we noticed that, although the slope was roughly the same in most cases, in each panel the SF amplitude corresponding to higher $M_{BH}$ values was systematically above the SF amplitude corresponding to lower $M_{BH}$ values. These results seemed to suggest that the sources with higher $M_{BH}$ are characterized by a larger variability amplitude, not affecting the slope of the linear region of the SF. Nonetheless, from Table \ref{tab:bel} we can see that the subsets with lower $M_{BH}$ are systematically characterized by a lower value of the  redshift. This suggested that we might be observing a correlation of the amplitude of variability with redshift, which, based on results from previous studies \citep[e.g.,][]{vdB,simm,paula17}, seems to be the effect of the anticorrelation with rest-frame wavelength. The maximum difference between the variability amplitudes (i.e., measured on the $y$-axis, $\log\mbox{SF}$) for the various pairs of subsets was measured in correspondence with the maximum difference between the  redshift values for a subset pair. All this led to the introduction of the above-mentioned V-correction to take into account the effect of wavelength. The panels related to $M_{BH}$ dependence in Fig. \ref{fig:bel_lin} show almost no difference between the variability amplitude for each pair of subsets analyzed in each panel, so we can state that we do not detect any dependence of variability on $M_{BH}$.

The figures and the information in Table \ref{tab:bel} corresponding to the analysis of $\lambda_E$ dependence show that, for each pair of subsets (low vs. high $\lambda_E$), the slopes of the linear region of the SF are consistent within the errors except for the MIR subsample, where the line corresponding to lower $\lambda_E$ values is steeper. The variability amplitude is systematically larger for each lower $\lambda_E$ subset, supporting the thesis of an anticorrelation of the variability amplitude with $\lambda_E$, consistent with several findings from previous works (see Section \ref{section:intro}).

The figures and the information in Table \ref{tab:bel} corresponding to the analysis of $L_{bol}$ dependence show conflicting results when we compare the slopes of the linear regions in each pair, while they seem to point towards an anticorrelation of the variability amplitude with $L_{bol}$, except for the sources in the \emph{main} sample. We point out that this is the most heterogeneous sample of AGN used in this study, and thus includes a significant fraction of Type II AGN with respect to the other samples of AGN analyzed. In particular, 31\% of the lower $L_{bol}$ subset consists of Type II AGN, while the percentages of Type II AGN in each of the other subsets in each pair is one order of magnitude lower. In Section \ref{section:sf_six} we discussed how, based on Fig. \ref{fig:sf}, Type II AGN seem to be responsible for a flattening of the SF; what we observe for the lower $L_{bol}$ subset corresponding to the \emph{main} sample is therefore consistent with this.

We point out that, in the case of the sources with $M_{BH}$, $\lambda_E$, and $L_{bol}$ estimates, the overlap among the various subsamples is almost total, leaving a few sources not overlapping any other samples. For this reason it is not possible to analyze them as separate sets as we did in Section \ref{section:overlap}.

\subsection{Redshift and obscuration dependence}
\label{section:z_NH}
\begin{figure*}[tbh]
 \centering
\subfigure
            {\includegraphics[width=5.2cm]{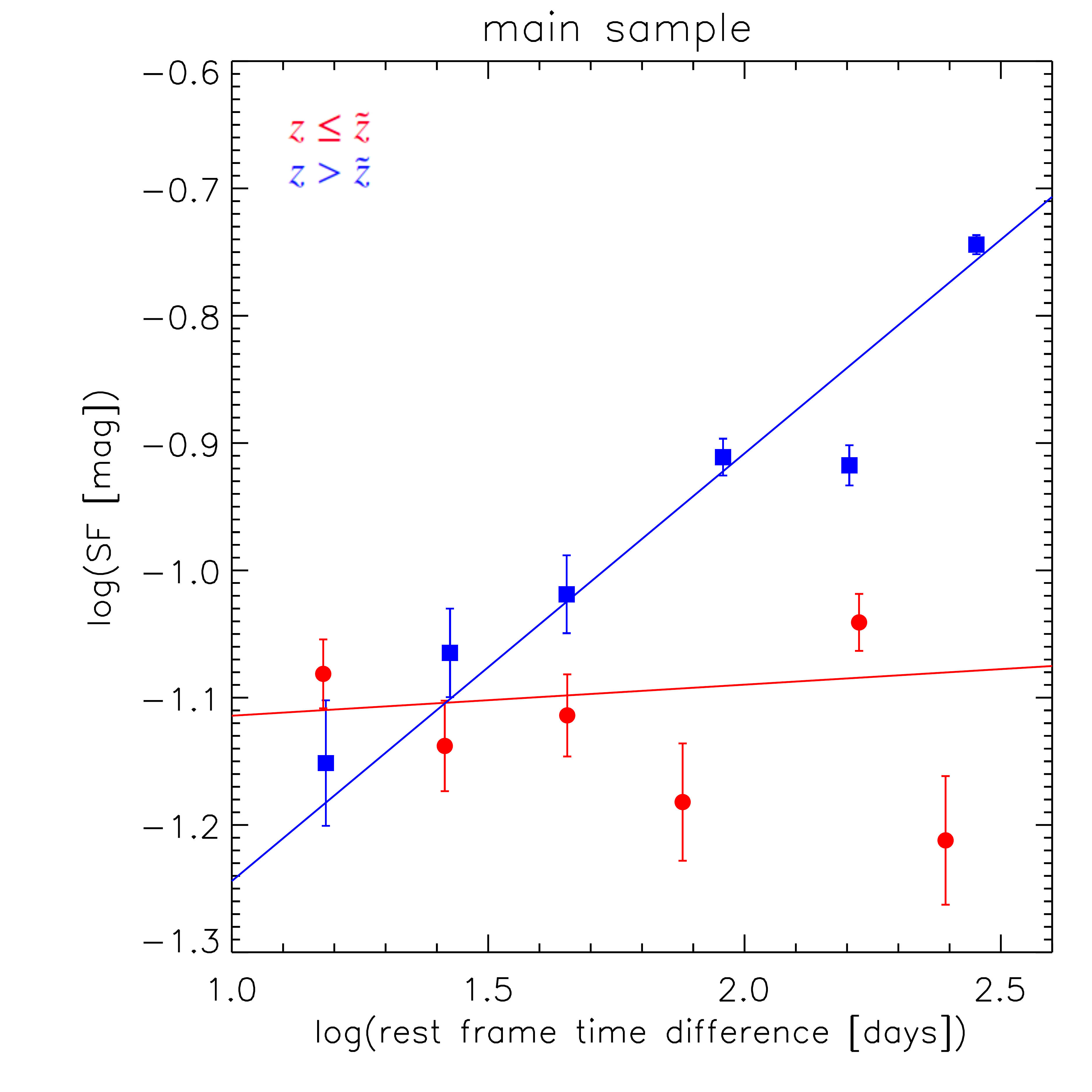}}
\subfigure
            {\includegraphics[width=5.2cm]{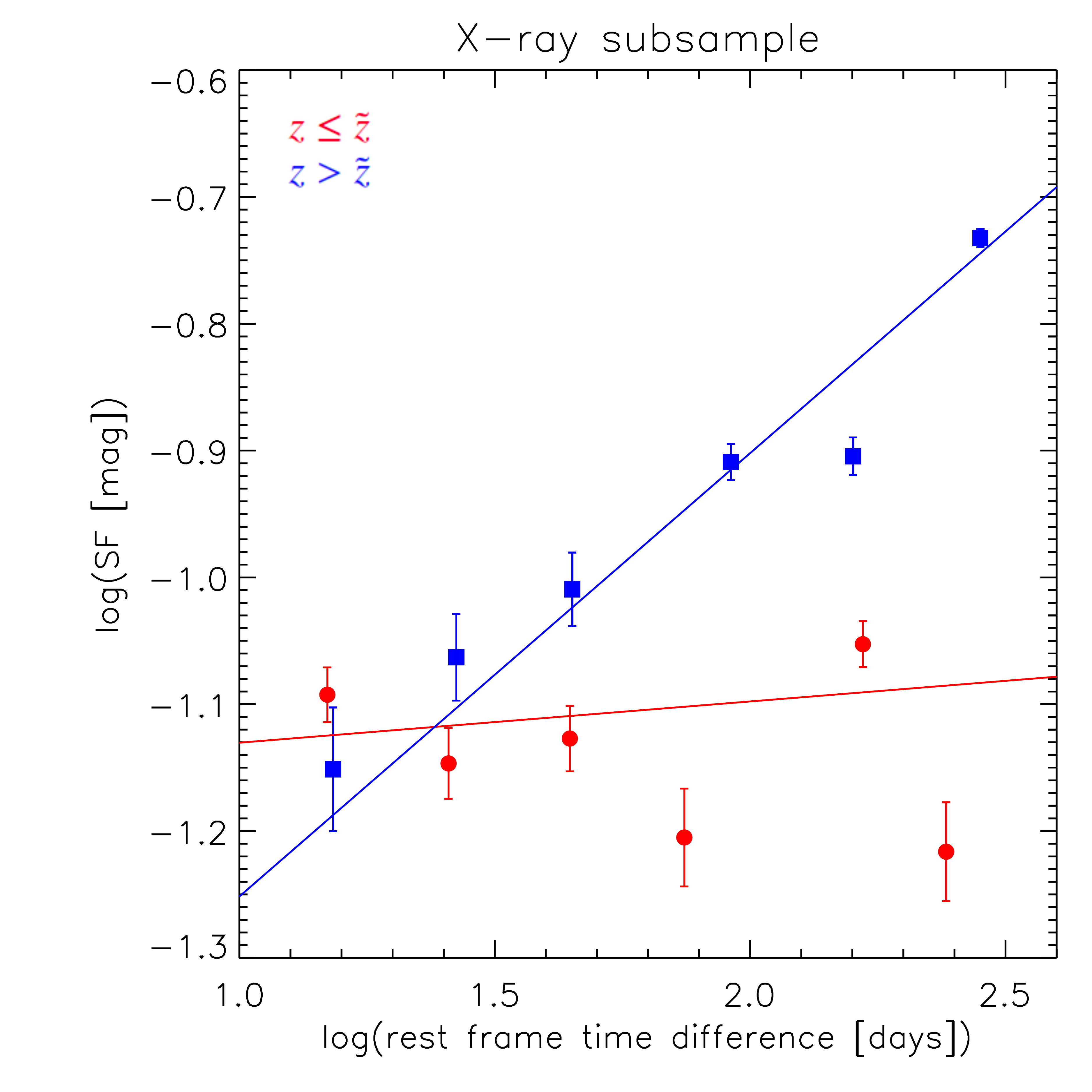}}
\subfigure
            {\includegraphics[width=5.2cm]{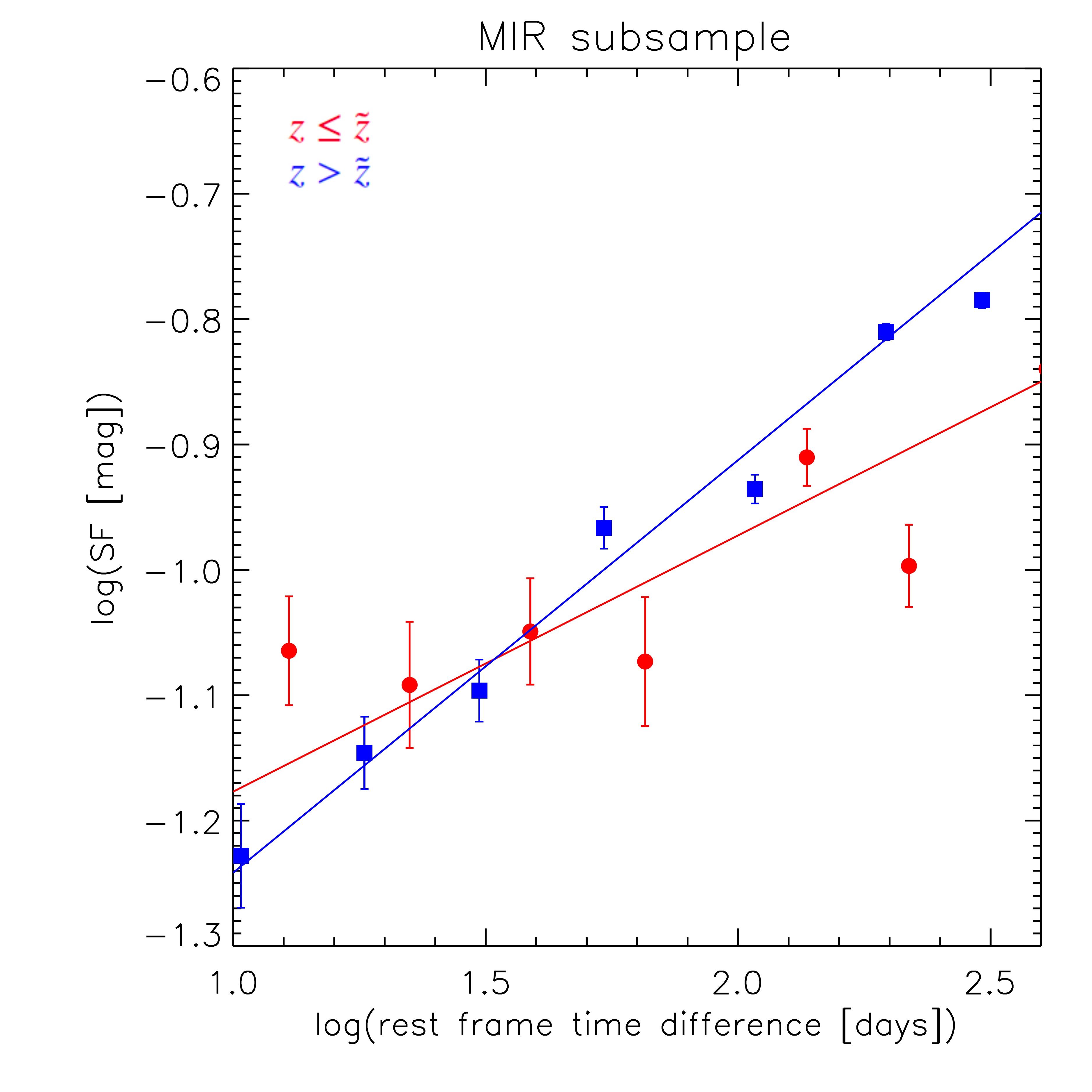}}
\subfigure
            {\includegraphics[width=5.2cm]{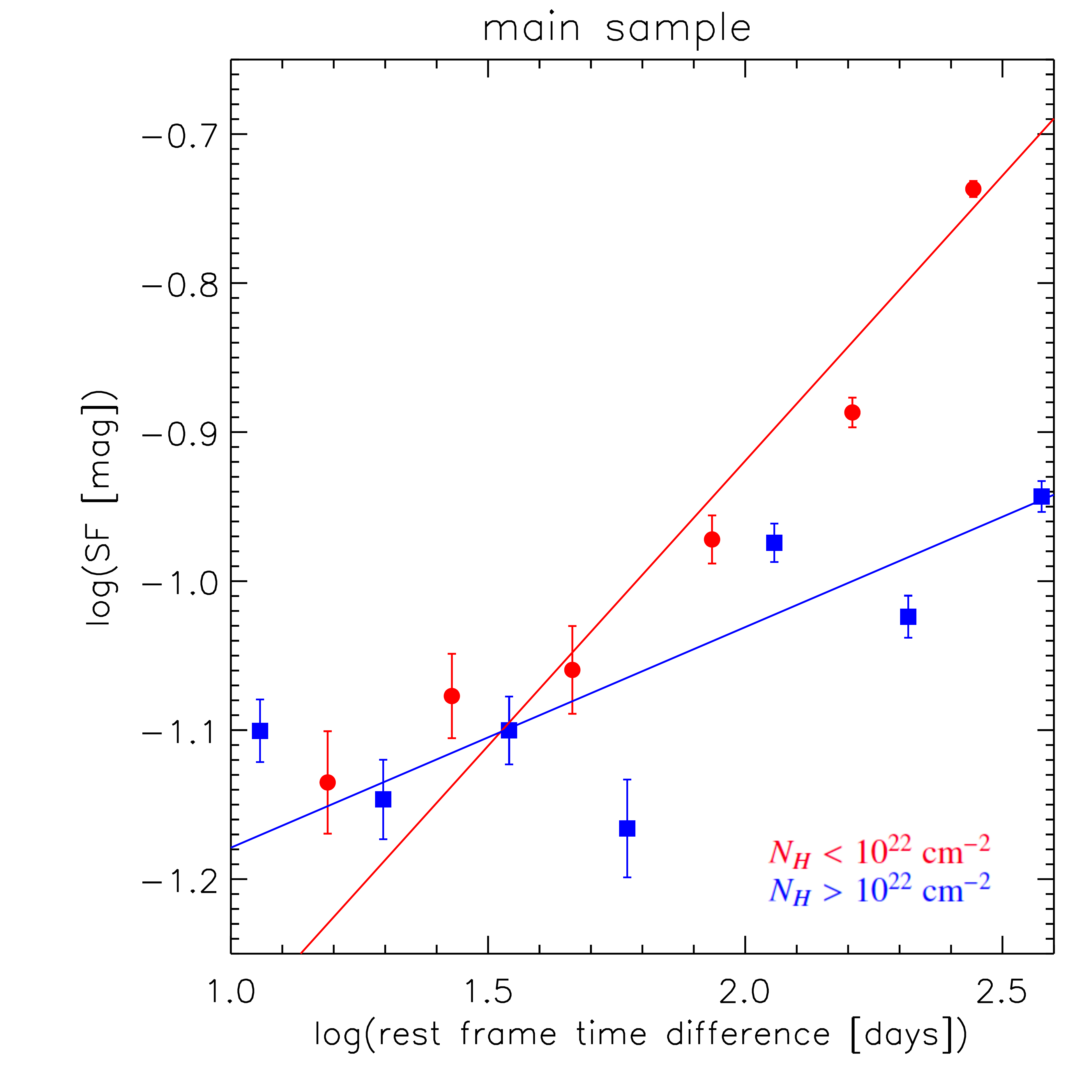}}
\subfigure
            {\includegraphics[width=5.2cm]{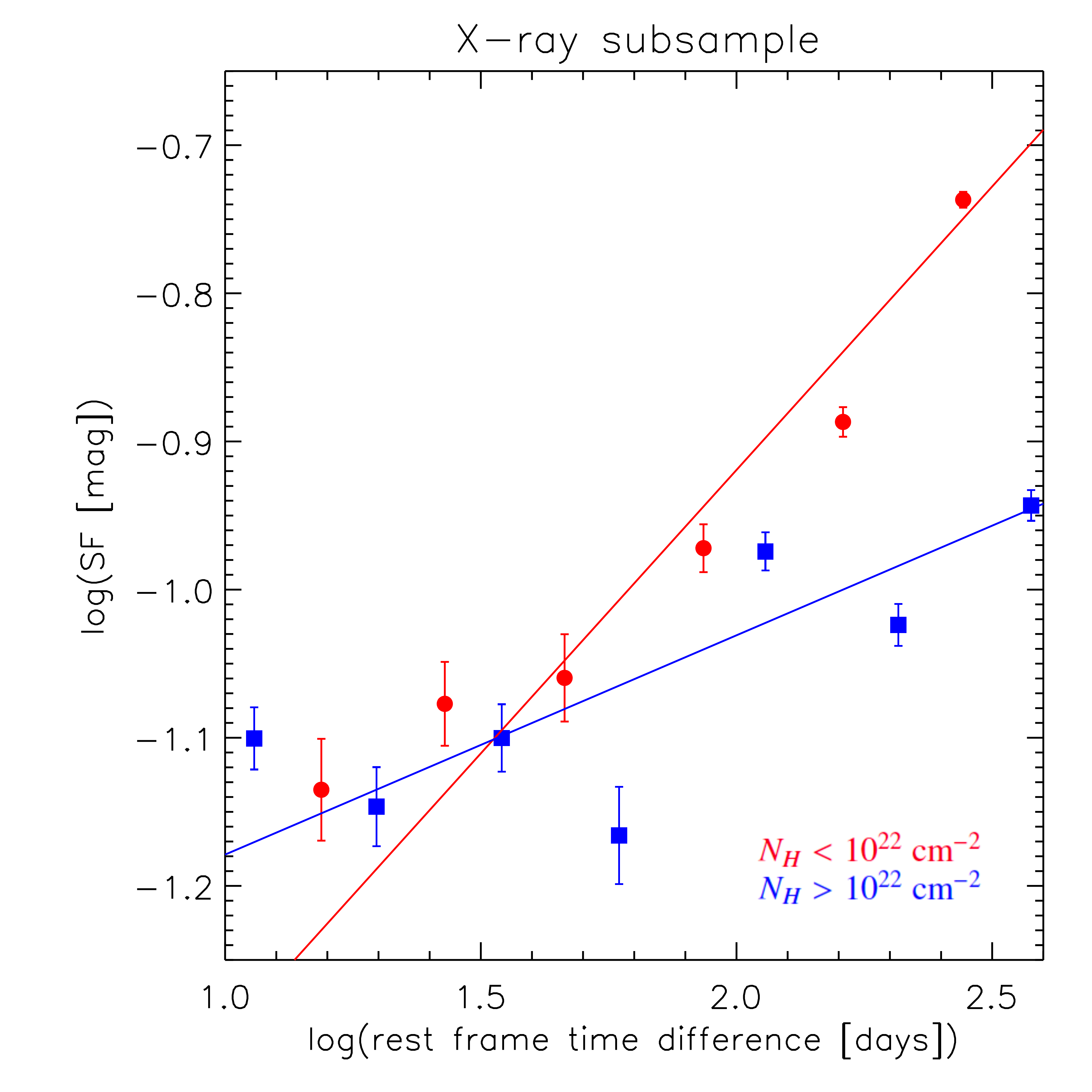}}
\subfigure
            {\includegraphics[width=5.2cm]{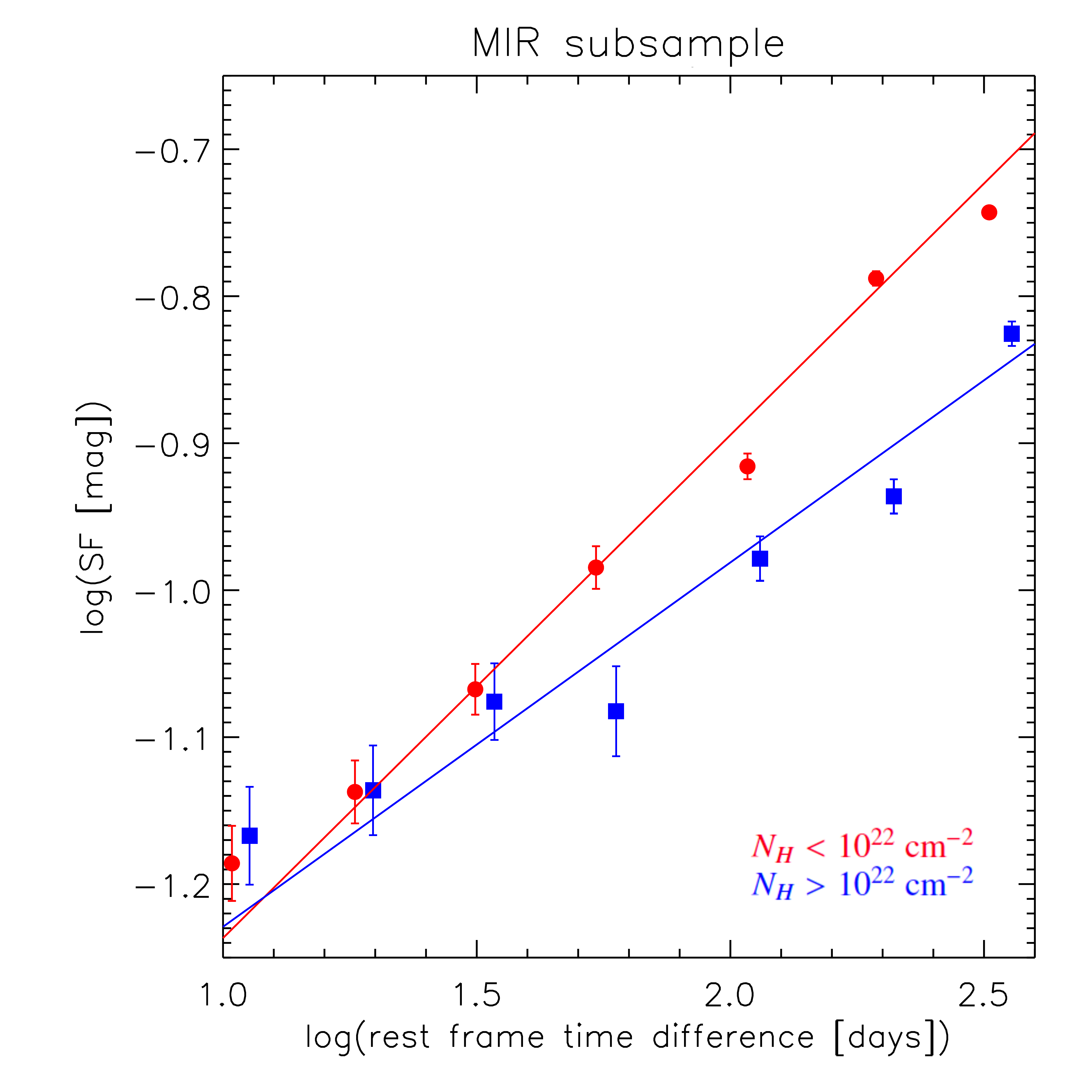}}
\subfigure
            {\includegraphics[width=5.2cm]{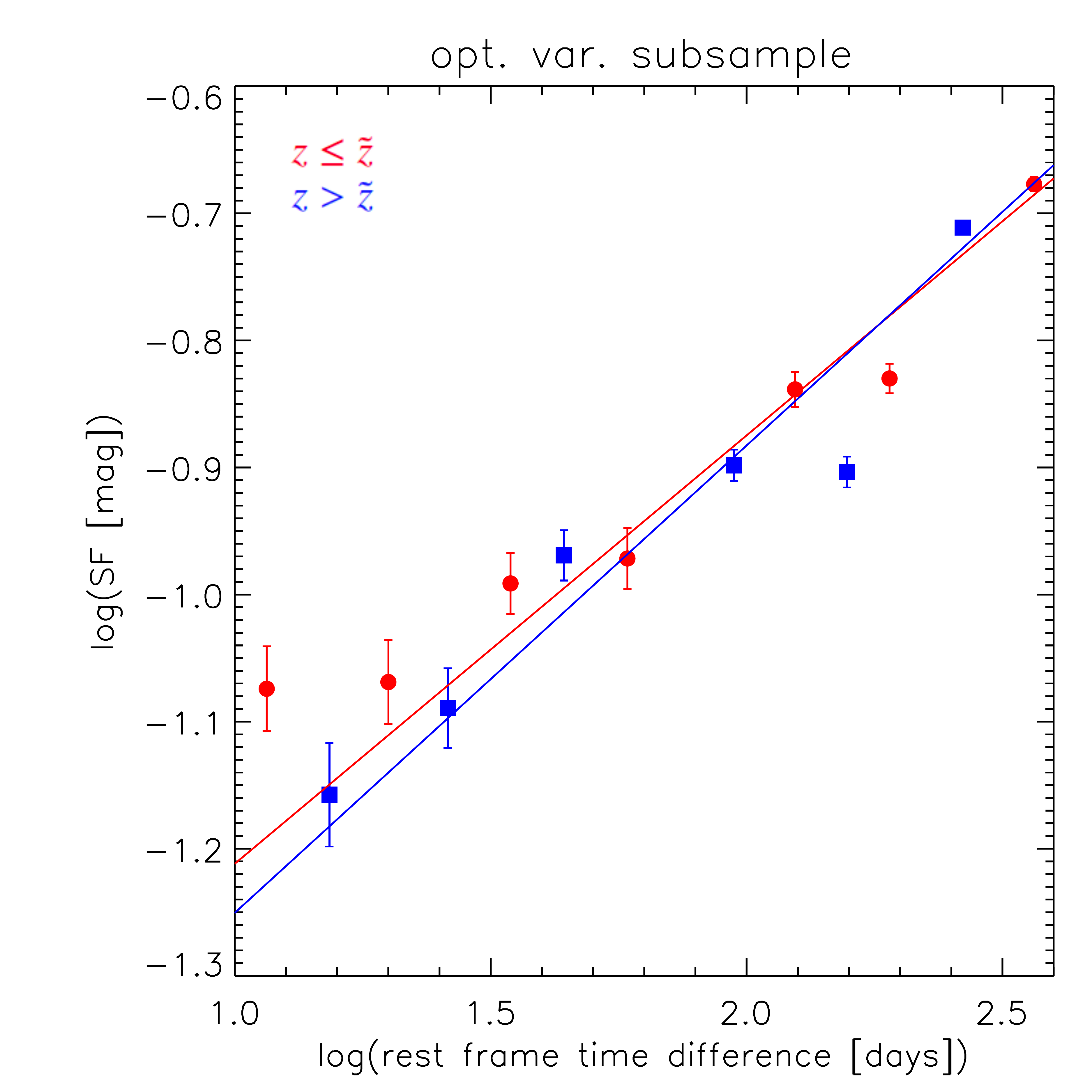}}
\subfigure
            {\includegraphics[width=5.2cm]{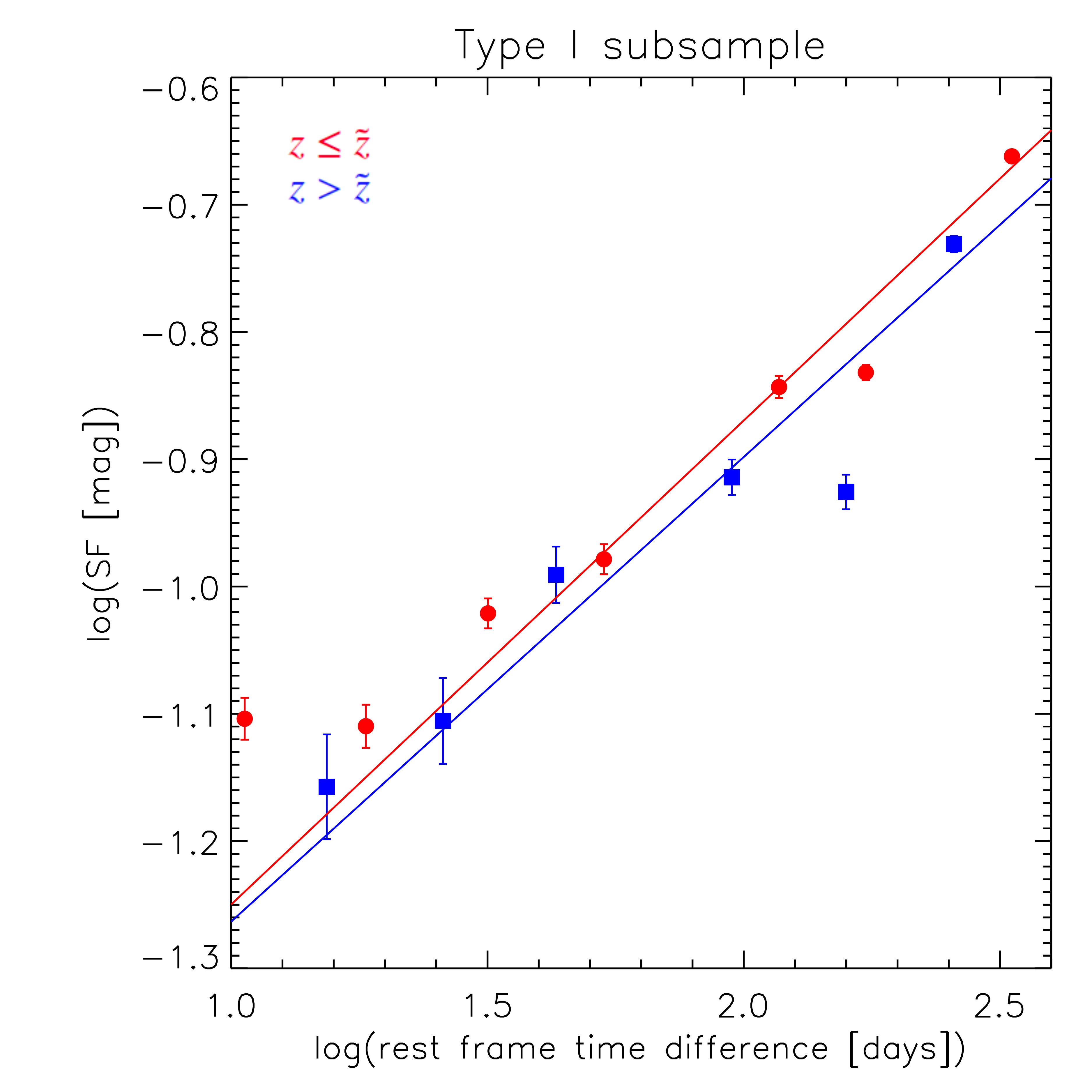}}
\subfigure
            {\includegraphics[width=5.2cm]{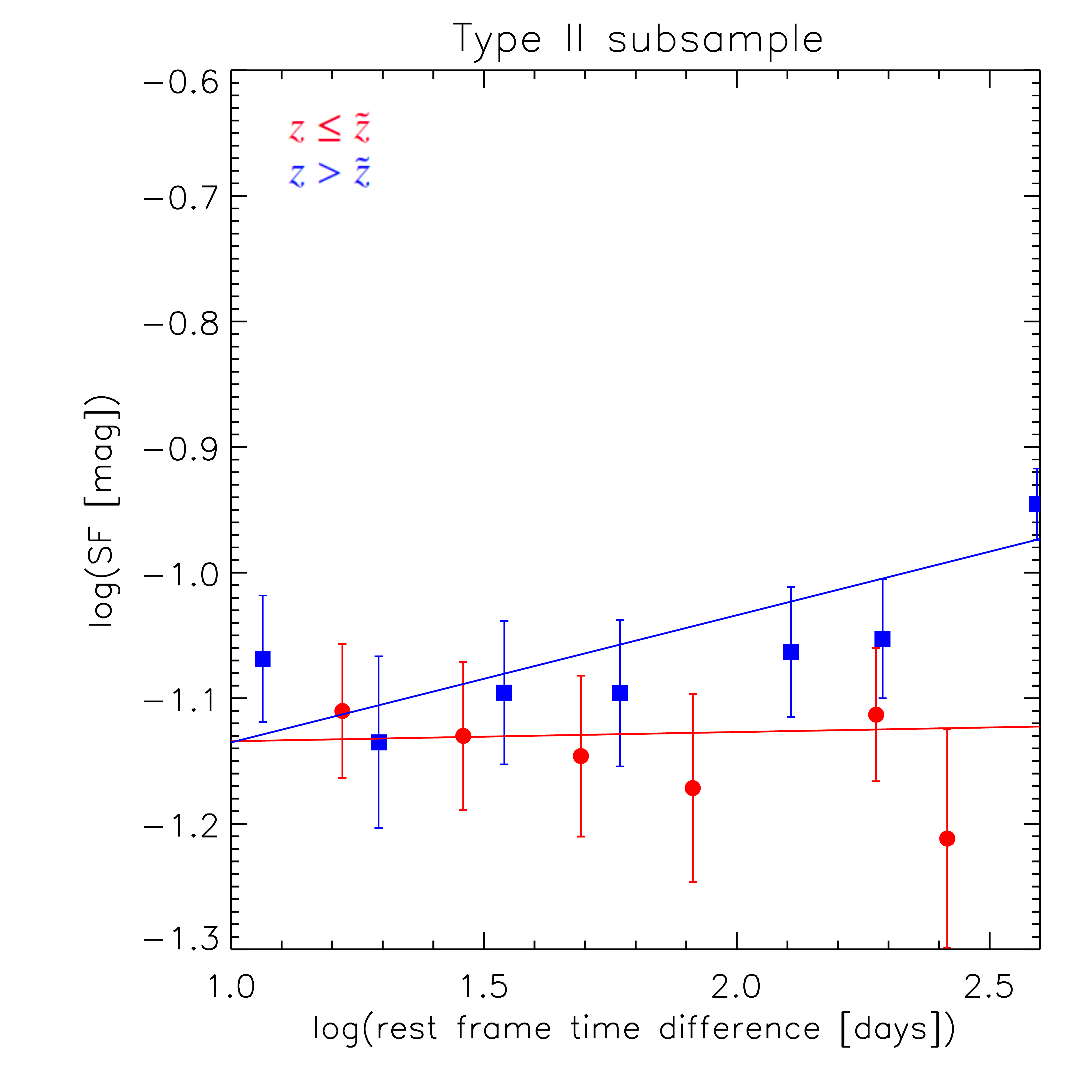}}
\subfigure
            {\includegraphics[width=5.2cm]{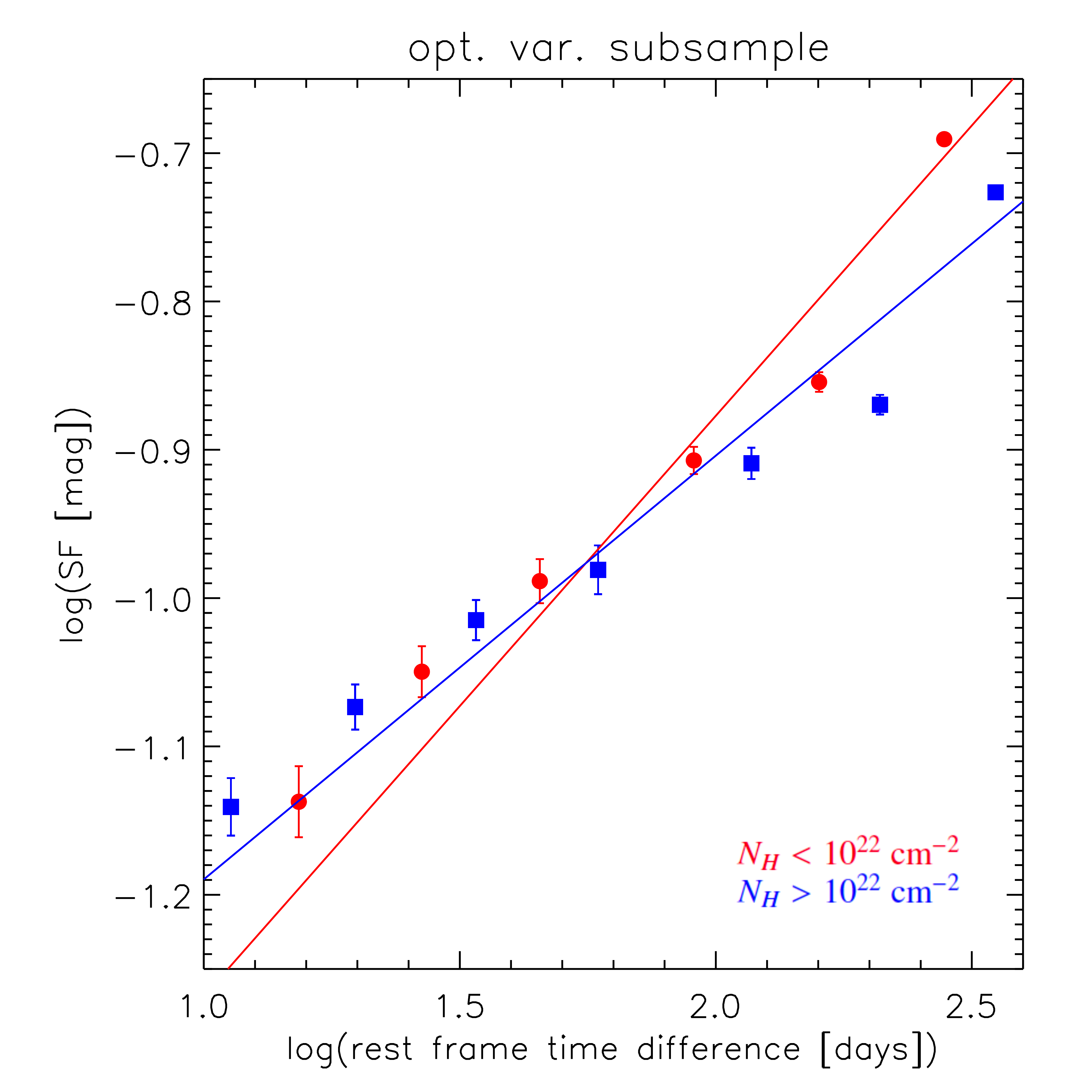}}
\subfigure
            {\includegraphics[width=5.2cm]{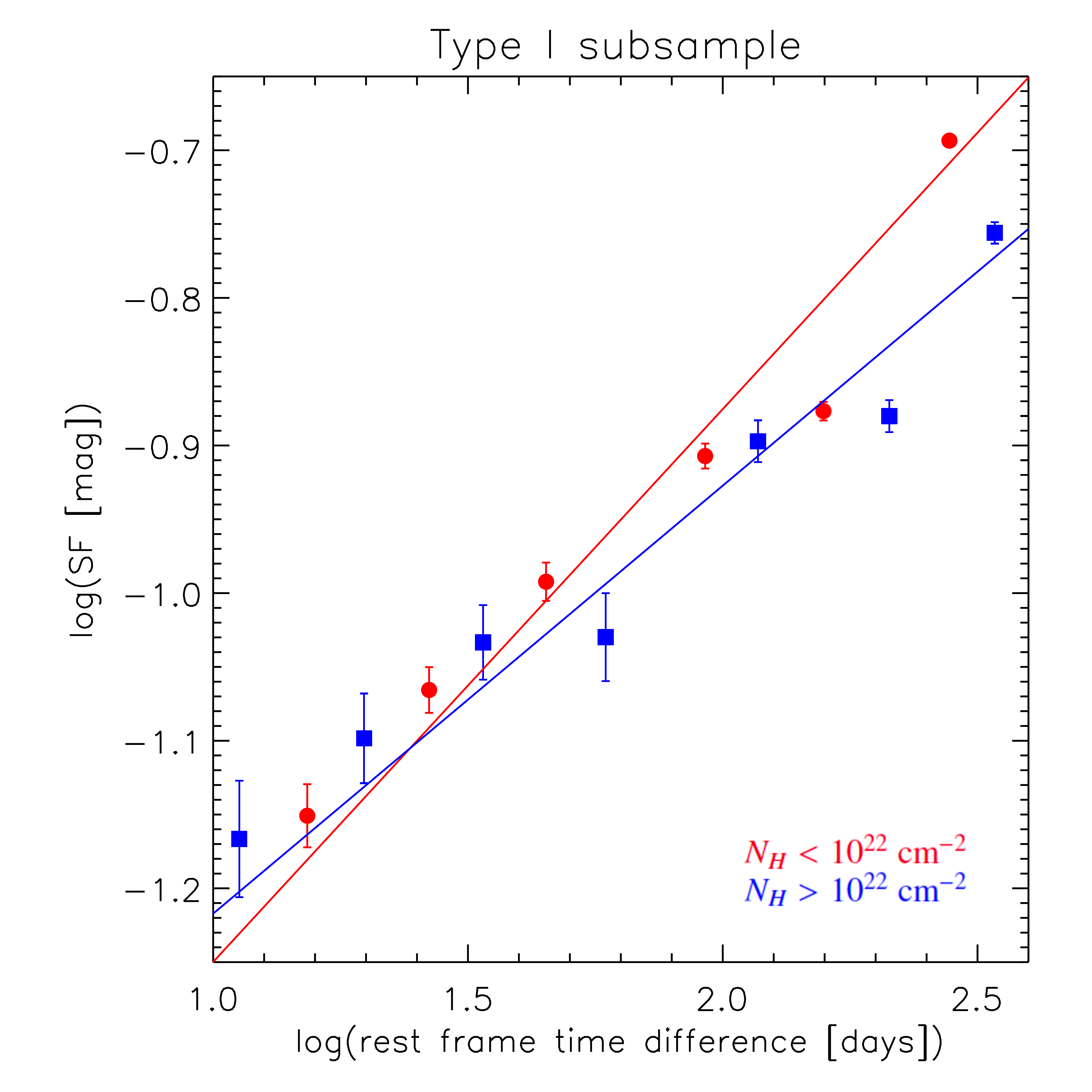}}
\subfigure
            {\includegraphics[width=5.2cm]{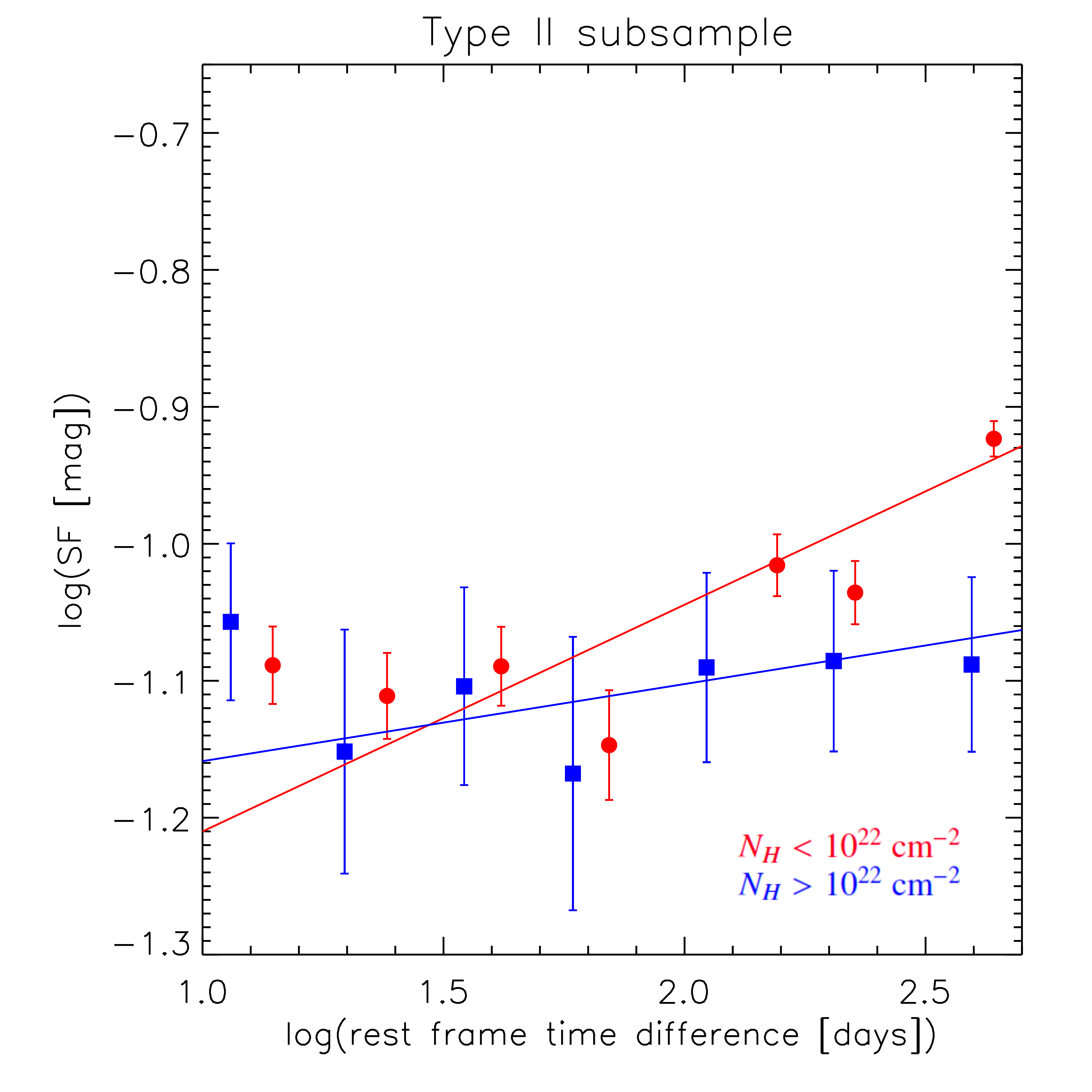}}\\
  \caption{Analysis of the dependence of variability on redshift and absorption (quantified by the hydrogen column density $N_H$), focusing on the linear regions of the SF for the six samples of AGN selected for this study: the two top lines show panels for the \emph{main} sample (\emph{left}), X-ray AGN (\emph{center}), and MIR AGN (\emph{right}); the two bottom lines show panels for optically variable AGN (\emph{left}), Type I AGN (\emph{center}), and Type II AGN (\emph{right}). Redshift (\emph{odd lines}) and absorption (\emph{even lines}) panels are coupled for each sample of sources, in order to ease the comparison of the results. Red dots indicate the samples of sources with lower redshift/absorption; blue squares indicate the sources with higher redshift/absorption values. As usual, the represented rest-frame baseline corresponds to $\approx 10-400$ days.}\label{fig:z_NH}
   \end{figure*}

In order to investigate the dependence of the SF on redshift, we consider our six original (i.e., independent of availability of $M_{BH}$ and $\lambda_E$ estimates) samples of AGN and split each of them in two based on their  redshift value, and analyzed the slopes and amplitudes of their SFs in the linear regions of each, following what we did in Section \ref{section:bel}. We report the slopes of the linear regions in each pair, delimited as usual, in the top section of Table \ref{tab:z_NH}, together with the  redshift values (which are the same reported in Table \ref{tab:samples}), and we show the SFs obtained for each pair of subsets in Figure \ref{fig:z_NH}, limited to their linear regions. We notice that the slopes in a pair are roughly consistent within the errors in the case of Type I AGN and optically variable AGN. Conversely, for the \emph{main} sample and the X-ray subsample, which largely overlap, we obtain the largest difference between the corresponding slopes. Mid-infrared AGN are in between, with a moderate difference between the two slopes, while Type II AGN, as usual, behave differently than all the other samples, and are characterized by almost flat SFs both for lower and higher redshift subsets, and also by larger errors on these estimates. We also notice that, apart from Type~II AGN, the slopes corresponding to higher redshifts are all roughly consistent within the errors, while this does not hold for the slopes corresponding to lower redshifts. Since we know that Type~II AGN are generally responsible for the flattening of the SF, we investigate the fraction of known Type I and Type~II AGN in each subset, for each pair. We find that, in general, Type I AGN dominate the various subsets corresponding to higher redshifts. They are also dominant in the subsets at lower redshifts corresponding to optically variable AGN (and of course, by construction they constitute the total subset for the subsample of Type I AGN). Conversely, in the \emph{main} sample and in the X-ray subsample Type II AGN are the ones dominating the lower redshift subsets, i.e., the ones characterized by a flattening with respect to the corresponding higher redshift subsets. The situation is intermediate for the MIR AGN subsets, which include a significant fraction of Type II AGN but are still dominated by Type I AGN. All this suggests that redshift itself does not affect the slope of the SF, but this is rather a consequence of the fraction of unobscured/obscured AGN in each sample. In addition, we should take into account that the basic classification capability likely changes with redshift; e.g., because of the different rest-frame lines available in optical spectra, the different rest-frame X-ray bands sampled, and so on. 

The analysis performed so far has shown how the AGN variability properties are different for Type I and Type II AGN. In order to test further the dependence on AGN obscuration, we investigate how our SFs change with absorption -- which in the X-rays is typically quantified by the hydrogen column density $N_H$ -- for the six samples of AGN. An estimate of this quantity is available for 71\% of the AGN in our \emph{main} sample from the \emph{Chandra} COSMOS Legacy Survey Multiwavelength Catalog \citep{marchesi}. We choose the value $10^{22}\mbox{ cm}^{-2}$ to split each sample in two as this is usually assumed as the limit between unabsorbed ($N_H < 10^{22}\mbox{ cm}^{-2}$) and absorbed ($N_H > 10^{22}\mbox{ cm}^{-2}$) sources. Once again, we estimate the slopes of the linear regions of the SFs of unabsorbed and absorbed sources in each sample. We report these estimates in the bottom section of Table \ref{tab:z_NH}, and we show the SFs obtained for each pair of subsets in Figure \ref{fig:z_NH}, limited to their linear regions. For each sample, the slope of the linear region corresponding to lower $N_H$ is always steeper than the one obtained for higher $N_H$. As usual, optical Type II AGN exhibit flatter SFs for both subsets than the rest of the analyzed samples. If we compare the lower $N_H$ slopes of the various samples with each other, we notice that they all are roughly consistent within the errors. When we do the same for higher $N_H$ slopes, we notice that the values obtained for Type I AGN, optically variable AGN, and MIR AGN are roughly consistent within the errors, and are higher than the values obtained for the \emph{main} sample and the X-ray subsample, which are in turn consistent with each other. From the plots in the figure we can see how the angle formed by the two lines in each pair decreases as we move from the \emph{main} sample towards Type I AGN (from top left to bottom central panels), i.e., as the contribution of Type II AGN becomes less relevant. All this suggests that absorption is responsible for a flattening in the SF.

\begin{table*}[htb]
\renewcommand\arraystretch{1.4}
\caption{Results from the investigation of possible dependencies of the SF on redshift (\emph{top section}) and obscuration (\emph{bottom section}) for the \emph{main} sample as well as the other five subsamples of AGN selected via different properties and analyzed in this study (listed in the top row).}
\label{tab:z_NH}
\begin{center}
\footnotesize
\begin{tabular}{c c c c c c c}
\toprule
\ & \textbf{\emph{main} sample} & \textbf{X-ray} & \textbf{MIR} & \textbf{opt. var.} & \textbf{spec. Type I} & \textbf{spec. Type II}\\
\midrule
\ & & & redshift dependence & & \\
\midrule
\ $z$ range & 0.121 to 3.715 & 0.121 to 3.715 & 0.350 to 3.715 & 0.346 to 3.715 & 0.346 to 3.715 & 0.121 to 1.447\\ 
\midrule
\vspace{1mm}
\ 1st quartile  & 0.530 & 0.620 & 0.704 & 0.961 & 1.171 & 0.520 \\
\vspace{1mm}
\ median ($\tilde{z}$) & 0.876 & 0.893 & 1.214 & 1.441 & 1.653 & 0.696\\
\ 3rd quartile  & 1.402 & 1.450 & 2.021 & 2.032 & 2.152 & 0.890 \\
\midrule
\ $z < \tilde{z}$ slope & $0.02\pm0.03$ & $0.03\pm0.02$ & $0.20\pm0.03$ & $0.34\pm0.01$ & $0.38\pm0.01$ & $0.01\pm0.04$\\
\ $z > \tilde{z}$ slope & $0.34\pm0.03$ & $0.35\pm0.02$ & $0.33\pm0.02$ & $0.37\pm0.01$ & $0.37\pm0.02$ & $0.10\pm0.03$\\
\midrule
\ & & & absorption dependence & & \\
\midrule
\ \makecell{$N_H$ range \\ ($\times 10^{22}\mbox{ cm}^{-2}$)} & 0.0 to 25.9 & 0.0 to 25.9 & 0.0 to 25.9 & 0.0 to 25.9 & 0.0 to 25.9 & 0.0 to 13.0\\ 
\ $N_H < 10^{22}\mbox{ cm}^{-2}$ slope & $0.38\pm0.02$ & $0.38\pm0.02$ & $0.34\pm0.01$ & $0.39\pm0.01$ & $0.40\pm0.01$ & $0.10\pm0.03$\\
\ $N_H > 10^{22}\mbox{ cm}^{-2}$ slope & $0.15\pm0.02$ & $0.15\pm0.02$ & $0.25\pm0.01$ & $0.29\pm0.01$ & $0.28\pm0.01$ & $0.04\pm0.07$\\
\bottomrule

\end{tabular}
\end{center}
\footnotesize{\textbf{Notes.} Absorption is quantified by the hydrogen column density $N_H$. For each quantity we report the corresponding range boundaries. In the redshift-related section of the table (\emph{top}) we report the 1st quartile, median, and 3rd quartile values of redshift. Each sample is divided in two subsets on the basis of the median redshift of the sample (\emph{top section}), or on the basis of the value $N_H = 10^{22}\mbox{ cm}^{-2}$ (\emph{bottom section}). For each subset we report the estimated slope value for the best-fit line, obtained as usual via a weighted least squares regression, approximating the corresponding set of points in the linear region of the SF, with its error.}
\end{table*}

\section{Discussion and conclusions}
\label{section:conclusions}
In this work we presented an analysis of the SF of 677 AGN in the VST-COSMOS area, defining a sample of AGN as a result of different diagnostics based on: spectroscopy, X-ray properties, MIR properties, selection via optical variability. We analyzed the SF of the \emph{main} sample of sources as well as the SF of the various subsamples defined by the various diagnostics. We investigated the properties of these samples at a depth that is approximately two orders of magnitude deeper than most of the similar analyses from the literature. This study therefore represents, to our knowledge, one of the few investigating the ensemble variability of fainter AGN than most datasets to date, thus exploring new ranges of luminosity and, therefore, of $M_{BH}$ and $\lambda_E$. On the other hand, this work suffers from two main limitations, these being the very irregular sampling and the small size of the sample of AGN for which estimates of the physical quantities of interest for this analysis are available. Indeed, we also examined the behavior of the SF in relation to two major properties of our AGN, i.e., the mass of the central SMBH and the accretion rate, the latter being quantified by $\lambda_E$ (both with estimates available for 264 sources), and also in relation to the AGN $L_{bol}$, redshift, and obscuration. 

Our catalog of 677 AGN is available in the electronic version of this paper as Table 6. In what follows we summarize our main findings.

The shape of the SF is affected by the sample of AGN used to build it: in Fig. \ref{fig:sf} we report the SFs obtained for the six samples of AGN used in this work, and it is apparent that, when the sample is dominated by Type I AGN, the shape of the SF shows a ``clean'' region with a fairly clear linear slope. This is consistent with the fact that Type I AGN are typically easier to select via optical variability. Conversely, the presence of a significant fraction of Type II AGN in the sample heavily affects the shape of the SF. Based on the comparison of the SFs shown in Fig. \ref{fig:sf} for the various samples of AGN used in this study, this effect is more relevant at timescales in the range $\approx10-65$ days, where the SF starts to rise for Type I AGN-dominated samples, while it is still rather flat otherwise. It is well known that Type II AGN are harder to detect via optical variability; clearly there will be a reprocessed component on larger physical scales, and hence it will also ``reprocess''/damp the timescales. If we assume that the origin of optical variability is the same for Type I and Type II AGN, then the main difference between the two types lies in the fact that for Type I AGN we detect the disk variability directly, while for Type II we do not as it is absorbed, but we detect a scattered component of this variability coming from the narrow line region. Since this is supposed to be located at larger distances from the central BH (typically 10-1000 pc, vs. the $\approx1$ pc radius of the accretion disk), this suggests that we need longer timescales in order to be able to detect larger variations for Type~II AGN. Indeed, our previous studies \citep{decicco15,decicco19} already showed how the fraction of Type II AGN detected via optical variability increases from 6\% to 18\% when the baseline is extended from five months to 3.3 yr, and hence we expect some improvement with a, e.g., 10 yr baseline (i.e., the full baseline for the LSST main survey, but also the baseline that we will be able to cover if new VST-COSMOS observations are carried on; see further. While the shape of the SF changes with the type of sources, it does not seem to change with the depth of the sample, as can be seen from Fig. \ref{fig:sf}, where two distinct magnitude thresholds ($r \le 23.5$ and $r \leq 22.5$) are tested.

Where a region of linearity can be identified, the slope of this region is fairly consistent with results from several past studies dedicated to the AGN SF. It is worth mentioning that the slope of the SF is highly affected by the way the noise contribution in the SF definition is estimated \citep[e.g.,][]{kozlowski}. 

The study of possible connections between the shape of the SF and the mass of the central SMBHs in our sample of AGN shows no significant relations for what concerns both the amplitude and the slope of the SF (see Figs. \ref{fig:bel_analysis} and \ref{fig:bel_lin}, and Table \ref{tab:bel}). The analysis of connections with $\lambda_E$, on the other side, suggest an anticorrelation with variability amplitude, based on Figs. \ref{fig:bel_analysis} and \ref{fig:bel_lin}, and Table \ref{tab:bel}. This is consistent with most results from past investigations on the topic, and it is important to stress once again that, with our study, we are extending this result to fainter sources. The investigation of possible relations with the $L_{bol}$ also points towards an anticorrelation of the variability amplitude with luminosity, consistent once again with previous findings. Following \citet{suberlak}, in Table \ref{tab:literature} we summarize the main results of the works mentioned in Section \ref{section:intro} analyzing the variability dependence on $M_{BH}$, $\lambda_E$, and $L_{bol}$, and compare them with results from this work. Based on it, for what concerns the variability amplitude the most controversial results are obtained for the dependence on $M_{BH}$, while all the works investigating a dependence on $\lambda_E$ or on $L_{bol}$ are consistent in showing an anticorrelation, in some cases very strong. The dependence of the variability timescale on these same quantities was investigated just in a few works, and so we cannot state anything significant about that.

We also analyzed the dependence of the SF on redshift, and found that it is affected by the type of AGN included in the various samples: we observe a strong correlation with redshift for our \emph{main} sample of sources as well as the X-ray subsample, but in both cases the lower redshift subsets turn out to be dominated by Type II AGN, which seem to be responsible for the corresponding flattening in the SF. Indeed, this correlation is weaker for MIR AGN, where Type I AGN start to outnumber Type II AGN in both lower and higher redshift subsets, while it is not observed at all in the subsample of Type I AGN, nor in the subsample of optically variable AGN, where Type I AGN constitute 75\% of the total. In addition we investigated the dependence on absorption, quantified by the hydrogen column density $N_H$, and our tests show that the slope of the linear region of the SF is shallower for absorbed sources. 

\begin{table*}[htb]
\renewcommand\arraystretch{1.4}
\caption{Comparison of our results with some findings from the literature (see Section \ref{section:intro}), where several methods to analyze the dependence of AGN variability on $M_{BH}$, $\lambda_E$, and $L_{bol}$ were used.}\label{tab:literature}
\begin{center}
\footnotesize
\begin{tabular}{c|c|c c c|c c c}
\toprule
\ Publication & Method & \multicolumn{3}{c}{Amplitude} & \multicolumn{3}{c}{Timescale}\\
\hline
\ & & $M_{BH}$ & $\lambda_E$ & $L_{bol}$ & $M_{BH}$ & $\lambda_E$ & $L_{bol}$ \\
\hline
\ this work & SF & N & A & A & - & - & - \\
\ \citet{wilhite} & SF & C & A & A & - & - & - \\
\ \citet{macleod10} & DRW & C & AA & AA & C & - & N \\
\ \citet{kelly13} & PSD & A & A & A & - & - & - \\
\ \citet{simm} & EV, PSD & N & A & A & N & N & N \\
\ \citet{caplar} & SF & C & - & AA & - & - & C \\
\ \citet{paula18} & SF & N & A & - & - & - & - \\
\ \citet{laurenti} & SF & C & A & AA & - & - & - \\
\bottomrule

\end{tabular}
\end{center}
\footnotesize{\textbf{Notes.} The table reports (\emph{left to right}): reference paper; method applied for the analysis; dependence of amplitude/timescale on $M_{BH}$, $\lambda_E$, and $L_{bol}$. \emph{C} = correlation (CC = strong); \emph{A} = anticorrelation (AA = strong); \emph{N} = no relation found; - = dependence not investigated.}
\end{table*}

As the other works in our series making use of VST data, this study can provide some forecasts for the LSST. The survey will include ultra-deep investigation of regions known as Deep-Drilling Fields (DDFs; they include the COSMOS area), widely surveyed in the past decades and therefore with valuable multiwavelength and spectroscopic coverage available, which makes them ideal for AGN science as well as training sets in the context of the main survey. The surveyed area will cover 9.6 sq. deg. per DDF. Based on the results from \citet{decicco21}, we expect to be able to identify at least $\approx 3,500$ AGN per DDF via optical variability combined with the colors that will be available for the LSST. 
We point out that this estimate depends on the characteristics of our VST-COSMOS survey; hence we should take into account that, while the single-visit depth is roughly the same for the two surveys, for the LSST we will have the chance to stack close visits together, as we expect observations with about a two-night cadence. This will return deeper images and, therefore, will allow us to identify fainter AGN and increase their number per square degree. We should also consider that the number of confirmed AGN per sq. deg. in the DDFs is expected to roughly double as the multiwavelength coverage returns other samples of AGN selected via other diagnostics. To give an idea, the estimated sky density limited to BLAGN and obtained from the XMM-SERVS survey of two other DDFs, namely W-CDF-S and ELAIS-S1, is $\approx200$ per sq. deg., which translates into $\approx 1,900$ BLAGN per DDF \citep{ni21}.

Such larger samples of AGN will allow us to improve our analysis. Indeed, our \emph{main} sample of 677 AGN reduced to 264 sources with available $M_{BH}$ and $\lambda_E$ estimates. This affected the whole analysis of the correlations of variability with AGN physical properties, as further selections in the space of the parameters under investigation led to scant samples of a few tens of sources, unsuitable for a statistically significant analysis. The LSST will therefore provide us with a suitable dataset for the exploration of multidimensional parameter dependence for the SF of AGN.

We mentioned that our light curves are characterized by two large gaps of one year and seven months plus eight months, as shown in Table \ref{tab:seasons}. From the table it is apparent that mean and median observed baseline values, computed for the sources in the \emph{main} sample for each season, are very close to the maximum observed baseline (when not exactly coincident with it) for the corresponding season. Similarly, the mean and median number of visits for each season are very close to the total number of visits (when not exactly coincident with it) for the corresponding season. This shows how, for individual seasons, our dataset can take advantage of a dense sampling, which plays a key role in the context of AGN detection efficiency, as shown in \citet{decicco19}.
Nevertheless, the two gaps affect the shape of our SF with inadequate sampling in correspondence with some timescales.
Sparse and/or irregular sampling is a very common issue in SF analysis \citep[e.g.,][]{dV03,peters,simm,sartori}. Indeed, there are works from the literature where the cadence is low but the sampling is regular: as an example, \citet{hawkins} uses quasar light curves from a long-term monitoring program with 24 yearly observations per source to investigate the origin of the emission mechanism in AGN. Nonetheless, one of the advantages in the use of the SF is its relative insensitivity to irregular sampling when sources are considered as an ensemble rather than individually \citep[e.g.,][]{hawkins07,kozlowski,sartori}. As mentioned in Section \ref{section:sf_six}, \citet{Bauer} analyze the effect of irregular sampling by means of simulations, and conclude that the turnover that is observed in the light curve is an effect of sparse sampling at longer timescales, and not a real feature in the SF of AGN. \citet{emmanoulopoulos} also resort to simulations in order to assess whether and to what extent the SF is robust against the presence of gaps in the light curves. They simulate a single light curve and then investigate the effect of three different gaps in the data, representing three different situations: almost periodic data gaps, dense and sparse sampling, and purely sparsely sampled data, corresponding to 57\%, 83\%, and 92\% of the data being removed from a single simulated light curve that is 2,000 time units long, respectively. They then use bootstrap to extract 1,000 light curves from each of the obtained light curves with gaps, and then compare the results obtained with and without gaps. While they find that the presence of gaps is responsible for the presence of wiggles and bends in the SF, from the right panels of their Figure 12 we can infer that these wiggles and bends do not alter the slope of the SF obtained from the light curve with no gaps. In this work we are not investigating the turnover as our baseline is not long enough (Section \ref{section:sf_six}); our analysis is instead focused on the linear region of the SF (and the possible dependence on physical quantities of interest), where the irregularity of the sampling does not constitute a major issue.

The gaps in our baseline are one of the reasons why we need long observing seasons and high-cadence observations for this kind of studies. For the LSST light curves we expect more regular sampling, and hence we expect the sampling issue to be a minor one. Indeed, for the DDFs a 2-day 2-filter cadence is planned \citep{ddf}. In addition, the 10 yr duration of the survey will allow us to probe longer timescales than our 3.3 yr (and to extend the redshift coverage), which has been done in some works from the literature, but not at these depths. In this way we can investigate the above-mentioned characteristic timescale $\tau$ where a turnover in the SF is generally observed, and explore its possible connections with $M_{BH}$ (see, e.g., \citealt{burke}) and other AGN physical properties.

Some of the above challenges can be overcome even without the Vera C. Rubin Observatory being operational. Indeed, a fourth season consisting of 13 observations of the COSMOS field with the VST was recently completed (ESO P108, Dec. 2021 - Mar 2022) and the data are currently being reduced. This can be combined with $\sim10$ yr of archival DECam monitoring of the central DDFs (which probably lacks the high cadence, but grants a long baseline and reasonable depth). This means that, while we wait for the Vera C. Rubin Observatory to be operational, we can take advantage of the data from this new observing season to extend our baseline from 3.3 yr to $\approx10$ yr, i.e., roughly the same baseline that we will have when the LSST will be completed. Including data from this new observing season means that the VST-COSMOS baseline will suffer from an additional, longer gap between its observing seasons; therefore we plan to investigate the effect of the presence of these gaps thoroughly in a forthcoming paper in this series. Nevertheless, the VST-COSMOS dataset still remains one of the few that, to date, can benefit from a high observing cadence in each of the observing seasons, together with a considerable depth. This will allow us to lead additional AGN studies focused on their optical variability, to pursue the above-mentioned goals in the context of SF analysis and -- last but not least -- it will favor the identification of fainter AGN, for which longer baselines are needed, as shown in, e.g., \citealt{decicco19}.\\

\begin{acknowledgements}
We thank D. J. Rosario and B. Trakhtenbrot for providing their catalog of black hole masses.
We acknowledge support from: ANID grants FONDECYT Postdoctorado Nº 3200222 (D.D.) and 3200250 (P.S.S.); PON R\&I 2021, CUP E65F21002880003 (D.D.); ANID - Millennium Science Initiative Program ICN12\_009 (F.E.B.); CATA-Basal AFB-170002 and FB210003 (F.E.B); FONDECYT Regular 1190818 and 1200495 (F.E.B.); NSF grant AST-2106990 and V.M. Willaman Endowment (W.N.B.); South Africa's DSI-NRF Grant Number 113121 \& 121291 (M.V.).
\end{acknowledgements}

\bibliographystyle{aa}
\bibliography{main}

\begin{thebibliography}{63}
\expandafter\ifx\csname natexlab\endcsname\relax\def\natexlab#1{#1}\fi

\bibitem[{{Aretxaga} \& {Terlevich}(1994)}]{a&t}
{Aretxaga}, I. \& {Terlevich}, R. 1994, \mnras, 269, 462

\bibitem[{{Bauer} {et~al.}(2009){Bauer}, {Baltay}, {Coppi}, {Ellman}, {Jerke},
  {Rabinowitz}, \& {Scalzo}}]{Bauer}
{Bauer}, A., {Baltay}, C., {Coppi}, P., {et~al.} 2009, \apj, 696, 1241

\bibitem[{{Brandt} {et~al.}(2018){Brandt}, {Ni}, {Yang}, {Anderson}, {Assef},
  {Barth}, {Bauer}, {Bongiorno}, {Chen}, {De Cicco}, {Gezari}, {Grier}, {Hall},
  {Hoenig}, {Lacy}, {Li}, {Luo}, {Paolillo}, {Peterson}, {Popovi{\'c}},
  {Richards}, {Shemmer}, {Shen}, {Sun}, {Timlin}, {Trump}, {Vito}, \&
  {Yu}}]{ddf}
{Brandt}, W.~N., {Ni}, Q., {Yang}, G., {et~al.} 2018, arXiv e-prints,
  arXiv:1811.06542

\bibitem[{Breiman(2001)}]{Breiman2001}
Breiman, L. 2001, Machine Learning, 45, 5, cited By 34434

\bibitem[{{Brusa} {et~al.}(2010){Brusa}, {Civano}, {Comastri}, {Miyaji},
  {Salvato}, {Zamorani}, {Cappelluti}, {Fiore}, {Hasinger}, {Mainieri},
  {Merloni}, {Bongiorno}, {Capak}, {Elvis}, {Gilli}, {Hao}, {Jahnke},
  {Koekemoer}, {Ilbert}, {Le Floc'h}, {Lusso}, {Mignoli}, {Schinnerer},
  {Silverman}, {Treister}, {Trump}, {Vignali}, {Zamojski}, {Aldcroft},
  {Aussel}, {Bardelli}, {Bolzonella}, {Cappi}, {Caputi}, {Contini},
  {Finoguenov}, {Fruscione}, {Garilli}, {Impey}, {Iovino}, {Iwasawa},
  {Kampczyk}, {Kartaltepe}, {Kneib}, {Knobel}, {Kovac}, {Lamareille},
  {Leborgne}, {Le Brun}, {Le Fevre}, {Lilly}, {Maier}, {McCracken}, {Pello},
  {Peng}, {Perez-Montero}, {de Ravel}, {Sanders}, {Scodeggio}, {Scoville},
  {Tanaka}, {Taniguchi}, {Tasca}, {de la Torre}, {Tresse}, {Vergani}, \&
  {Zucca}}]{Brusa}
{Brusa}, M., {Civano}, F., {Comastri}, A., {et~al.} 2010, \apj, 716, 348

\bibitem[{{Burke} {et~al.}(2021){Burke}, {Shen}, {Blaes}, {Gammie}, {Horne},
  {Jiang}, {Liu}, {McHardy}, {Morgan}, {Scaringi}, \& {Yang}}]{burke}
{Burke}, C.~J., {Shen}, Y., {Blaes}, O., {et~al.} 2021, Science, 373, 789

\bibitem[{{Capaccioli} \& {Schipani}(2011)}]{VST}
{Capaccioli}, M. \& {Schipani}, P. 2011, The Messenger, 146, 2

\bibitem[{{Caplar} {et~al.}(2017){Caplar}, {Lilly}, \& {Trakhtenbrot}}]{caplar}
{Caplar}, N., {Lilly}, S.~J., \& {Trakhtenbrot}, B. 2017, \apj, 834, 111

\bibitem[{{Collier} \& {Peterson}(2001)}]{cp01}
{Collier}, S. \& {Peterson}, B.~M. 2001, \apj, 555, 775

\bibitem[{{De Cicco} {et~al.}(2021){De Cicco}, {Bauer}, {Paolillo}, {Cavuoti},
  {S{\'a}nchez-S{\'a}ez}, {Brandt}, {Pignata}, {Vaccari}, \&
  {Radovich}}]{decicco21}
{De Cicco}, D., {Bauer}, F.~E., {Paolillo}, M., {et~al.} 2021, \aap, 645, A103

\bibitem[{{De Cicco} {et~al.}(2015){De Cicco}, {Paolillo}, {Covone}, {Falocco},
  {Longo}, {Grado}, {Limatola}, {Botticella}, {Pignata}, {Cappellaro},
  {Vaccari}, {Trevese}, {Vagnetti}, {Salvato}, {Radovich}, {Brandt},
  {Capaccioli}, {Napolitano}, \& {Schipani}}]{decicco15}
{De Cicco}, D., {Paolillo}, M., {Covone}, G., {et~al.} 2015, \aap, 574, A112

\bibitem[{{De Cicco} {et~al.}(2019){De Cicco}, {Paolillo}, {Falocco},
  {Poulain}, {Brandt}, {Bauer}, {Vagnetti}, {Longo}, {Grado}, {Ragosta},
  {Botticella}, {Pignata}, {Vaccari}, {Radovich}, {Salvato}, {Covone},
  {Napolitano}, {Marchetti}, \& {Schipani}}]{decicco19}
{De Cicco}, D., {Paolillo}, M., {Falocco}, S., {et~al.} 2019, \aap, 627, A33

\bibitem[{{de Vries} {et~al.}(2003){de Vries}, {Becker}, \& {White}}]{dV03}
{de Vries}, W.~H., {Becker}, R.~H., \& {White}, R.~L. 2003, \aj, 126, 1217

\bibitem[{{de Vries} {et~al.}(2005){de Vries}, {Becker}, {White}, \&
  {Loomis}}]{deVries05}
{de Vries}, W.~H., {Becker}, R.~H., {White}, R.~L., \& {Loomis}, C. 2005, \aj,
  129, 615

\bibitem[{{di Clemente} {et~al.}(1996){di Clemente}, {Giallongo}, {Natali},
  {Trevese}, \& {Vagnetti}}]{diClemente}
{di Clemente}, A., {Giallongo}, E., {Natali}, G., {Trevese}, D., \& {Vagnetti},
  F. 1996, \apj, 463, 466

\bibitem[{{Donley} {et~al.}(2012){Donley}, {Koekemoer}, {Brusa}, {Capak},
  {Cardamone}, {Civano}, {Ilbert}, {Impey}, {Kartaltepe}, {Miyaji}, {Salvato},
  {Sanders}, {Trump}, \& {Zamorani}}]{donley}
{Donley}, J.~L., {Koekemoer}, A.~M., {Brusa}, M., {et~al.} 2012, \apj, 748, 142

\bibitem[{{Emmanoulopoulos} {et~al.}(2010){Emmanoulopoulos}, {McHardy}, \&
  {Uttley}}]{emmanoulopoulos}
{Emmanoulopoulos}, D., {McHardy}, I.~M., \& {Uttley}, P. 2010, \mnras, 404, 931

\bibitem[{{Falocco} {et~al.}(2015){Falocco}, {Paolillo}, {Covone}, {De Cicco},
  {Longo}, {Grado}, {Limatola}, {Vaccari}, {Botticella}, {Pignata},
  {Cappellaro}, {Trevese}, {Vagnetti}, {Salvato}, {Radovich}, {Hsu},
  {Capaccioli}, {Napolitano}, {Brandt}, {Baruffolo}, {Cascone}, \&
  {Schipani}}]{falocco}
{Falocco}, S., {Paolillo}, M., {Covone}, G., {et~al.} 2015, \aap, 579, A115

\bibitem[{{Gonz{\'a}lez-Mart{\'\i}n} \& {Vaughan}(2012)}]{gonzalez-martin}
{Gonz{\'a}lez-Mart{\'\i}n}, O. \& {Vaughan}, S. 2012, \aap, 544, A80

\bibitem[{{Graham} {et~al.}(2014){Graham}, {Djorgovski}, {Drake}, {Mahabal},
  {Chang}, {Stern}, {Donalek}, \& {Glikman}}]{Graham}
{Graham}, M.~J., {Djorgovski}, S.~G., {Drake}, A.~J., {et~al.} 2014, \mnras,
  439, 703

\bibitem[{{Hawkins}(2002)}]{hawkins}
{Hawkins}, M.~R.~S. 2002, \mnras, 329, 76

\bibitem[{{Hawkins}(2007)}]{hawkins07}
{Hawkins}, M.~R.~S. 2007, \aap, 462, 581

\bibitem[{{Hook} {et~al.}(1994){Hook}, {McMahon}, {Boyle}, \& {Irwin}}]{Hook}
{Hook}, I.~M., {McMahon}, R.~G., {Boyle}, B.~J., \& {Irwin}, M.~J. 1994,
  \mnras, 268, 305

\bibitem[{{Ivezi{\'c}} {et~al.}(2019){Ivezi{\'c}}, {Kahn}, {Tyson}, {Abel},
  {Acosta}, {Allsman}, {Alonso}, {AlSayyad}, {Anderson}, {Andrew}, {Angel},
  {Angeli}, {Ansari}, {Antilogus}, {Araujo}, {Armstrong}, {Arndt}, {Astier},
  {Aubourg}, {Auza}, {Axelrod}, {Bard}, {Barr}, {Barrau}, {Bartlett}, {Bauer},
  {Bauman}, {Baumont}, {Bechtol}, {Bechtol}, {Becker}, {Becla}, {Beldica},
  {Bellavia}, {Bianco}, {Biswas}, {Blanc}, {Blazek}, {Bland ford}, {Bloom},
  {Bogart}, {Bond}, {Booth}, {Borgland}, {Borne}, {Bosch}, {Boutigny},
  {Brackett}, {Bradshaw}, {Brand t}, {Brown}, {Bullock}, {Burchat}, {Burke},
  {Cagnoli}, {Calabrese}, {Callahan}, {Callen}, {Carlin}, {Carlson}, {Chand
  rasekharan}, {Charles-Emerson}, {Chesley}, {Cheu}, {Chiang}, {Chiang},
  {Chirino}, {Chow}, {Ciardi}, {Claver}, {Cohen-Tanugi}, {Cockrum}, {Coles},
  {Connolly}, {Cook}, {Cooray}, {Covey}, {Cribbs}, {Cui}, {Cutri}, {Daly},
  {Daniel}, {Daruich}, {Daubard}, {Daues}, {Dawson}, {Delgado}, {Dellapenna},
  {de Peyster}, {de Val-Borro}, {Digel}, {Doherty}, {Dubois},
  {Dubois-Felsmann}, {Durech}, {Economou}, {Eifler}, {Eracleous}, {Emmons},
  {Fausti Neto}, {Ferguson}, {Figueroa}, {Fisher-Levine}, {Focke}, {Foss},
  {Frank}, {Freemon}, {Gangler}, {Gawiser}, {Geary}, {Gee}, {Geha}, {Gessner},
  {Gibson}, {Gilmore}, {Glanzman}, {Glick}, {Goldina}, {Goldstein}, {Goodenow},
  {Graham}, {Gressler}, {Gris}, {Guy}, {Guyonnet}, {Haller}, {Harris},
  {Hascall}, {Haupt}, {Hernand ez}, {Herrmann}, {Hileman}, {Hoblitt},
  {Hodgson}, {Hogan}, {Howard}, {Huang}, {Huffer}, {Ingraham}, {Innes},
  {Jacoby}, {Jain}, {Jammes}, {Jee}, {Jenness}, {Jernigan}, {Jevremovi{\'c}},
  {Johns}, {Johnson}, {Johnson}, {Jones}, {Juramy-Gilles}, {Juri{\'c}},
  {Kalirai}, {Kallivayalil}, {Kalmbach}, {Kantor}, {Karst}, {Kasliwal},
  {Kelly}, {Kessler}, {Kinnison}, {Kirkby}, {Knox}, {Kotov}, {Krabbendam},
  {Krughoff}, {Kub{\'a}nek}, {Kuczewski}, {Kulkarni}, {Ku}, {Kurita}, {Lage},
  {Lambert}, {Lange}, {Langton}, {Le Guillou}, {Levine}, {Liang}, {Lim},
  {Lintott}, {Long}, {Lopez}, {Lotz}, {Lupton}, {Lust}, {MacArthur}, {Mahabal},
  {Mand elbaum}, {Markiewicz}, {Marsh}, {Marshall}, {Marshall}, {May},
  {McKercher}, {McQueen}, {Meyers}, {Migliore}, {Miller}, {Mills}, {Miraval},
  {Moeyens}, {Moolekamp}, {Monet}, {Moniez}, {Monkewitz}, {Montgomery},
  {Morrison}, {Mueller}, {Muller}, {Mu{\~n}oz Arancibia}, {Neill}, {Newbry},
  {Nief}, {Nomerotski}, {Nordby}, {O'Connor}, {Oliver}, {Olivier}, {Olsen},
  {O'Mullane}, {Ortiz}, {Osier}, {Owen}, {Pain}, {Palecek}, {Parejko},
  {Parsons}, {Pease}, {Peterson}, {Peterson}, {Petravick}, {Libby Petrick},
  {Petry}, {Pierfederici}, {Pietrowicz}, {Pike}, {Pinto}, {Plante}, {Plate},
  {Plutchak}, {Price}, {Prouza}, {Radeka}, {Rajagopal}, {Rasmussen},
  {Regnault}, {Reil}, {Reiss}, {Reuter}, {Ridgway}, {Riot}, {Ritz}, {Robinson},
  {Roby}, {Roodman}, {Rosing}, {Roucelle}, {Rumore}, {Russo}, {Saha},
  {Sassolas}, {Schalk}, {Schellart}, {Schindler}, {Schmidt}, {Schneider},
  {Schneider}, {Schoening}, {Schumacher}, {Schwamb}, {Sebag}, {Selvy},
  {Sembroski}, {Seppala}, {Serio}, {Serrano}, {Shaw}, {Shipsey}, {Sick},
  {Silvestri}, {Slater}, {Smith}, {Smith}, {Sobhani}, {Soldahl},
  {Storrie-Lombardi}, {Stover}, {Strauss}, {Street}, {Stubbs}, {Sullivan},
  {Sweeney}, {Swinbank}, {Szalay}, {Takacs}, {Tether}, {Thaler}, {Thayer},
  {Thomas}, {Thornton}, {Thukral}, {Tice}, {Trilling}, {Turri}, {Van Berg},
  {Vanden Berk}, {Vetter}, {Virieux}, {Vucina}, {Wahl}, {Walkowicz}, {Walsh},
  {Walter}, {Wang}, {Wang}, {Warner}, {Wiecha}, {Willman}, {Winters},
  {Wittman}, {Wolff}, {Wood-Vasey}, {Wu}, {Xin}, {Yoachim}, \&
  {Zhan}}]{ivezich19}
{Ivezi{\'c}}, {\v{Z}}., {Kahn}, S.~M., {Tyson}, J.~A., {et~al.} 2019, \apj,
  873, 111

\bibitem[{{Ivezi{\'c}} \& {MacLeod}(2014)}]{ivezic13}
{Ivezi{\'c}}, {\v{Z}}. \& {MacLeod}, C. 2014, in Multiwavelength AGN Surveys
  and Studies, ed. A.~M. {Mickaelian} \& D.~B. {Sanders}, Vol. 304, 395--398

\bibitem[{{Kawaguchi} {et~al.}(1998){Kawaguchi}, {Mineshige}, {Umemura}, \&
  {Turner}}]{kawaguchi}
{Kawaguchi}, T., {Mineshige}, S., {Umemura}, M., \& {Turner}, E.~L. 1998, \apj,
  504, 671

\bibitem[{{Kelly} {et~al.}(2009){Kelly}, {Bechtold}, \&
  {Siemiginowska}}]{kelly09}
{Kelly}, B.~C., {Bechtold}, J., \& {Siemiginowska}, A. 2009, \apj, 698, 895

\bibitem[{{Kelly} {et~al.}(2013){Kelly}, {Treu}, {Malkan}, {Pancoast}, \&
  {Woo}}]{kelly13}
{Kelly}, B.~C., {Treu}, T., {Malkan}, M., {Pancoast}, A., \& {Woo}, J.-H. 2013,
  \apj, 779, 187

\bibitem[{{Kimura} {et~al.}(2020){Kimura}, {Yamada}, {Kokubo}, {Yasuda},
  {Morokuma}, {Nagao}, \& {Matsuoka}}]{kimura}
{Kimura}, Y., {Yamada}, T., {Kokubo}, M., {et~al.} 2020, \apj, 894, 24

\bibitem[{{Koz{\l}owski}(2016)}]{kozlowski}
{Koz{\l}owski}, S. 2016, \apj, 826, 118

\bibitem[{{Koz{\l}owski}(2017)}]{kozlowski17}
{Koz{\l}owski}, S. 2017, \aap, 597, A128

\bibitem[{{Laurenti} {et~al.}(2020){Laurenti}, {Vagnetti}, {Middei}, \&
  {Paolillo}}]{laurenti}
{Laurenti}, M., {Vagnetti}, F., {Middei}, R., \& {Paolillo}, M. 2020, \mnras,
  499, 6053

\bibitem[{{LSST Science Collaboration} {et~al.}(2009){LSST Science
  Collaboration}, {Abell}, {Allison}, {Anderson}, {Andrew}, {Angel}, {Armus},
  {Arnett}, {Asztalos}, {Axelrod}, \& et~al.}]{lsst}
{LSST Science Collaboration}, {Abell}, P.~A., {Allison}, J., {et~al.} 2009,
  ArXiv e-prints, arXiv:0912.0201

\bibitem[{{Lusso} {et~al.}(2012){Lusso}, {Comastri}, {Simmons}, {Mignoli},
  {Zamorani}, {Vignali}, {Brusa}, {Shankar}, {Lutz}, {Trump}, {Maiolino},
  {Gilli}, {Bolzonella}, {Puccetti}, {Salvato}, {Impey}, {Civano}, {Elvis},
  {Mainieri}, {Silverman}, {Koekemoer}, {Bongiorno}, {Merloni}, {Berta}, {Le
  Floc'h}, {Magnelli}, {Pozzi}, \& {Riguccini}}]{lusso12}
{Lusso}, E., {Comastri}, A., {Simmons}, B.~D., {et~al.} 2012, \mnras, 425, 623

\bibitem[{{Maccacaro} {et~al.}(1988){Maccacaro}, {Gioia}, {Wolter}, {Zamorani},
  \& {Stocke}}]{Maccacaro}
{Maccacaro}, T., {Gioia}, I.~M., {Wolter}, A., {Zamorani}, G., \& {Stocke},
  J.~T. 1988, \apj, 326, 680

\bibitem[{{MacLeod} {et~al.}(2010){MacLeod}, {Ivezi{\'c}}, {Kochanek},
  {Koz{\l}owski}, {Kelly}, {Bullock}, {Kimball}, {Sesar}, {Westman}, {Brooks},
  {Gibson}, {Becker}, \& {de Vries}}]{macleod10}
{MacLeod}, C.~L., {Ivezi{\'c}}, {\v{Z}}., {Kochanek}, C.~S., {et~al.} 2010,
  \apj, 721, 1014

\bibitem[{{MacLeod} {et~al.}(2012){MacLeod}, {Ivezi{\'c}}, {Sesar}, {de Vries},
  {Kochanek}, {Kelly}, {Becker}, {Lupton}, {Hall}, {Richards}, {Anderson}, \&
  {Schneider}}]{macleod12}
{MacLeod}, C.~L., {Ivezi{\'c}}, {\v{Z}}., {Sesar}, B., {et~al.} 2012, \apj,
  753, 106

\bibitem[{{Marchesi} {et~al.}(2016){Marchesi}, {Civano}, {Elvis}, {Salvato},
  {Brusa}, {Comastri}, {Gilli}, {Hasinger}, {Lanzuisi}, {Miyaji}, {Treister},
  {Urry}, {Vignali}, {Zamorani}, {Allevato}, {Cappelluti}, {Cardamone},
  {Finoguenov}, {Griffiths}, {Karim}, {Laigle}, {LaMassa}, {Jahnke}, {Ranalli},
  {Schawinski}, {Schinnerer}, {Silverman}, {Smolcic}, {Suh}, \&
  {Trakhtenbrot}}]{marchesi}
{Marchesi}, S., {Civano}, F., {Elvis}, M., {et~al.} 2016, \apj, 817, 34

\bibitem[{{McHardy} {et~al.}(2006){McHardy}, {Koerding}, {Knigge}, {Uttley}, \&
  {Fender}}]{mcHardy}
{McHardy}, I.~M., {Koerding}, E., {Knigge}, C., {Uttley}, P., \& {Fender},
  R.~P. 2006, \nat, 444, 730

\bibitem[{{Morganson} {et~al.}(2014){Morganson}, {Burgett}, {Chambers},
  {Green}, {Kaiser}, {Magnier}, {Marshall}, {Morgan}, {Price}, {Rix},
  {Schlafly}, {Tonry}, \& {Walter}}]{morganson14}
{Morganson}, E., {Burgett}, W.~S., {Chambers}, K.~C., {et~al.} 2014, \apj, 784,
  92

\bibitem[{{Ni} {et~al.}(2021){Ni}, {Brandt}, {Chen}, {Luo}, {Nyland}, {Yang},
  {Zou}, {Aird}, {Alexander}, {Bauer}, {Lacy}, {Lehmer}, {Mallick}, {Salvato},
  {Schneider}, {Tozzi}, {Traulsen}, {Vaccari}, {Vignali}, {Vito}, {Xue},
  {Banerji}, {Chow}, {Comastri}, {Del Moro}, {Gilli}, {Mullaney}, {Paolillo},
  {Schwope}, {Shemmer}, {Sun}, {Timlin}, \& {Trump}}]{ni21}
{Ni}, Q., {Brandt}, W.~N., {Chen}, C.-T., {et~al.} 2021, \apjs, 256, 21

\bibitem[{{Padovani} {et~al.}(2017){Padovani}, {Alexander}, {Assef}, {De
  Marco}, {Giommi}, {Hickox}, {Richards}, {Smol{\v c}i{\'c}}, {Hatziminaoglou},
  {Mainieri}, \& {Salvato}}]{padovani}
{Padovani}, P., {Alexander}, D.~M., {Assef}, R.~J., {et~al.} 2017, \aapr, 25, 2

\bibitem[{{Paolillo} {et~al.}(2017){Paolillo}, {Papadakis}, {Brandt}, {Luo},
  {Xue}, {Tozzi}, {Shemmer}, {Allevato}, {Bauer}, {Comastri}, {Gilli},
  {Koekemoer}, {Liu}, {Vignali}, {Vito}, {Yang}, {Wang}, \&
  {Zheng}}]{paolillo17}
{Paolillo}, M., {Papadakis}, I., {Brandt}, W.~N., {et~al.} 2017, \mnras, 471,
  4398

\bibitem[{{Peters} {et~al.}(2015){Peters}, {Richards}, {Myers}, {Strauss},
  {Schmidt}, {Ivezi{\'c}}, {Ross}, {MacLeod}, \& {Riegel}}]{peters}
{Peters}, C.~M., {Richards}, G.~T., {Myers}, A.~D., {et~al.} 2015, \apj, 811,
  95

\bibitem[{{Poulain} {et~al.}(2020){Poulain}, {Paolillo}, {De Cicco}, {Brandt},
  {Bauer}, {Falocco}, {Vagnetti}, {Grado}, {Ragosta}, {Botticella},
  {Cappellaro}, {Pignata}, {Vaccari}, {Schipani}, {Covone}, {Longo}, \&
  {Napolitano}}]{poulain}
{Poulain}, M., {Paolillo}, M., {De Cicco}, D., {et~al.} 2020, \aap, 634, A50

\bibitem[{{Rakshit} {et~al.}(2020){Rakshit}, {Stalin}, \&
  {Kotilainen}}]{rakshit}
{Rakshit}, S., {Stalin}, C.~S., \& {Kotilainen}, J. 2020, \apjs, 249, 17

\bibitem[{{Rengstorf} {et~al.}(2006){Rengstorf}, {Brunner}, \&
  {Wilhite}}]{rengstorf}
{Rengstorf}, A.~W., {Brunner}, R.~J., \& {Wilhite}, B.~C. 2006, \aj, 131, 1923

\bibitem[{{Rosario} {et~al.}(2013){Rosario}, {Trakhtenbrot}, {Lutz}, {Netzer},
  {Trump}, {Silverman}, {Schramm}, {Lusso}, {Berta}, {Bongiorno}, {Brusa},
  {F{\"o}rster-Schreiber}, {Genzel}, {Lilly}, {Magnelli}, {Mainieri},
  {Maiolino}, {Merloni}, {Mignoli}, {Nordon}, {Popesso}, {Salvato}, {Santini},
  {Tacconi}, \& {Zamorani}}]{rosario}
{Rosario}, D.~J., {Trakhtenbrot}, B., {Lutz}, D., {et~al.} 2013, \aap, 560, A72

\bibitem[{{S{\'a}nchez} {et~al.}(2017){S{\'a}nchez}, {Lira}, {Cartier},
  {P{\'e}rez}, {Miranda}, {Yovaniniz}, {Ar{\'e}valo}, {Milvang-Jensen},
  {Fynbo}, {Dunlop}, {Coppi}, \& {Marchesi}}]{paula17}
{S{\'a}nchez}, P., {Lira}, P., {Cartier}, R., {et~al.} 2017, \apj, 849, 110

\bibitem[{{S{\'a}nchez-S{\'a}ez} {et~al.}(2018){S{\'a}nchez-S{\'a}ez}, {Lira},
  {Mej{\'\i}a-Restrepo}, {Ho}, {Ar{\'e}valo}, {Kim}, {Cartier}, \&
  {Coppi}}]{paula18}
{S{\'a}nchez-S{\'a}ez}, P., {Lira}, P., {Mej{\'\i}a-Restrepo}, J., {et~al.}
  2018, \apj, 864, 87

\bibitem[{{Sartori} {et~al.}(2019){Sartori}, {Trakhtenbrot}, {Schawinski},
  {Caplar}, {Treister}, \& {Zhang}}]{sartori}
{Sartori}, L.~F., {Trakhtenbrot}, B., {Schawinski}, K., {et~al.} 2019, \apj,
  883, 139

\bibitem[{{Schulze} {et~al.}(2018){Schulze}, {Silverman}, {Kashino}, {Akiyama},
  {Schramm}, {Sanders}, {Kartaltepe}, {Daddi}, {Rodighiero}, {Renzini},
  {Arimoto}, {Nagao}, {Puglisi}, {Trakhtenbrot}, {Civano}, \& {Suh}}]{Schulze}
{Schulze}, A., {Silverman}, J.~D., {Kashino}, D., {et~al.} 2018, \apjs, 239, 22

\bibitem[{{Scoville} {et~al.}(2007){Scoville}, {Aussel}, {Brusa}, {Capak},
  {Carollo}, {Elvis}, {Giavalisco}, {Guzzo}, {Hasinger}, {Impey}, {Kneib},
  {LeFevre}, {Lilly}, {Mobasher}, {Renzini}, {Rich}, {Sanders}, {Schinnerer},
  {Schminovich}, {Shopbell}, {Taniguchi}, \& {Tyson}}]{scoville07b}
{Scoville}, N., {Aussel}, H., {Brusa}, M., {et~al.} 2007, \apjs, 172, 1

\bibitem[{{Serafinelli} {et~al.}(2017){Serafinelli}, {Vagnetti}, {Chiaraluce},
  \& {Middei}}]{serafinelli17}
{Serafinelli}, R., {Vagnetti}, F., {Chiaraluce}, E., \& {Middei}, R. 2017,
  Frontiers in Astronomy and Space Sciences, 4, 21

\bibitem[{{Shakura} \& {Sunyaev}(1976)}]{shakura_sunyaev}
{Shakura}, N.~I. \& {Sunyaev}, R.~A. 1976, \mnras, 175, 613

\bibitem[{{Simm} {et~al.}(2016){Simm}, {Salvato}, {Saglia}, {Ponti},
  {Lanzuisi}, {Trakhtenbrot}, {Nandra}, \& {Bender}}]{simm}
{Simm}, T., {Salvato}, M., {Saglia}, R., {et~al.} 2016, \aap, 585, A129

\bibitem[{{Simonetti} {et~al.}(1985){Simonetti}, {Cordes}, \&
  {Heeschen}}]{Simonetti}
{Simonetti}, J.~H., {Cordes}, J.~M., \& {Heeschen}, D.~S. 1985, \apj, 296, 46

\bibitem[{{Suberlak} {et~al.}(2021){Suberlak}, {Ivezi{\'c}}, \&
  {MacLeod}}]{suberlak}
{Suberlak}, K.~L., {Ivezi{\'c}}, {\v{Z}}., \& {MacLeod}, C. 2021, \apj, 907, 96

\bibitem[{{Vagnetti} {et~al.}(2016){Vagnetti}, {Middei}, {Antonucci},
  {Paolillo}, \& {Serafinelli}}]{vagnetti16}
{Vagnetti}, F., {Middei}, R., {Antonucci}, M., {Paolillo}, M., \&
  {Serafinelli}, R. 2016, \aap, 593, A55

\bibitem[{{Vanden Berk} {et~al.}(2004){Vanden Berk}, {Wilhite}, {Kron},
  {Anderson}, {Brunner}, {Hall}, {Ivezi{\'c}}, {Richards}, {Schneider}, {York},
  {Brinkmann}, {Lamb}, {Nichol}, \& {Schlegel}}]{vdB}
{Vanden Berk}, D.~E., {Wilhite}, B.~C., {Kron}, R.~G., {et~al.} 2004, \apj,
  601, 692

\bibitem[{{Wilhite} {et~al.}(2008){Wilhite}, {Brunner}, {Grier}, {Schneider},
  \& {vanden Berk}}]{wilhite}
{Wilhite}, B.~C., {Brunner}, R.~J., {Grier}, C.~J., {Schneider}, D.~P., \&
  {vanden Berk}, D.~E. 2008, \mnras, 383, 1232

\bibitem[{{York} {et~al.}(2000){York}, {Adelman}, {Anderson}, {Anderson},
  {Annis}, {Bahcall}, {Bakken}, {Barkhouser}, {Bastian}, {Berman}, {Boroski},
  {Bracker}, {Briegel}, {Briggs}, {Brinkmann}, {Brunner}, {Burles}, {Carey},
  {Carr}, {Castander}, {Chen}, {Colestock}, {Connolly}, {Crocker}, {Csabai},
  {Czarapata}, {Davis}, {Doi}, {Dombeck}, {Eisenstein}, {Ellman}, {Elms},
  {Evans}, {Fan}, {Federwitz}, {Fiscelli}, {Friedman}, {Frieman}, {Fukugita},
  {Gillespie}, {Gunn}, {Gurbani}, {de Haas}, {Haldeman}, {Harris}, {Hayes},
  {Heckman}, {Hennessy}, {Hindsley}, {Holm}, {Holmgren}, {Huang}, {Hull},
  {Husby}, {Ichikawa}, {Ichikawa}, {Ivezi{\'c}}, {Kent}, {Kim}, {Kinney},
  {Klaene}, {Kleinman}, {Kleinman}, {Knapp}, {Korienek}, {Kron}, {Kunszt},
  {Lamb}, {Lee}, {Leger}, {Limmongkol}, {Lindenmeyer}, {Long}, {Loomis},
  {Loveday}, {Lucinio}, {Lupton}, {MacKinnon}, {Mannery}, {Mantsch}, {Margon},
  {McGehee}, {McKay}, {Meiksin}, {Merelli}, {Monet}, {Munn}, {Narayanan},
  {Nash}, {Neilsen}, {Neswold}, {Newberg}, {Nichol}, {Nicinski}, {Nonino},
  {Okada}, {Okamura}, {Ostriker}, {Owen}, {Pauls}, {Peoples}, {Peterson},
  {Petravick}, {Pier}, {Pope}, {Pordes}, {Prosapio}, {Rechenmacher}, {Quinn},
  {Richards}, {Richmond}, {Rivetta}, {Rockosi}, {Ruthmansdorfer}, {Sandford},
  {Schlegel}, {Schneider}, {Sekiguchi}, {Sergey}, {Shimasaku}, {Siegmund},
  {Smee}, {Smith}, {Snedden}, {Stone}, {Stoughton}, {Strauss}, {Stubbs},
  {SubbaRao}, {Szalay}, {Szapudi}, {Szokoly}, {Thakar}, {Tremonti}, {Tucker},
  {Uomoto}, {Vanden Berk}, {Vogeley}, {Waddell}, {Wang}, {Watanabe},
  {Weinberg}, {Yanny}, {Yasuda}, \& {SDSS Collaboration}}]{sdss}
{York}, D.~G., {Adelman}, J., {Anderson}, John~E., J., {et~al.} 2000, \aj, 120,
  1579

\bibitem[{{Zu} {et~al.}(2013){Zu}, {Kochanek}, {Koz{\l}owski}, \&
  {Udalski}}]{zu}
{Zu}, Y., {Kochanek}, C.~S., {Koz{\l}owski}, S., \& {Udalski}, A. 2013, \apj,
  765, 106

\end{thebibliography}

\end{document}